\documentclass[12pt]{iopart}

\usepackage{amsmath}
\usepackage{parskip}
\usepackage{graphicx}
\usepackage{booktabs}
\usepackage[colorlinks=true, allcolors=blue]{hyperref}
\usepackage{framed}
\usepackage{caption}
\usepackage{comment}
\usepackage{soul}

\begin{document}

    \title[The Interplay of Polar and Nematic Order in Active Matter]{The Interplay of Polar and Nematic Order in Active Matter: Implications for Non-Equilibrium Physics and Biology}

\author{Varun Venkatesh, Niels de Graaf Sousa, \\and Amin Doostmohammadi}
\address{Niels Bohr Institute, University of Copenhagen, 2100 Copenhagen, Denmark}
\ead{doostmohammadi@nbi.ku.dk}
\vspace{10pt}

\begin{indented}
\item[]December 2024
\end{indented}

\begin{abstract}
Active matter has played a pivotal role in advancing our understanding of non-equilibrium systems, leading to a fundamental shift in the study of biophysical phenomena. The foundation of active matter research is built on assumptions regarding the symmetry of microscopic constituents. While these assumptions have been validated extensively, instances of mixed or joint symmetries are prevalent in biological systems. This review explores the coexistence of polar and nematic order in active matter, emphasizing the theoretical and experimental challenges associated with these systems. By integrating insights from recent studies, we highlight the importance of considering mixed symmetries to accurately describe biological processes. This exploration not only benefits the field of biology but could also open new horizons for non-equilibrium physics, offering a comprehensive framework for understanding complex behavior in active matter.
\end{abstract}

%
%
\submitto{\JPA}
%
%
%
\section{Introduction}
Active matter represents a transformative paradigm in the study of nonequilibrium systems, describing a diverse class of materials powered by internal energy dissipation. These systems derive their unique properties from their ability to autonomously convert energy from the environment into mechanical work, resulting in complex behavior such as collective motion, pattern formation, and self-organization~\cite{Pedley_kessler,lauga2009hydrodynamics}. Unlike equilibrium systems, which are governed by thermodynamic laws and strive to minimize free energy, active systems remain far from equilibrium, generating novel emergent phenomena that challenge traditional frameworks in physics ~\cite{bechinger2016active,gompper20202020,marchetti_hydrodynamics_2013,Prost_nature_2015,Amin_Active_Nematics_2018,Chat2020}. This growing field draws inspiration from biology, where active processes are fundamental to life, while also offering opportunities for designing synthetic materials with unprecedented functionalities~\cite{bechinger2016active,gompper20202020,needleman2017active}.

A central organizing principle in active matter is symmetry, which governs the interactions and dynamical behavior of its constituent elements~\cite{Symmetry_Theormodynamics_2022_Marchetti}. Symmetry considerations not only define the structural properties of these systems but also determine their macroscopic manifestations, such as collective flows and defect structures. Among the symmetries relevant to active matter, polar and nematic symmetries stand out for their ubiquity and significance. 

Polar active matter refers to systems where constituents possess a well-defined direction, breaking head-tail symmetry and enabling motion along a preferred direction, with examples ranging from bird flocks and fish schools to synthetic self-propelled colloids and bacteria ~\cite{Kilobots_robotic_swarm,Zttl2016,vischek_1995,Dombrowski_2004}. In contrast, nematic active matter describes systems where constituents are head-tail symmetric, interacting through their relative orientation without directional preference. Nematic symmetry is observed in systems ranging from cytoskeletal filaments and liquid crystals to suspensions of rod-like particles ~\cite{Tim_Sanchez_Microtubule_Motors,RuiZhang_2018_Actin_Nematic}.

These symmetry classes showcase disparities in critical phenomena such as flocking ~\cite{vischek_1995,Flocking_Chiral_Activa_Matter_Benno_Liebchen_2017,toner2024physics}, swarming ~\cite{Swarming_Buhl_2006}, motility induced phase separation ~\cite{Cates2015}, active turbulence ~\cite{Dombrowski_2004}, defect dynamics ~\cite{Giomi_2014_defect_dynamics} and spontaneous flows ~\cite{Spontaneous_Flow_Patterns_J_Yeomans_2009,Wioland_2013}, which are pivotal for understanding processes such as tissue morphogenesis and cellular reorganization ~\cite{Thuan_Saw_Tissue_Nematics,BenoitLadoux_TissueMorphogenesis_Active_Nematic}. In this vein it is important to note that many cellular systems including bacteria and epithelial cells are inherently endowed with polar order associated with their motility. Yet, the phenomenology of active turbulence and spontaneous flow generation are often modelled and characterized using continuum equations of active nematics~\cite{Energy_spectra_3D_bacterial_suspensions,Spontaneous_flow_nem}. Furthermore, despite the polar symmetry, from the topological aspect, collection of such motile cells show half-integer topological defects (see Box. 1) that is typically characteristic of nematic systems (lowest energy defects in polar systems are of full-integer charge). Resolving this discrepancy requires detailed analyses of the cross-talk between polar and nematic symmetries in living matter.

While the study of polar and nematic active matter has yielded significant insights, real-world systems often exhibit a coexistence of these symmetries. Biological tissues, for example, frequently combine polar migratory behavior of cells with nematic alignment within the cytoskeleton ~\cite{Pedley_kessler,CellMigration_Doxzen_2013,cell_extrusion}. In synthetic systems, polar and nematic components can coexist dynamically, as in the case of hybrid colloidal assemblies or microtubule-motor complexes ~\cite{Coexisting_ordered_states_Frey}. Despite their prevalence, these hybrid systems remain poorly understood, as most studies focus on polar or nematic systems in isolation, neglecting the rich interplay that arises when both symmetries are present simultaneously.

The interplay between polar and nematic symmetries introduces a host of novel phenomena that demand a unified theoretical and experimental framework. For instance, transitions between polar and nematic phases may occur spontaneously in response to changes in density, activity, or confinement. Topological defects in nematic systems, such as $+1/2$ and $-1/2$ disclinations, acquire unique dynamics when coupled with polar fields, influencing processes like cell ordering, morphogenesis and cell extrusion ~\cite{ruider2024topological,PauGuillamatIntegerTopologicalDefectsTissueMorphogenesis,EpithelialTissue_Carlos_2019}. Conversely, polar systems can exhibit nematic-like alignment under dense or anisotropic conditions, blurring the boundaries between these symmetry classes ~\cite{Symmetry_Breaking_Polar_Nem_Frey,meacock2021bacteria}.

Recent theoretical models provide a foundation for exploring these coupled systems. Continuum descriptions that incorporate both polar and nematic order parameters offer insights into defect interactions, collective modes, and symmetry-breaking transitions ~\cite{Coexisting_ordered_states_Frey,Amiri_2022}. These models also reveal how gradients in activity or external fields can stabilize hybrid states, enabling precise control over symmetry interplay. Experimental advancements, including high-resolution imaging and micromanipulation techniques, now allow direct observation of these phenomena in living tissues and synthetic systems. For example, epithelial monolayers and bacterial aggregates display macroscopic nematic order alongside polar motility at the single-cell level, highlighting the multiscale nature of these interactions ~\cite{cell_extrusion,han2023local,wheeler2024individual}.

Beyond its scientific implications, understanding polar-nematic coexistence has transformative potential for applications. In active metamaterials, combining polar and nematic components could lead to tunable properties such as adaptive rigidity, programmable flow, and dynamic reconfigurability~\cite{Metamaterials}. In biological contexts, insights into symmetry interplay could inform advances in tissue engineering, cancer treatment, and the design of biomimetic systems. Similarly, integrating polar and nematic principles in robotics and soft matter could enable the creation of multifunctional devices with self-healing, shape-changing, or locomotory capabilities~\cite{robotics1,Harrison2022}.

This review aims to provide a concise overview of polar and nematic active matter, emphasizing their coexistence and coupling. We begin by summarizing the fundamental principles underlying these symmetry classes and their individual behavior. We start by discussing discrete models of individual polar particles together with studies of emergent polar and nematic order in discrete models of collectives of active agents. We then culminate on the continuum theories of polar and nematic active materials, discussing their common and distinct features as well as their implications for experimental observations. Next, we explore recent efforts to model and understand their interplay, with a focus on theoretical frameworks and experimental realizations. Finally, we discuss open challenges and future directions, highlighting the potential for a unified perspective to drive innovations across physics, biology, and engineering.

\newpage
\begin{framed}

\textbf{Box 1}
\label{box:defects}
\begin{center}
    \includegraphics[width=0.95\linewidth]{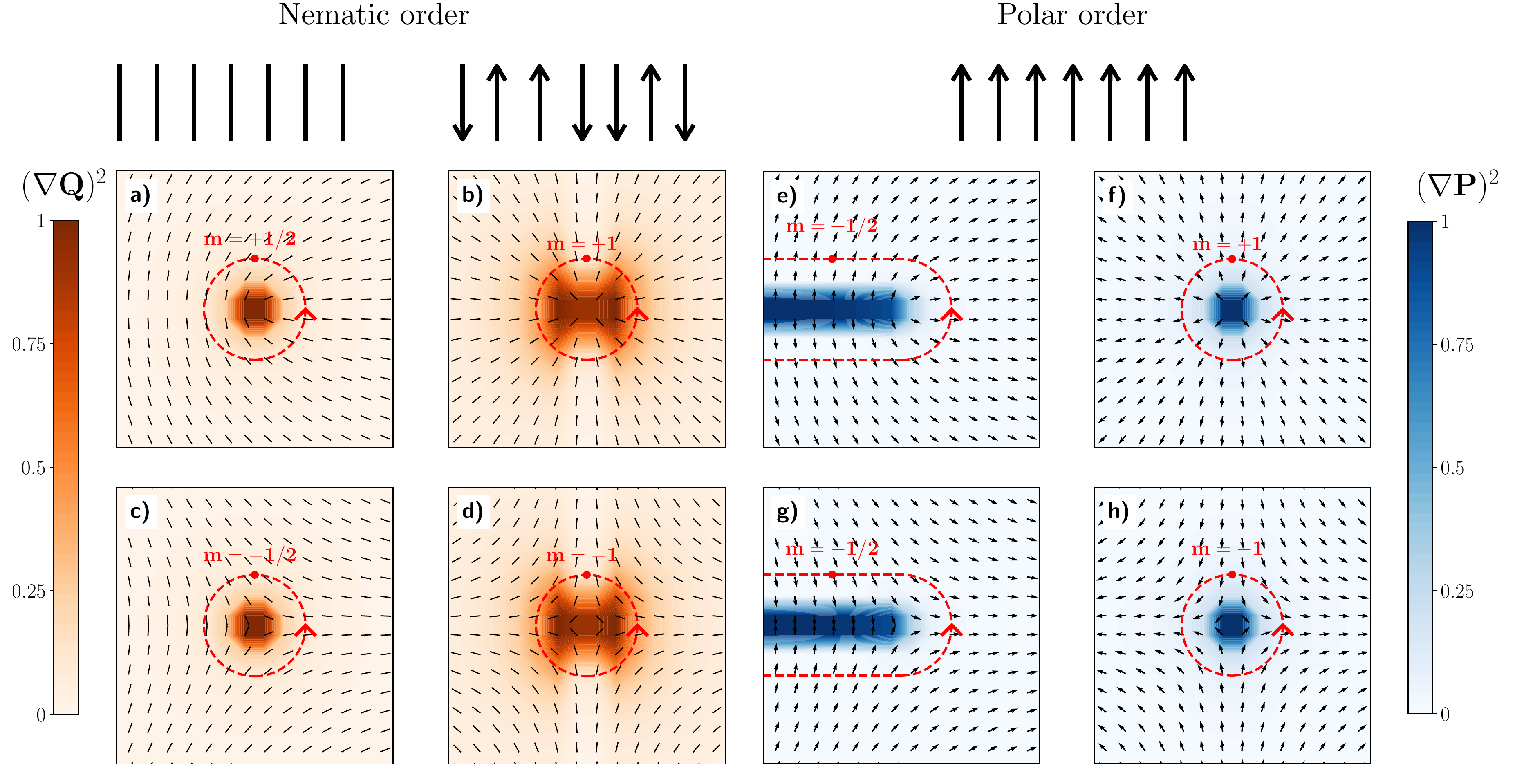}
\end{center}
\noindent\textbf{Figure: Polar and nematic order and their associated topological defects.} Shown here are half and full integer defects for the nematic director and polarization with normalized elastic deformation energy. Panels (a-d) show defect structures and associated energetic cost for nematic molecules. Panels (e-h) show the same structures and cost for polar molecules. The topological charge $m$, highlighted in red, is determined as multiples of $\pi$, corresponding to the angle a particle rotates along a closed contour around the defect. This figure shows energetically why polar active matter displays integer defects while nematic display half-integer.

Topological defects play a key role in active polar and nematic matter~\cite{Shankar_2022}. They are defined as singularities in the orientations fields and have been shown to play an important role in biological functions, from driving active nematic turbulence~\cite{pearce2024topologicaldefectsleadenergy}, to playing a central role in morphogenesis or apoptotic extrusion~\cite{cell_extrusion,PauGuillamatIntegerTopologicalDefectsTissueMorphogenesis}. Key in many of their functions is the elastic energy they intrinsically possess. The energy in a defect is directly related to the symmetry of underlying molecules. Take for example a polar molecule attempting to exhibit a half-integer defect; there is a large energetic cost directly due to not having a head-tail symmetry. Molecules with nematic symmetry do not have such a problem. The emergent defect structure is therefore directly a result of the underlying symmetry as a means to minimize energetic cost.
\end{framed}

\section{Polar and nematic active particles at the discrete level}

\subsection{Polar and nematic activity at the individual particle level}
We first discuss discrete models without momentum conservation, followed by individual-based models that account for hydrodynamic flows.

\begin{figure}[h]
    \centering
    \includegraphics[width=0.9\linewidth]{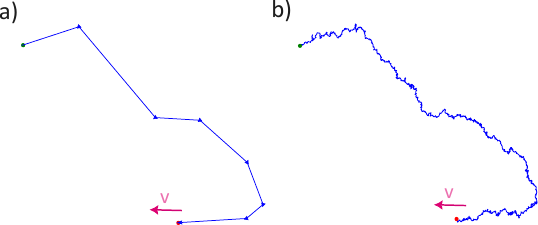}
    \caption{{\bf Self-propulsion mechanisms of individual active particles.} Schematics depicting typical trajectories of (a) run and tumble particles (R$\&$T) and (b) active Brownian particles (ABP)~\cite{Cates2015}. Run and tumble particles undergo sharp abrupt turns at a rate $\alpha$. Panel (a) depicts these changes through arrows along the particles velocity vector that abruptly change though the trajectory. By contrast, active Brownian particles turn through rotational diffusion, as shown in panel (b) which follows the same rough trajectory as (a) however without sharp changes instead diffusely.  On short time scales there is an apparent distinction of particles traversing the same path, while at large time scales both are equivalent.}
    \label{fig:singleparticle1}
\end{figure}

\subsubsection{Active particles without momentum conservation}
~\\
In systems without momentum conservation, the modeling of active particles typically involves solving for the trajectory and orientation of individual particles or collections of particles through stochastic ordinary differential equations. The position of the center of mass evolves according to a Newtonian or Langevin equation, which accounts for both deterministic motion and random fluctuations. These systems are often modeled as self-propelled particles that move in a direction defined by their orientation, with a set of rules determining how this orientation changes over time. This self-propulsion direction defines the polarity of individual particles.

The orientation dynamics can be driven by internal mechanisms (such as intrinsic alignment or rotational diffusion) or external interactions (such as alignment due to neighboring particles or environmental factors). The most common model for particles in these systems is the active Brownian particle (ABP) model~\cite{Erdmann2000}, where each particle experiences persistent motion in the direction of its orientation, which changes due to rotational diffusion or discrete reorientation events, such as those seen in run-and-tumble models~\cite{cates_diffusive_2012} (see Fig.~\ref{fig:singleparticle1}).

We consider a single particle that utilizes energy from its environment and converts it into motion. In discrete models the inherent polarity of the particles manifests in their direction of motion. A significant characteristic of most active matter systems is the presence of random fluctuations in the motion of individual active particles. This randomness can stem from various sources, such as environmental factors or internal fluctuations due to the intrinsic stochasticity of the processes driving motion. The randomness can be accounted for by introducing stochastic forces in the modeling, converting a simple ordinary differential equation into a stochastic one.

To capture these dynamics, a generic version of a stochastic active particle equation is commonly expressed as a pair of Langevin equations describing the motion of a particle~\cite{Basu2018}. The first equation represents the overdamped motion of the particle’s velocity:

\begin{equation} \label{eq: 1}
\dot{\mathbf{r}}(t) = f(\mathbf{v}, \theta, t) + \xi(t) \quad \text{with} \quad \langle \xi_i(t) \xi_j(t') \rangle = 2D_t \delta_{ij} \delta(t - t').
\end{equation}

Here, the white noise term $\xi(t)$ is delta-correlated, representing the random fluctuations in the motion of the particle. The second equation governs the mechanisms that cause the particle to change its direction. This is where the distinction between Active Brownian particles (ABPs) and Run-and-Tumble particles (R$\&$T) becomes apparent. Active Brownian particles undergo rotational diffusion described by:

\begin{equation}
\dot{\theta}(t) = \sqrt{2D_r} \eta_{\theta}(t) \quad \text{with} \quad \langle \eta_{\theta}(t) \eta_{\theta}(t') \rangle = \delta(t - t'),
\end{equation}

where $ \eta_{\theta}(t)$ is a Gaussian white noise term governing rotational diffusion with a diffusion constant $D_r$. In contrast, Run-and-Tumble particles undergo discrete tumbling events at a rate $\alpha$, and the mean time between two tumbles follows an exponential distribution $P(t) = \alpha e^{-\alpha t}$~\cite{cates_diffusive_2012,tailleur_statistical_2008}.

The clear distinction between these two types of models is that ABP orientations varies continuously while R$\&$T particles change abruptly. However, they employ similar styles with fixed velocity magnitudes to update positions and only changing directions. It is possible however, to consider fluctuations in the velocity magnitude  that are Gaussian white correlated $\langle v_{i\alpha}(t)v_{j\beta}(0) \rangle= \delta_{ij}\delta_{\alpha \beta} \Gamma(t)$. Where $\Gamma(t) = D \frac{e^{-|t|/\tau}}{\tau}$ and $\tau$ is a persistence time. Ornstein-Uhlenbeck processes are examples for this, were the velocity is obtained from 
$$
\tau \dot{\mathbf{v}}_i = -\mathbf{v}_i + \sqrt{2D} \, \eta_i,
$$
with $\eta_i$ being a zero-mean unit-variance Gaussian white noise controlled with noise magnitude D~\cite{ Fodor2016, Bonilla2019, Martin2021, Maggi2015, Uhlenbeck1930}. Such an approach has been termed active Ornstein-Uhlenbeck particles (AOUP) and has been an area of active research recently with successes in modeling the motion of passive tracers in bacterial suspensions~\cite{Koumakis2014} as well as for the collective dynamics of cells~\cite{Deforet2014, Hakim2017}.

All models are equivalent in the relevant limit; for example, in the limit of vanishing persistence time $\tau$, AOUP recovers the equations for Brownian motion. Although ABP and R $\&$ T models differ in their dynamics, they are known to exhibit equivalent long-term behavior. However, in the short term, these differences can manifest in interesting ways, such as in the separation of particles in mazes, as observed in experiments~\cite{Khatami2016}. For further reading, see~\cite{Romanczuk2012}.

While these models are typically applied to particles with polar symmetry—where all particles align their motion along a defined orientation—the mathematical framework can be generalized to include particles with nematic or even higher-order symmetries, such as $p$-atic symmetries, where $p=2$ and $p=6$ are nematic and hexatic symmetries. These extensions open up new avenues of research, particularly in the study of complex collective phenomena that emerge in systems with more intricate symmetry structures.

\subsubsection{Active particles with momentum conservation}
\begin{figure}[h]
    \centering
    \includegraphics[width=0.8\linewidth]{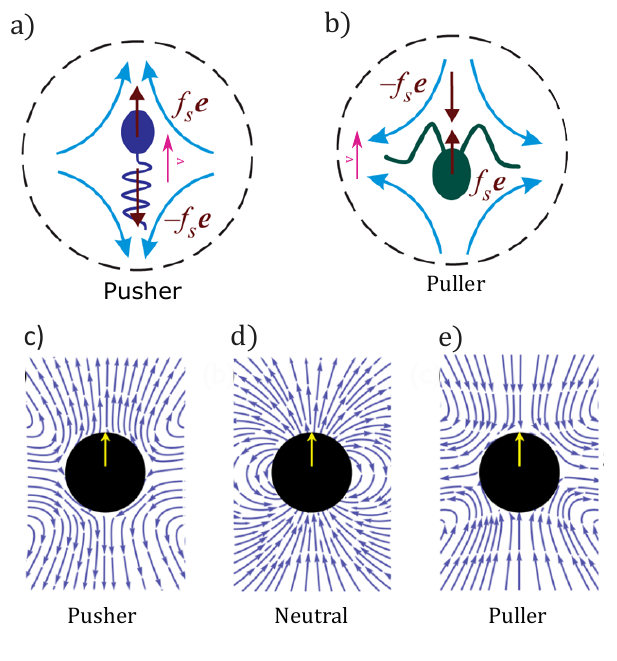}
    \caption{{\bf Self-propulsion mechanisms of active particles with momentum conservation.} Schematics depicting the flow fields generated by active particles interacting with the fluid, which play a significant role in their collective dynamics. Pullers (a) and pushers (b) exhibit distinct flow patterns that influence their interactions with each other and the surrounding fluid~\cite{Mirzakhanloo2020}. The particle moves in the direction of the pink arrow, with the fluid motion shown by the blue. One pole is located at the centre of the particle and the that other is either behind (pusher) or ahead (pusher) of the particle.  Squirmers can either generate surface flows for pushers (c), neutral (d) or pullers (e)~\cite{Evans2011}. Here, the particle moves along the arrow with puller and pushers having different flows. Figures adapted from~\cite{Mirzakhanloo2020} and~\cite{Evans2011}}
    \label{fig:singleparticle2}
\end{figure}

~\\
In the presence of momentum conservation the effects of hydrodynamics should be incorporated, where the motion of the particles generates a flow in the surrounding medium. In this class of systems, the particle dynamics are governed not only by the self-propulsion of the particles but also by the fluid dynamics of the surrounding environment, leading to momentum transfer between the particles and the fluid. The swimming mechanism in microswimmers involves non-reciprocal movements that, when averaged over time, create flow fields in the surrounding medium. In biological systems, such as bacterial swimmers or algae cells, these movements are typically driven by hair-like organelles called cilia or flagella. The movement of these organelles generates complex fluid flow patterns, which can be approximated at large distances by a dipolar force distribution~\cite{Higdon1979} that has a nematic symmetry. This means that microswimmers generate a flow in the surrounding fluid that is characterized by either a push or pull, depending on the nature of the force distribution~\cite{lauga2009hydrodynamics} (see Fig.~\ref{fig:singleparticle2}a,b).

Swimmers are categorized into two main types based on the nature of the forces they produce: pushers and pullers, Fig.~\ref{fig:singleparticle2}a,b, respectively. Pusher particles, like \textit{Escherichia coli} or \textit{Pseudomonas
aeruginosa} bacteria, expel fluid from their sides to the front and rear, creating a repulsive force that drives the swimmer forward~\cite{Deng2023}. In contrast, puller particles, such as \textit{Chlamydomonas reinhardtii}, generate forces that draw fluid from the front and rear of the swimmer, leading to an attractive flow that propels the swimmer forward~\cite{lauga2009hydrodynamics}. These flow patterns have analogues in the behaviour of nematic active fluids, where extensile and contractile properties define the collective behaviour of the system. 

One of the most powerful and widely used frameworks for studying active particles in the presence of momentum conservation is the squirmer model, first introduced by Lighthill~\cite{Lighthill1952}. Squiermers are often used to model microorganisms and artificial microswimmers, whose motion is influenced by both internal propulsion mechanisms and external hydrodynamic interactions. To capture these hydrodynamic effects, a pair of Langevin equations is often used~\cite{Pessot2018, Makino2004, Reinken2024}. A major contribution to the forces in these equations is from the surface velocities and torques. An established formulation describes the tangential surface velocity ($v_{sq}$) at surface point ($\mathbf{r}_s$) and causing propulsion in the direction of the squirmer’s instantaneous orientation ($\hat{e}$)~\cite{Blake1971}:
\begin{equation}
v_{sq}(\mathbf{r_s}, \hat{e}) = \sum^{\infty} _{n=1} B_n \frac{2}{n(n+1)}  \left( \frac{\mathbf{r_s}}{R} \frac{\hat{e}\cdot \mathbf{r_s}}{R} -\hat{e}\right) P'_n \left( \frac{\hat{e}\cdot \mathbf{r_s}}{R} \right),
\end{equation}
where $R$ is the radius of the sphere, $P'_n$ is the derivative of the \textit{n'th} Legendre polynomial, and $B_n$ are amplitudes corresponding to the $n^{th}$-modes of the surface velocity. This formulation accounts for the forces acting on the particle as a result of its own motion and the resulting fluid flow. The total local surface velocity of the squirmer is given by:
\begin{equation}
    v_s(\mathbf{r} - \mathbf{r}_c, \hat{e}) = \mathbf{v}_c + \mathbf{v}_{sq}(\mathbf{r} - \mathbf{r}_c, \hat{e}) + \boldsymbol{\Omega} \times (\mathbf{r} - \mathbf{r}_c),
\end{equation}
where \(\mathbf{r}_c\), \(\mathbf{v}_c\), and \(\boldsymbol{\Omega}\) are the sphere's position, velocity, and angular velocity, respectively. This model incorporates the dynamics of the squirmer, accounting for both translational and rotational motion, as well as the generated flow. By adjusting the ratio \(\beta = B_2/B_1\), one can modify the characteristics of the model, transitioning from pullers (\(\beta < 0\)) or neutral \(\beta =0\) to pushers (\(\beta > 0\)) (see Fig.~\ref{fig:singleparticle2}c,d,e)~\cite{Gtze2010, ISHIKAWA2006}. Importantly, in addition to generating flows with dipolar, nematic, symmetry in the far-field, the squirmer model captures particle's motility and as such is inherently polar. This makes the squirmer a well-suited model system for exploring emergent symmetries at the collective level.

Recent studies have expanded on these models to explore the collective behavior of squirmers, such as clustering, swarming, and the formation of dynamic patterns in fluid environments~\cite{Long_range_nem_Benoit}. These behavior are crucial for understanding the complex dynamics of active matter systems and have implications for designing synthetic active materials with tailored properties. The hydrodynamic interactions lead to complex behaviours that cannot be captured by simpler models without momentum conservation, highlighting the role of fluid-mediated interactions in active matter systems.
\begin{figure}
    \centering
    \includegraphics[width=1.0\linewidth]{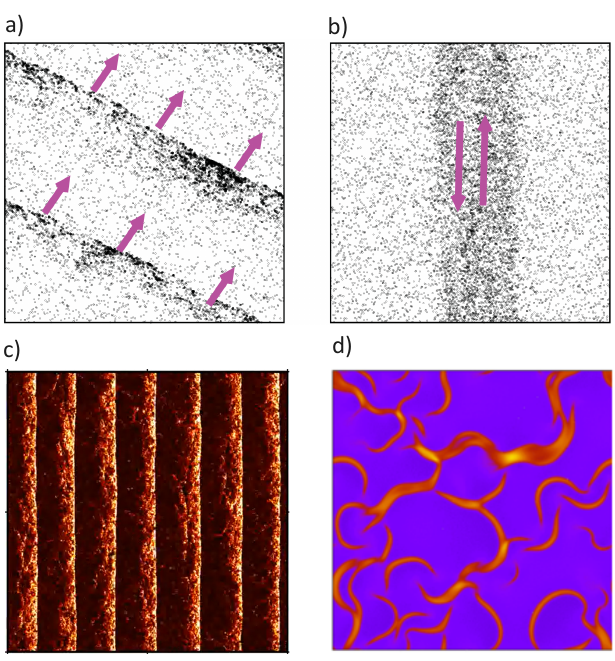}
    \caption{{\bf Distinct collective patterns of polar and nematic point particles without momentum conservation.} Shown here are simulation results for particle models ranging from point polar particles (a) and point nematic particles (b) to coarse-grained density fields of repulsive polar particles (c) and repulsive nematic particles (d). Panel (c) highlights features of polar particles such as smectic like bands, while panel (d) shows the formation of nematic band chaos. Figures adapted from~\cite{fig3panelab, fig3panelc, Ngo2014}.}
    \label{fig:vicsek-nem}
\end{figure}
\subsection{Polar and nematic order in collectives of active particles}

Much work has been done on the statistics of the problems described above and to understand the flow profiles. A striking aspect of active systems, however, involves situations of collective motion driven by individual dynamics. Agent-based models, like those previously described, offer a minimalistic yet powerful approach to studying active systems by explicitly considering interactions between individual agents and their neighbours. These models help understand how local interactions can lead to emergent collective behavior. A detailed review of aligning active particles in the absence of momentum conservation and their features can be found in Ref.~\cite{Chat2020}.

\subsubsection{Point particles}
~\\
The first step of most physical models is to consider point particles. A collection of such active Brownian particles that interact with neighbours already shows the emergent phenomenon of flocking as a phase transition (see Fig.~\ref{fig:vicsek-nem}a). This phenomenon was first studied by Vicsek~\cite{vischek_1995} and is defined as:

\begin{equation}
\mathbf{x}_i(t + \Delta t) = \mathbf{x}_i(t)   +v_0 \mathbf{\hat{e}}_i^{\theta}(t) \Delta t
\end{equation}
where the direction of the particle $\mathbf{\hat{e}}_i^{\theta}(t) = [\cos(\theta(t)), \sin(\theta(t))]$ changes according to:
\begin{equation}
\theta(t + 1) = \langle \theta(t) \rangle_r + \Delta \theta,.
\end{equation}
$\langle \theta(t) \rangle_r$ denotes the average direction of the velocities of particles within a circle of radius $r$ surrounding the given particle, and $\Delta \theta$ is a random change in $\theta$ drawn from a uniform distribution.

A nematic version of this model was developed by~\cite{Chat2006} where the change in orientation was dictated by a nematic director, where each particle $i$ is endowed with an orientation $\theta_i$ and moves along $\theta_i$ or $\theta_i + \pi$ with equal probabilities. The orientation is found at each time from the largest eigenvector of the nematic tensor, averaged over neighbours, 
$$\mathbf{Q}_{ij} = \begin{pmatrix}
\left\langle \cos^2 \theta_k \right\rangle - \frac{1}{2}  \left\langle \cos \theta_k \sin \theta_k \right\rangle \\
\left\langle \cos \theta_k \sin \theta_k \right\rangle \left\langle \sin^2 \theta_k \right\rangle - \frac{1}{2}
\end{pmatrix}.$$
Typical of active nematics, this model exhibits the characteristic giant number fluctuations, where the variance of the particle number N in a subregion scales super-linearly with its average $
\langle (\Delta N)^2 \rangle \sim \langle N \rangle^{\alpha},$ $\alpha>1$. Nematically aligning models have additionally been shown to exhibit large-scale spatio-temporal chaos, characterized by interacting dense, ordered, band-like structures~\cite{Ngo2014} and long-range nematic order~\cite{Long_range_nem_Benoit} (see Fig.~\ref{fig:vicsek-nem}b). At the point particle level a direct comparison between Vicsek-like polar particles and these nematic particles would be of significant interest, as it is evident that by simply altering the symmetry of the underlying particles, the global phenomena transition from polar flocks to nematic order, bands, and giant number fluctuations. Indeed, in terms of density-driven phase separation, nematic lane formation and polar flocking can be regarded as analogous to motility-induced phase separation (MIPS) observed in spherically symmetric particles. While MIPS arises due to a feedback mechanism between particle motility and local density, leading to the coexistence of dense and dilute phases, nematic and polar order introduce distinct symmetry constraints that shape the emergent collective behavior. In nematic systems, particles align head-to-tail, giving rise to lane-like structures, whereas in polar systems, collective motion results in cohesive flocking states. Despite these differences, all three phenomena reflect a fundamental interplay between density variations and self-organization in active matter.

We end this section with a brief discussion of a minimal model for dilute suspensions of motile particles that interact hydrodynamically. In this case, the orientation evolves according to
\begin{equation}
    \dot{\hat{\mathbf{e}}}_i = \left[(\mathbf{I} - \hat{\mathbf{e}}_i\hat{\mathbf{e}}_i^T )\otimes ( \mathbf{\Omega} \left( \mathbf{x}_i \right) + B \mathbf{E} \left( \mathbf{x}_i \right) \right)] \hat{\mathbf{e}}_i,
\end{equation}

where $\mathbf{\Omega}$ and $\mathbf{E}$ are the vorticity and rate-of-strain tensors and B is an aspect ratio depended parameter. In addition to pure self-propulsion, the particle also gets advected, $ \partial_t \mathbf{r}_i = v_0 \hat{\mathbf{e}}_i  + \mathbf{ u_i} $, 
 by $\mathbf{u_i}$, the velocity of the fluid at the position of the particle. The full model, including the calculation of the flow field is found in Ref.~\cite{kultty2020}. This model still uses point particles but by interacting through flow fields, it does not necessitate a specific alignment rule and provides a method  to understand\st{ing} many-body hydrodynamic interactions and their role in the transition to collective motion in active matter systems with momentum conservation. 


\subsubsection{Particles with volume} 
~\\
Alignment-based interactions are inspired by systems such as schools of fish or flocks of birds, making the use of a polar model quite natural. A natural extension to the defined model is to move from point particles to those that collide with each other, while still adapting their orientation based on that of their neighbours. A slight simplification is to first consider particles that only change orientation based on collision dynamics. The microscopic model is once again a set of Langevin equations. The first is for position:
\begin{equation}
\gamma \partial_t \mathbf{r}_i = -\sum\limits_{j \ne i} \nabla_i U(r_{ij}) + f_a \hat{\mathbf{e}}_i +\sqrt{2D} \xi(t), \label{eq:mips-langevin}
\end{equation}
where $U(r_{ij})$ is typically the Weeks-Andersen potential. By writing $f(\theta,\mathbf{v})= -\sum\limits_{j \ne i} \nabla_i U(r_{ij}) + f_a \hat{\mathbf{e}}_i$ we recover equation \ref{eq: 1}.

Even without re-orienting through neighbour interaction, such a model still displays rich density-related phenomena such as clustering and phase separation. The particle in this situation is propelled proportional to the active force $f_a$ in the direction of the particle polarity $\hat{e}_i = [\cos(\theta), \sin(\theta)]$ which changes according to:
\begin{equation}
\dot{\theta}_i(t) = \sqrt{2D_r} \eta_i(t).
\end{equation}
Clustering is a prominent phenomenon in systems of active polar particles, where self-propulsion and inter-particle interactions lead to the formation of dense aggregates~\cite{Fily2012}. This behaviour is markedly different from passive clustering, as active particles can form stable clusters even in the absence of explicit attractive interactions, purely due to their self-propulsion and collision dynamics. This phenomenon, known as motility-induced phase separation (MIPS)~\cite{Fily2012, Cates2015}, reveals how activity can drive phase separation in a manner distinct from equilibrium systems.

MIPS has shown a remarkable amount of truly non-equilibrium phenomena such as hexatic phases~\cite{Digregorio2022}, negative interface surface tension~\cite{bialke2015negative,Fausti2021,Hermann2019}, and temperature difference based on density~\cite{Mandal2019}. While MIPS has been extensively studied using microscopic models, coarse-grained continuum theories provide a broader perspective. For instance, recent work has extended the theory from polar particles to nematic and higher-order symmetries, highlighting the versatility of MIPS across different types of active matter~\cite{Lee2022}. These continuum models help bridge the gap between microscopic dynamics and macroscopic phenomena, offering insights into the collective behaviour of active systems.

For readers interested in a more comprehensive understanding of MIPS, we refer to ~\cite{Fily2012} and to a review that delves into the theoretical and experimental advancements in this area~\cite{Cates2015}. These works provide a detailed overview of the mechanisms driving MIPS and its implications for the study of active matter.

\textit{Active particles with aligning interactions}

Aligning interactions play a crucial role in the ordered phases they induce and enter through the orientation equation:
\begin{equation} 
\dot{\theta }_i= \sum _{j \ne i} \varphi_{ij} + \sqrt{2D_r} \eta _i, 
\end{equation}
where $\varphi$ is a generic alignment function. An alignment function of importance to this discussion is $ \varphi_{ij} = \sin(p(\theta_j - \theta_i))$, where $p=1$ can be referred to as polar (see Fig.~\ref{fig:vicsek-nem}c) and $p=2$ as nematic (see Fig.~\ref{fig:vicsek-nem}d). For polar aligning particles, an increase in alignment strength enhances the tendency of particles to aggregate into clusters, thereby promoting the complete de-mixing of the system into distinct low and high-density phases~\cite{SesSansa2018}. Nematic particles, on the other hand, are found to either induce or prevent phase separation~\cite{Spera2024} through fluctuations which enhance polar order. It is interesting that even at the level of discrete particles, a change in the polar or nematic aligning interaction can lead to fundamentally different phase behaviours such as particle clustering and phase separation.

Global polar order is additionally attained in such systems through weak attraction between particles~\cite{Caprini2023} or by incorporating non-reciprocal orientation interactions~\cite{Kneevi2022} through $\varphi_{ij} = H (R - |{\mathbf {r}}_{ij}|) \left( \hat{\mathbf{e}}_i \times \frac{{\mathbf {r}}_{ij}}{|{\mathbf {r}}_{ij}|} \right)$, which has revealed new ordered phases of MIPS, including static flocks and dynamic flocks. $H$ is the Heaviside function, $r_{ij}=r_i-r_j$ is the two particle distance, $\mathbf{e}$ the orientation of the particle and the indices $i$ and $j$ refer to the $i$'th and $j$'th particles. These findings were motivated by the steric interactions of the flagella of \textit{Chlamydomonas} with neighbouring particles, demonstrating how specific biological features can influence collective behaviour. The study of non-reciprocal interactions, which has shown a plethora of interesting behaviours~\cite{ 	
Fruchart2021, Dinelli2023, nonreciprocal3}, opens up new avenues for understanding how asymmetries in interactions can lead to novel collective states in active matter.

Finally, hydrodynamic interactions provide a generic mechanism for emergent alignment dynamics in microswimmer suspensions, without imposing any specific alignment. In particular, studies using suspensions of squirmers have revealed how the pusher-puller activity of the particles can lead to an emergent attractive-repulsive interactions between the swimmers~\cite{ishikawa2008coherent}.  A report has examined how the ratio of self-propelling to attractive forces, combined with the strength of active stress, influences giant number fluctuations in squirmer suspensions~\cite{Alarc_n_2017}. The giant number fluctuations displays asymmetry between pusher and puller particles, with pullers showing larger scaling exponents.  Recent work on a collection of confined swimmers showed pullers form dense clusters, while pushers displayed short wavelength instability~\cite{Brdfalvy2024}. These results highlight a non-trivial impact of the form of active stress (puller versus pusher type) on the core structure of the collectives, and emphasize the fundamental differences that arise when comparing systems with and without hydrodynamic interactions. 

\begin{figure}
    \centering
    \includegraphics[width=0.9\linewidth]{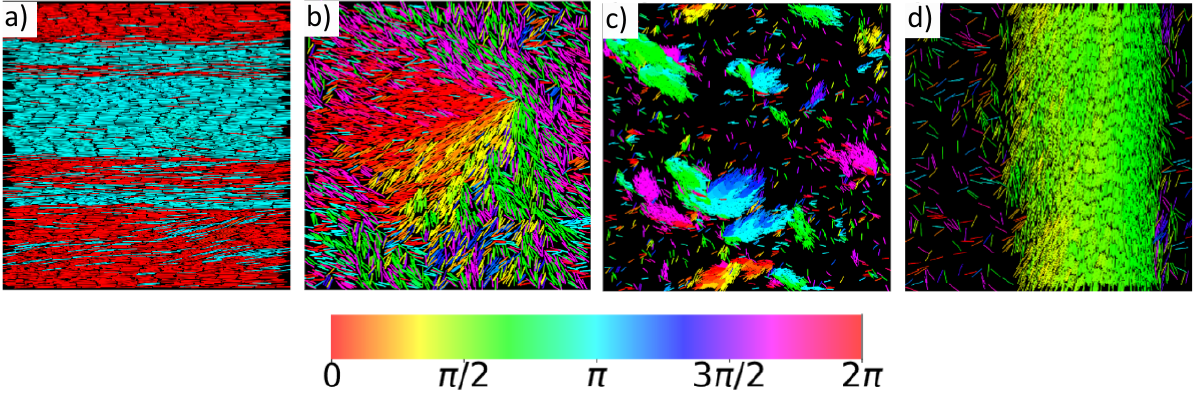}
    \caption{{\bf Emergent polar and nematic features in self-propelled hard rods.} Depiction of phases shown by polar self-propelled rods: (a) nematic lanes,  (b) large scale nematic defect structures, (c) polar clusters, and  (d) polar flocking. Figure adapted from~\cite{sprimages}.}
    \label{fig:spr_phases}
\end{figure}

\subsection{Non-spherical particles\label{sec:elongated}}
Our discussion so far has focused on systems and models where active particles themselves have an isotorpic shapes, and as such, polar or nematic symmetry is defined based on their direction of self-propulsion and its alignment dynamics. Most real biological systems, however, are characterized by shape anisotropy, for example bacteria~\cite{meacock2021bacteria} or sperm cells~\cite{Riedel2005} . In the presence of such an anisotropy, the orientation associated with the shape of the particles also gains significance and results in an intricate interplay between nematic and polar order.

\subsubsection{Self-propelled rods}
~\\
Extending microscopic particle models to incorporate elongation leads to what are known as self-propelled rods (SPR). Unlike spherical particles, these rods align along their body axes. The alignment direction does not have head or tail and therefore has nematic symmetry. Therefore, these rod-shaped particles are inherently endowed with nematic order, in addition to the their polarity due to self-propulsion. The equations governing their behavior are similar to previous models~\cite{SelfPropelledRods}:
\begin{equation}
\dot{\mathbf{r}}_i = v_0 \mathbf{e}_i(\theta_i, t) + \mu  \sum\limits_{j} \mathbf{f}_2 \left(\mathbf{r}_i - \mathbf{r}_j, \theta_i, \theta_j\right) + \boldsymbol{\eta}(t),
\end{equation}
\begin{equation}
\dot{\theta}_i = \mu_{\theta} \sum_{j} m_2 \left(\mathbf{r}_i - \mathbf{r}_j, \theta_i, \theta_j\right) + \sqrt{2D_{\theta}} \xi_i(t),
\end{equation}
where $\mu$ is mobility parameter, $\mathbf{f_2}$ and $m_2$ are anisotropic repulsion and alignment, \st{and} $v_0$ is the self-propulsion of the particle, and $\xi$ and $\mathbf{\eta}$ are white noise. A key new parameter is the aspect ratio of the long and short axes. When this ratio is 1, the model simplifies to a disk displaying motility-induced phase separation (MIPS). A phase field particle model, employing the usual equations of SPR, observed the transition from MIPS to local nematic order to polar domain and finally to global polar bands as the aspect ratio was changed from a disk to an elongated rod~\cite{Gromann2020}. 

The repulsive interactions between particles, the torques, dissipation, and fluctuations are all shape-dependent and exhibit uniaxial symmetry. This symmetry is only broken by polar driving. The basis for this class of particle models is binary collisions. When rods collide at acute angles, anisotropic repulsion and self-propulsion cause them to align their motion in parallel. The speed of the rods remains unchanged before and after the collision, but their aligned motion results in a net change in combined momentum, leading to non-conserved momentum. Multiple such collisions lead to local order and cluster formation. SPRs display a variety of collective behavior, including polar clusters, giant aggregates, polar bands, swarms, and laning (see Fig.~\ref{fig:spr_phases})~\cite{sprimages, Weitz2015, Wensink2008, Abkenar2013, Yang2010, McCandlish2012}. In the absence of self-propulsion, these rods behave like dry nematic rods with giant number fluctuations. Rods can further be studied as nematic particles by switching the direction of self-propulsion equally along the long axis. An example appears in~\cite{Shi2013} where self-driven hard elliptic rods were used to study defects without hydrodynamic interactions, showing that breaking detailed balance leads to defect pair creation and unbinding and lead to defect-mediated collective motion. 

Self-propelled rods are by far the model with the most potential to exbibit both polar and nematic features. When two rods collide and align, they either do so parallel or anti-parallel. Parallel alignment leads to polar features while interactions that align particles regardless of their polarity create an ordered state with nematic symmetry. This versatility has led to wide spread adoption and many extensions from rods with hydrodynamic interactions~\cite{Zantop2022, Qi2022}, to bead-spring models that allow the `rod' to be flexible~\cite{Venkatesh2022}. As computational power increases, continuing research on SPR and its extensions  will continue to provide more accurate and detailed understanding of complex biophysical processes. For a comprehensive review of SPRs, refer to~\cite{SelfPropelledRods}.


\subsubsection{Flexible Filaments}
~\\
\begin{figure}
    \centering  \includegraphics[width=1.0\linewidth]{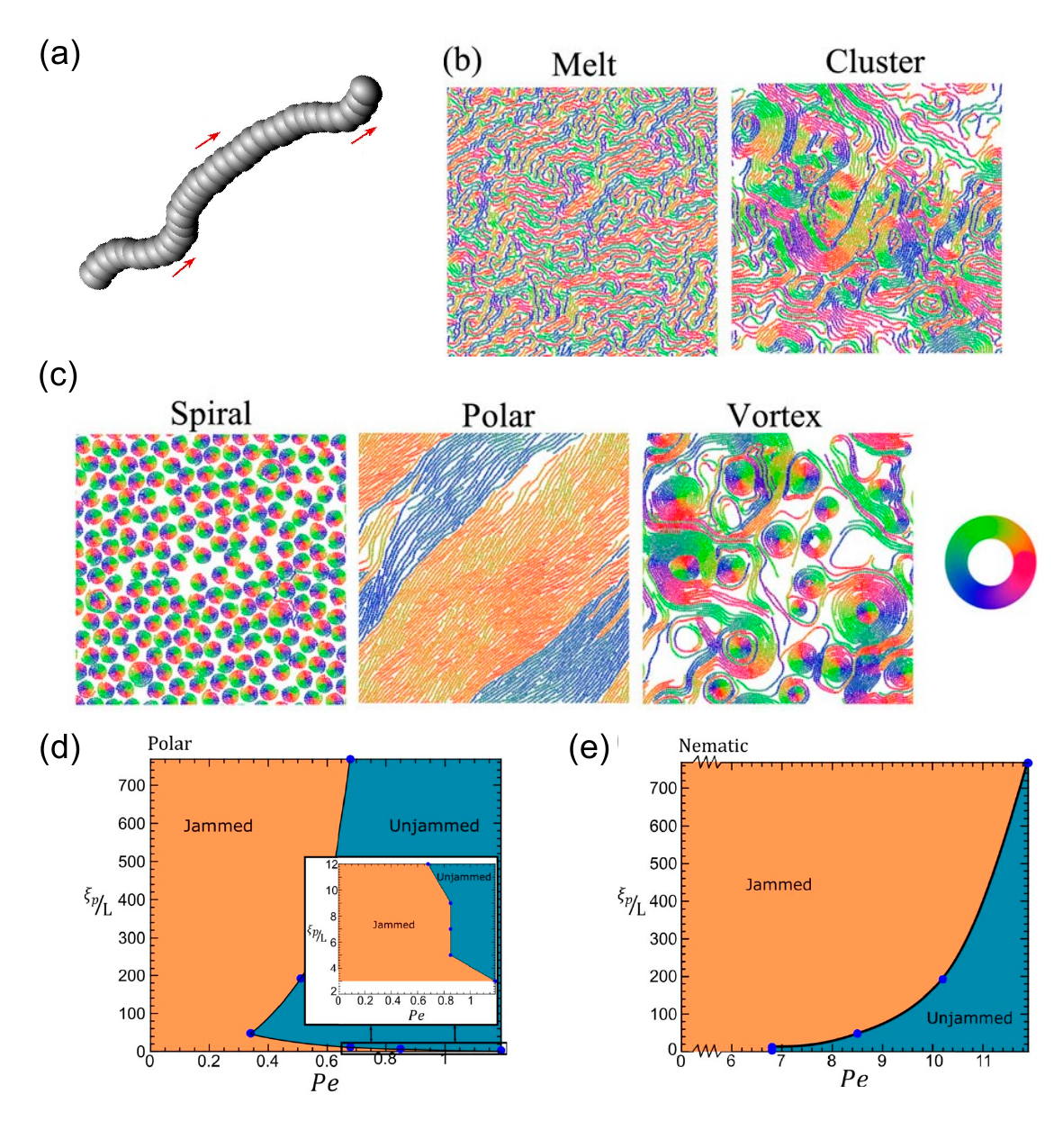}
    \caption{{\bf Polar and nematic features in patterns induced by flexible active filaments.} Panel (a) depicts a schematic of a bead-spring model where the red arrows show the self-propulsion acting on each bead tangentially to the filament. Panels (b) and (c) show various structures and types of ordering showed by a collective of such flexible filaments with distinct behaviour of polar (d) and nematic (e) filaments in the context of jamming-unjamming where the panels are phase diagrams with self-propulsion on the x-axis and increasing} rigidity on the y-axis where $\xi/L$ is proportional to persistence length. Figures adapted from~\cite{Zhao2024} and~\cite{Venkatesh2022}.
    \label{fig:fill}
\end{figure}
Extensions to the self-propelled rod model have been done by either introducing flexibility or deformability of an ellipse. Both ideas are similar as they both look to relax the condition of rigid rods. For a review on deformable active particles, readers can refer to~\cite{Ohta2017}. We briefly describe flexible filaments further as modeling filaments is often done with only a single Langevin equation for the position:
\begin{equation}
\gamma \dot{\mathbf{r}}_i = - \mathbf{\nabla}_i U + \mathbf{F}_i^a + \xi,
\end{equation}
where $\mathbf{F}_i^a$ is the active self-population  and acts tangentially along the axis of the filament.

Filament bead-spring models (see Fig.~\ref{fig:fill}a) typically employ potentials for excluded volume, flexibility, and bonds between beads to form the filament. Flexible self-propelled filaments have been studied as both nematic~\cite{Joshi2019} and polar~\cite{Duman2018} in simulations. Characteristic of polar particles, flexible filaments too display flocking behaviour. The onset of flocking was studied recently for such particles by~\cite{Huber2021} where it was found that global polar order emerges by nucleation and growth of polar clusters through cycles of growth and fragmentation. 

Flexible filaments have had great success when compared with experiments. Examples include observations of band formation, bending, buckling, defects~\cite{Vliegenthart2020}, implications of confinement~\cite{Peterson2021}, and nematic phases in which both integer and half-integer defects were observed in the same system~\cite{Dunajova2023}. Simulations of polar filaments reveal a broad range of behaviours from polar flocks, vortices, and spirals (see Fig.~\ref{fig:fill}b,c)~\cite{Zhao2024} to jamming-unjamming transitions~\cite{Venkatesh2022}.  

The jamming-unjamming transition was studied for both polar and nematic filaments where the filament flexibility was found to be especially important. As expected, making the filaments more flexible for nematic filaments proved to enhance the systems ability to unjam. For polar filaments, however, there was a re-entrant transition where extremally flexible filaments instead remained jammed as the activity was increased (see Fig.~\ref{fig:fill}d,e)~\cite{Venkatesh2022}. These filaments showed smectic-like ordering. This distinction highlights the importance of comparing nematic and polar systems in the same condition and a fundamental impact polar and nematic symmetries could have on shifting phase transition boundaries in active matter. 

\subsubsection{Deformable particles: Phase field models}
~\\
\begin{figure}
    \centering \includegraphics[width=0.9\linewidth]{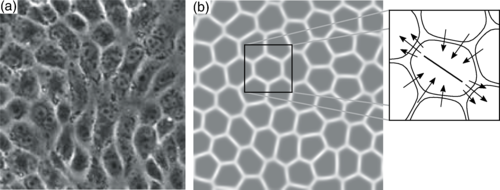}
    \caption{{\bf Orientational order in deformable cell collectives.} An example showing the comparison between a cell layer (a) and the phase field model emulating a cell layer (b), with an insert schematic depicting the intercellular forces}.  Figure adapted from~\cite{Mueller2019}.
    \label{fig:celadro}
\end{figure}
At the crossroad of continuum active models and particulate active models lie phase field models that aim to emulate collective cell behaviour in tissues. A distinctive features of such models is their ability to account for deformability of active particles. This is crucial, for example in studying cellular monolayers where cells constantly change shape due to adhesion forces they experience from neighbouring cells and in interaction with their environment~\cite{ladoux2017mechanobiology,balasubramaniam2021investigating}. As such, both the cell orientation and cell polarity can be continuously evolving with the cell shape changes. Here, we present only the key modeling idea behind mulit-phase field approach. For a detailed description of phase field models of active matter we refer the reader to~\cite{Mueller2019,mueller2021phase,Zhang2020, Ardaeva2022, Hopkins2022, Chiang2024,monfared2023mechanical}. 

Multi-phase field models consist of (N) phase field cells (see Fig.~\ref{fig:celadro}) , represented by a scalar field ranging from 0 to 1, where a value of 1 indicates the presence of a cell. While such models exist for single-cell dynamics~\cite{Tiribocchi2023}, the collective motion of phase field cells is unique as it follows the dynamics of discrete particle models, where individual cell positions evolve according to force balance, given by:
\begin{equation} \xi \mathbf{v}^{(i)}(\mathbf{x}) = \mathbf{f}_{\text{passive}}^{(i)}(\mathbf{x}) + \mathbf{f}_{\text{active}}^{(i)}(\mathbf{x}), \end{equation}
where $\xi$ is a friction constant and $\mathbf{f}_{\text{passive}}$ forces result from the divergence of stresses, including contributions from cell-cell interactions, cell-wall interactions, volume conservation, shape operators, and phase field free energy relaxation. The deformability of the cells can lead to interesting consequences; for example, both $+1/2$ and $-1/2$ defects were found to be diffuse with a mean squared displacement (MSD) slope close to one, differing from continuum nematic theory where the $+1/2$ defect is ballistic~\cite{Ardaeva2022}.

Active self-propulsion can either be polar~\cite{Zhang2020, Chiang2024} and proportional to the velocity magnitude, or arise from the divergence of an active nematic stress $\zeta \mathbf{Q}$~\cite{Mueller2019, Ardaeva2022,Zhang2023}. In previous active polar particle models, the direction of the polarization vector coincided with either the velocity vector or the body long axis of the particle, as in self-propelled rods (SPR). In contrast, here, the cell polarization vector is free to evolve independently through internal mechanisms.

This phase field model provides the opportunity to directly compare the effects of nematic versus polar symmetry, as well as the emergent hexatic order. Such a system, rich in symmetries, is unique and has the potential to explore a variety of non-equilibrium phenomena specific to active matter systems.

\section{Polar and nematic active systems at the continuum level}

Continuum models complement discrete systems by representing the macroscopic behaviour through continuous fields. This method simplifies the analytical analysis and simulation of complex systems, making it easier to study intricate interactions and dynamic processes. Furthermore, the use of macroscopic fields provides a natural bridge for comparing theoretical predictions with experimental observations.

This section delves into the continuum formalism of both nematic and polar active matter and highlights the significant differences between them. Differentiating the two primary formalisms that emerge based on the particle-solvent interaction: systems without momentum conservation, where frictional forces dominate, and systems with momentum conservation, where hydrodynamic interactions are significant. Separate sections are introduced placing emphasis on phenomena where the polar and nematic formalisms diverge: the formation of topological defects, active turbulence and motile droplets.

\subsection{Systems Without Momentum Conservation}

Systems without momentum conservation refer to those in which the interaction between particles and the substrate is influenced by friction, leading to the non-conservation of momentum. Such systems are commonly observed in various biological settings, including animal flocks~\cite{bird_flocks}, fish schools~\cite{fish_schools}, and migrating cell layers~\cite{migrting_cells_Sknepnek}.

\subsubsection{Vicsek to Toner Tu}
~\\
The constitutive equation for systems without momentum conservation can be obtained either through symmetry arguments or by utilizing probability distributions originating from the Vicsek model. To clearly demonstrate how polarity effects manifest at the continuum level we illustrate here the derivation of the Toner Tu equation from the original, discrete, Vicsek model, drawing heavily from the research presented in~\cite{Chat2020,Vicsek_to_Hydro, Bertin_2009_Vicsek_Toner_2009}. Introducing probability densities is essential for transitioning from a microscopic to a macroscopic model. The key distribution is the phase space evolution distribution of an individual particle denoted as $f(\mathbf{r},\theta,t)$, which represents the probability of a particle being at position $\mathbf{r}$ with a velocity $v_0$, assumed to be constant for all particles, along the $\theta$ angle. The evolution of this distribution over time is governed by the Boltzmann equation:

\begin{equation}
    \label{eq: boltzmann equation}
    \frac{\partial}{\partial t} f(\mathbf{r},\theta,t) + v_0 \, \mathbf{e}(\theta) \cdot \nabla f(\mathbf{r},\theta,t) = I_{dif}[f] + I_{col}[f].
\end{equation}

Where $\mathbf{e}(\theta)$ is the unit vector pointing in the direction $\theta$. The terms on the right-hand side represent self-diffusion and collision, respectively, and integrate the Vicsek model. To establish the continuum equations, the density and velocity fields are introduced as follows:

\begin{equation}
    \label{hydrodynamic fields}
    \rho(\mathbf{r},t) = \int d \theta f(\mathbf{r},\theta,t), \quad \mathbf{v}(\mathbf{r},t) = \frac{v_0}{\rho(\mathbf{r},t)} \int d \theta f(\mathbf{r},\theta,t) \mathbf{e}(\theta).
\end{equation}

By integrating Eq.~\ref{eq: boltzmann equation} with respect to $\theta$, we obtain the dynamical equation for density:

\begin{equation}
    \frac{\partial \rho}{\partial t} + \nabla \cdot (\rho \mathbf{v}) = 0,
\end{equation}

which essentially represents the conservation of particles. Utilizing the Fourier transform of the phase space distribution, defined as $\hat{f}_k(\mathbf{r},t) = \int_{-\pi}^{\pi} d \theta f(\mathbf{r},\theta,t) e^{ik\theta}$, and assuming the near-isotropic nature of $f(\mathbf{r},\theta,t)$, which implies a scaling of $\hat{f}_k(\mathbf{r},t) = \mathcal{O}(\epsilon^{\lvert k \rvert})$, we can derive the hydrodynamic equation for momentum. Retaining terms up to the third order $\epsilon^{3}$, we obtain the dynamic equation of momentum $\mathbf{w} = \rho \mathbf{v}$,

\begin{equation}
    \label{eq: TonerTu written as momentum}
    \frac{\partial \mathbf{w}}{\partial t} + \gamma (\mathbf{w} \cdot \nabla) \mathbf{w} = (\mu - \xi \mathbf{w}^2) \mathbf{w} + \nu \nabla^2 \mathbf{w} - \frac{v_0^2}{2} \nabla \rho + \frac{\kappa}{2} \nabla \mathbf{w}^2 - \kappa (\nabla \cdot \mathbf{w}) \mathbf{w}.
\end{equation}

Where the coupling between density and momentum gradients has been neglected, implying that the $\nu$ coefficient does not depend on the density of the system, $\rho$. Eq.~\ref{eq: TonerTu written as momentum} shows great resemblance with the Navier-Stokes equation. For a detailed comparison between this equation and the Navier-Stokes equation, see~\cite{marchetti_hydrodynamics_2013, Bertin_2009_Vicsek_Toner_2009}. The first term on the right hand side of Eq.~\ref{eq: TonerTu written as momentum} is the \underline{Landau-de Gennes free energy term} that drives the order-disorder transition. The second term accounts for the energy expense for deviations from the aligned state, this term is analogous to the \underline{Frank free energy term} in liquid crystals that penalizes gradients in the orientation field. The third and forth term are considered an \underline{effective pressure term}, $P_{\text{eff}}=\frac{1}{2}(v_0^2 \rho -\kappa \mathbf{w})$. The last term accounts for the \underline{non-linear contributions of compressible systems}. These last three terms
account for spontaneous splay and can provide local alignment for $\mathbf{P}$. An important aspect of Eq.~\ref{eq: TonerTu written as momentum} is the inherent assumption that the velocity field is analogous to the polarity field of the particles. While such an assumption could be valid at the limit of low density (dilute) of active particles, care must be taken in considering identical polarity and velocity fields at higher densities, where particle interactions become dominant.

\subsection{Systems With Momentum Conservation}

Systems with momentum conservation involve significant hydrodynamic interactions with the substrate, leading to the conservation of momentum. These systems are often observed in fluid environments, such as bacterial suspensions, algae colonies, and synthetic microswimmers.

\subsubsection{Toner Tu continuum model} 

~\\
The Toner Tu equation was initially derived using symmetry arguments~\cite{Toner_1998} and is compatible with the derivation used in the last sections that incorporated the Vicsek model (Eq.~\ref{eq: TonerTu written as momentum}). The Toner Tu equation reads:
\begin{equation}
    \partial_t \mathbf{P} + \lambda(\mathbf{P} \cdot \nabla) \mathbf{P} = [\alpha(\rho) - \beta \lvert \mathbf{P} \lvert^2] \mathbf{P} + K \nabla^2 \mathbf{P} - \nu_1 \nabla \frac{\rho}{\rho_0} + \frac{\lambda}{2} \nabla \lvert \mathbf{P} \lvert^2 - \lambda \mathbf{P} (\nabla \cdot \mathbf{P}).
    \label{eq: TonerTu equation}
\end{equation}
The first term on the right-hand side accounts for the phase transition that occurs from an isotropic ordered state with no polarity, $\mathbf{P}=0$, to an ordered flock with polarity, $\mathbf{P}=\sqrt{\frac{\alpha}{\beta}}$. The second term represents the energy cost for inhomogeneities in the polarization field, with $K$ being the Frank constant. The last three terms account for spontaneous splay and can provide local alignment for $\mathbf{P}$.

In this model, particle velocity and polarity are treated as indistinguishable, implying that particles always move along their polar axis. While this assumption is appropriate for systems like bird flocks or fish schools, it does not necessarily apply to bacteria or cell layers, where particles do not always align their movement with their polar axis~\cite{han2023local,Fluid_Dynamics_Bact_2013}. While recent studies on confluent human keratinocyte cells have revealed that full-integer topological defects in the velocity field drive flow ordering~\cite{Topology_Polar_Ordering_cell_exp}, a detailed analysis of the relationship between the flow field and the cell polarity is lacking. In this vein, earlier experiments have indeed shown that in cell monolayers, the polarity direction could be related to the total force experienced by the cells~\cite{peyret2019sustained}, highlighting the need for further investigation into both the flow field and the orientation parameter independently.

One of the most significant characteristics of the Toner Tu equations is banding~\cite{SelfPropelledRods,Mishra_2010}, which appears close to the mean field transition. Bands are high-density periodic stripes that move along a certain direction against a low-density background. These waves occur due to the particle self-propulsion and are also found in discrete systems~\cite{Vliegenthart2020}.



\subsubsection{Self-propelled hard rods}
~\\
Motivated by the giant number fluctuations that active systems display~\cite{Granular_nem_2007}, self-propelled hard rods models were developed~\cite{Hydrodynamics_of_self_propelled_hard_rods}. Self-propelled hard rod models are also divided between systems without momentum conservation and systems with momentum conservation~\cite{SelfPropelledRods}. The former comprises particles that self-propel along a specific axis, making them polar, and interact through hard-core collisions. This model generally aligns particles in a nematic/apolar manner and importantly, is one of the few continuum models capable of predicting nematic alignment with polar particles. The continuum representation of hard rod models is detailed in~\cite{Self_propelled_rods_2015,Peshkov_SPR} and links the polarization with the nematic tensor and particle density $\{ \mathbf{P}, \mathbf{Q}, \rho \}$.

This model successfully describes the transition from nematic to isotropic phases and reveals that self-propulsion speed boosts nematic ordering. Additionally, this model accurately anticipates the formation of bands, a phenomenon also observed in discrete systems. Recent contributions~\cite{SPR_reversal_speed} have introduced a new term that accounts for particle velocity reversal, leading to the formation of arches, global polar smectic patterns, at high velocity reversal rates.

Systems with momentum conservation account for the particle-fluid interactions, and as the systems usually have a small Reynolds number, the Stokes equation is used to model the fluid flow. The incompressible Toner-Tu equation or Toner–Tu–Swift–Hohenberg equation~\cite{IncombressibleTonerTu} is considered a hybrid model between systems without and with momentum conservation and predicts active turbulence well matched with experimental realizations of bacterial suspensions~\cite{Meso_scale_turbulence}. The model treats the particle velocity field and polarity as indistinguishable, assuming the motion always occurs along the particle polarization and the temporal evolution is decoupled from the hydrodynamic field. However, it incorporates phenomenological constants that relate to the hydrodynamics of the flow, such as differentiating if the particles are contractile or extensile.


\subsubsection{Active swimmer suspensions}
~\\
Kinetics based models for active suspensions have been developed~\cite{Santillan_2008}, where the probability distribution function $\psi(\mathbf{r},\mathbf{P},t)$ depends on the particle position $\mathbf{r}$ and orientation $\mathbf{P}$. This model builds on the Smoluchowski equation and assumes that the flux of velocities, $\Dot{\mathbf{r}}$, is affected by the particle self-propulsion $V_s$, the local background fluid velocity $\mathbf{v}(\mathbf{r},t)$ and the translational diffusion $D$. The equation reads,
\begin{equation}
    \label{eq: Active suspensions Saintillan}
    \Dot{\mathbf{r}}=V_s \mathbf{P} + \mathbf{v} -D \nabla_\mathbf{r} \ln \psi.
\end{equation}
The rotational velocity of the particle  $\mathbf{\Dot{P}}$, accounts for the rotational diffusivity $d$ and the rotation of an anisotropic particle in the local flow, which follows Jeffrey equation~\cite{jeffery1922}:
\begin{equation}
    \label{eq: rotational velocity Saintillan}
    \Dot{\mathbf{P}}=(\mathbf{I} -\mathbf{P}\mathbf{P}) \cdot (\beta \mathbf{E}+\mathbf{\Omega}) \cdot \mathbf{P} -d \nabla_\mathbf{P} \ln \psi,
\end{equation}
where the parameter $\beta$ differentiates the individual particle geometry and $\mathbf{E} = \frac{1}{2}(\nabla \mathbf{v} + (\nabla \mathbf{v})^T)$ is the strain rate tensor and $\mathbf{\Omega} = \frac{1}{2}(\nabla \mathbf{v} - (\nabla \mathbf{v})^T)$ the vorticity. The fluid flow dynamics follow the Stokes equation with the active stress contribution, $\sigma^a$, typically modelled as:
\begin{equation}
    \label{eq: active stress}
    \sigma^a =- \zeta \left( \mathbf{P} \mathbf{P}^T - \mathbf{P}^2\frac{\mathbf{I}}{2}  \right),
\end{equation}
where $\zeta$ is the activity coefficient, $\mathbf{P}$ is the polarization field, and $\mathbf{I}$ is the identity matrix. This form of active stress was first introduced in the works of \textit{Pedley and Kessler}~\cite{Pedley_kessler} in the context of bioconvection flows, providing both the expression for the active stress and the relative importance with respect to other hydrodynamic stress contributions. The classification between the two types of swimmers, pullers and pushers, is introduced~\cite{Santillan_2008} as the sign of the activity parameter, $\zeta >0$ for pushers and $\zeta <0$ for pullers. The kinetic model discussed here is based on a dilute approximation, and in order to account for the collective behaviour at high densities an effective steric torque is introduced in the rotational velocity, $\Dot{\mathbf{P}}$~\cite{Ezhilan2013,saintillan2015theory}. The interplay between the ratio of the steric alignment torque to rotational diffusion leads to an isotropic-nematic system transition by augmenting the steric alignment torque, suggesting that this steric torque favours nematic symmetry in dense systems of polar particles. An important aspect of this approach is that the polarity and velocity fields are no longer treated as the same, which makes the model more appropriate for studying systems such as swimming bacteria and collectives of sperm cells.

\subsubsection{Hydrodynamic theories}
~\\
The continuum description of systems with momentum conservation typically involves the coupling of the Navier-Stokes equations with additional fields representing the active components. The hydrodynamic theories for these systems are built upon the principles of fluid dynamics, incorporating the effects of active stresses and flows generated by the self-propulsion of particles~\cite{JULICHER20073}. An important feature of these hydrodynamic theories is accounting for the distinction between velocity and polarity fields by coupling the evolution of the polarity to the Navier-Stokes equations. As such, they provide an important modelling framework for describing such active systems. 

One of the foundational models in this context is the active gel theory~\cite{Prost_nature_2015}, which describes the behavior of active polar gels. This theory extends the classical hydrodynamics of passive gels by including active stress terms that account for the forces generated by the active components. The governing equations for active gels are given by:
\begin{equation}
    \label{eq: NavierStokes}
    \rho \left( \frac{\partial \mathbf{v}}{\partial t} + \mathbf{v} \cdot \nabla \mathbf{v} \right) = -\nabla p + \eta \nabla^2 \mathbf{v} + \nabla \cdot \sigma^a,
\end{equation}
where $\mathbf{v}$ is the velocity field, $p$ is the pressure, $\eta$ is the viscosity, and $\sigma^a$ is the active stress tensor (Eq.~\ref{eq: active stress}). This formulation captures the essential features of active matter, including the generation of spontaneous flows and the formation of complex patterns.

Recent studies have expanded on these models to explore the interplay between polar and nematic order in active fluids. For instance, the coupling of polar and nematic order parameters has been shown to give rise to novel phenomena such as active turbulence, defect dynamics, and spontaneous flow patterns~\cite{ Long_range_nem_Benoit,Nem_polar_fluid_surf}. These studies highlight the rich dynamics that emerge from the interaction between different types of order in active matter systems.

\subsubsection{Flow effects on system dynamics}
~\\
In order to consider hydrodynamic interactions, the temporal evolution of both the velocity field and polarity is handled separately. The evolution of the polarity director is described by the following equation:
\begin{equation}
    \label{eq: dynamical equation polar wet}
    \partial_t \mathbf{P} + \mathbf{v} \cdot \nabla \mathbf{P} + \mathbf{\Omega} \cdot \mathbf{P} = \frac{1}{\gamma} \mathbf{h} + \lambda_1 \Delta \mu \mathbf{P} - \lambda \mathbf{E} \cdot \mathbf{P}.
\end{equation}
A rigorous derivation employing the Onsager approach is detailed in~\cite{Jülicher_2018}. Equation \ref{eq: dynamical equation polar wet} assumes that the fluid is incompressible, denoted by $\nabla \cdot \mathbf{v} = 0$. The terms on the left-hand side of Equation \ref{eq: dynamical equation polar wet} represent the components of the comoving and corotational derivative of polarization, $\frac{D \mathbf{P}}{D t} = \partial_t \mathbf{P} + \mathbf{v} \cdot \nabla \mathbf{P} + \mathbf{\Omega} \cdot \mathbf{P}$.

The term $\lambda_1 \Delta \mu \mathbf{P}$ in Eq.~\ref{eq: dynamical equation polar wet} serves as an active term that, when positive, aligns the polarization. One commonly accepted approximation is to treat this term as a Lagrange multiplier to ensure $\mathbf{P}^2 = 1$, thereby reducing the number of independent variables and facilitating theoretical analysis~\cite{Jülicher_2018}. However, certain experimental systems, like cell layers, do not adhere to this constraint as the magnitude of polarity varies, necessitating the consideration of this term. The parameters $\gamma$ and $\lambda$ correspond to the rotational viscosity and the flow alignment parameter, respectively, which can be determined from passive nematic liquid crystals.

The molecular field $\mathbf{h}$ is the functional derivative of the Frank free energy $\mathcal{F}_\mathbf{P}$ (Eq.~\ref{eq: Frank free energy}), which accounts for the energetic cost of inhomogeneities in the order phase. The three types of possible deformations are splay, bend, and twist with elastic constants $K_i$ with $i \in \{1,2,3\}$ respectively. A commonly used approximation is to neglect the anisotropy in the three Frank elastic constants and combine them into a single term $\frac{K}{2}(\nabla \mathbf{P})^2$. Furthermore, to incorporate the isotropic-ordered transition, mean-field terms (Landau expansion) are added to the Frank Free energy,
\begin{equation}
    \label{eq: Frank free energy}
    \mathcal{F}_\mathbf{P} = \int d \mathbf{r} \left[ A \left( -\frac{\mathbf{P}^2}{2} + \frac{\mathbf{P}^4}{2} \right) + \frac{K_1}{2} (\nabla \cdot \mathbf{P})^2 + \frac{K_2}{2} [\mathbf{P} \cdot (\nabla \times \mathbf{P})]^2 + \frac{K_3}{2} [\mathbf{P} \times (\nabla \times \mathbf{P})]^2 \right].
\end{equation}
The fluid dynamics are governed by the Navier-Stokes equation (Eq.~\ref{eq: NavierStokes}) with three stress contributions appearing from three distinct forces, obtained from the Onsager approach: the viscous force (Eq.~\ref{eq: stress viscous}), the elastic force (Eq.~\ref{eq: stress elastic}), and the active stress (Eq.~\ref{eq: active stress}) which accounts for the dipolar stress created by single particles generated by the consumption of molecular energy transducers such as ATP.
\begin{align}
\sigma^{vis} &= 2 \eta \mathbf{E}, \label{eq: stress viscous} \\
\sigma^{ela} &= \frac{\lambda+1}{2} \mathbf{P} \mathbf{h} + \frac{\lambda-1}{2} \mathbf{h} \mathbf{P} - \frac{\lambda}{2} (\mathbf{P} \cdot \mathbf{h}) \mathbf{I}. \label{eq: stress elastic}
\end{align}
Contrary to systems without momentum conservation, the hydrodynamic interactions are introduced by coupling the dynamical equations of polarity and velocity. The rotation and translation of the particle along the fluid flow is accounted for by the comoving and corotational derivative of polarity in Eq.~\ref{eq: dynamical equation polar wet}. The coupling between the strain rate and polarity is introduced proportional to the flow alignment coefficient $\lambda$. Moreover, both the elastic stress, well known from liquid crystals, and the active stress couple the polarity to the flow field.

\subsection{Active nematics}
The continuum approaches discussed thus far all describe the evolution of the polarity field of active particles. As discussed in section~\ref{sec:elongated} for discrete particles, active particles with anisotropic shapes are characterized by an orientation field with nematic symmetry (no head or tail direction) that could interact with the existing polarity field, set by the self-propulsion dynamics.

In the case of nematic suspensions, to describe the particles the second rank nematic tensor is introduced $\mathbf{Q}$, substituting the polarization $\mathbf{P}$, such that the nematic symmetry, $\mathbf{\hat{n}}=-\mathbf{\hat{n}}$, is satisfied. The dynamical equation is given by~\cite{beris1994thermodynamics},
\begin{equation}
    \label{eq: nematic continuum}
    \partial_t \mathbf{Q}+ \mathbf{v}\cdot \nabla \mathbf{Q}- \mathbf{S}= \frac{1}{\gamma} \mathbf{h}.
\end{equation}
This equation closely resembles the dynamical equation of polarization obtained using the Onsaguer approach, Eq.~\ref{eq: dynamical equation polar wet}. The corotational term, $\mathbf{S}$, introduces the coupling between the fluid flow and the nematic tensor, see~\cite{Amin_Active_Nematics_2018}. The free energy functional has the following form,
\begin{equation}
    \label{eq: nematic free functional}
    \mathcal{F}_\mathbf{Q} = \int d \mathbf{r} \left[ \frac{A}{2}\mathbf{Q}^2+\frac{B}{3}\mathbf{Q}^3+\frac{C}{4}\mathbf{Q}^4 + \frac{K}{2}(\nabla \mathbf{Q})^2 \right].
\end{equation}
Preserving the main characteristics of the polar formalism: the Landau-de Gennes expansion, to account for the isotropic-nematic transition, and the Frank free energy term that accounts for inhomogeneities in the order parameter. The fluid dynamics are coupled to the same stress components as for the polar; the viscous the elastic and the active, which reads $\sigma^{a}=-\zeta \nabla \mathbf{Q}$. For a complete expression of the stress components see~\cite{Amin_Active_Nematics_2018}. It is worth noting that this formalism can be redefined and described by two scalar fields, easing the analytical analysis~\cite{Oza_2016}.

This set of equations satisfies nematic symmetry and, as a result, cannot account for naturally polar phenomena, such as cell migration or particle self-propulsion, that drives bacterial and cell motility. Instead particles are modelled as immobile shakers that change their orientation and generate dipolar flows. Despite lacking polarity, this model is well-suited to describe dense microtubule-kinesin mixtures~\cite{Microtubule_nematic}, as they exhibit nematic symmetry and lack intrinsic polarity.

\subsection{Physical implications of polar and nematic active matter}
Having reviewed the common frameworks for studying polar and nematic active matter at the continuum level, we next discuss the physical implications of these models and distinct features that they represent at the structural level of active materials. To this end, we focus on three fundamental characteristics of active systems: (i) structure and order of the material, (ii) flow fields, and (iii) active interfaces. In each case we discuss the common and distinguishing features of polar and nematic representations of active matter.
\begin{figure}
    \centering
    \includegraphics[width=1.0\linewidth]{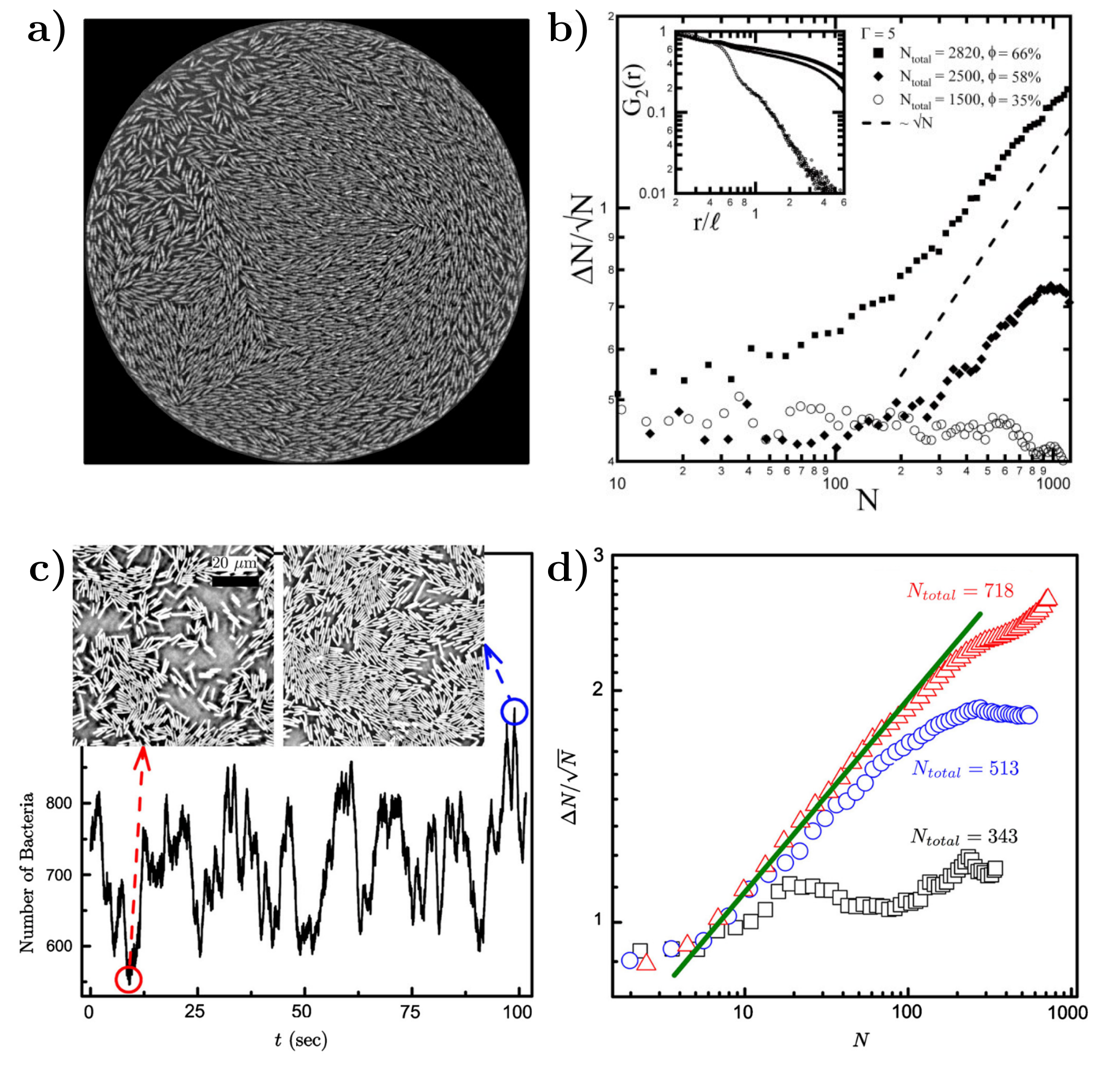}
    \caption{{\bf Giant density fluctuations for distinct experimental realizations.} (a) Experimental snapshot of shaken granular rods, top left displays a large density fluctuation. (b,d) Number fluctuations magnitude, $\Delta N / \sqrt{N}$, as a function of the average number of particles, $N$, for shaken granular rods and bacteria, respectively. (b) Plotted for different total number of rods, $N_{total}$, and cell fraction, $\phi$. Inset plot displays the nematic-order correlation function against the spatial separation. (c) number of bacteria as a function of time, with associated images, in the field of view, for large and small density magnitudes. (d) For different total number of bacteria. Green line has a slope of 0.25 indicating a giant density fluctuations of exponent $\alpha=0.75$. Figures adapted from~\cite{Granular_nem_2007} and~\cite{bacterial_colony_gnf}.}
    \label{fig: GNF}
\end{figure}
\subsubsection{Structure, order and topological defects}
~\\
In two dimensions Mermin-Wagner theorem prohibits long-range order in passive system~\cite{mermin1966absence}. This is, however, not true for active systems, where breaking of detailed-balance at the local particle scale can lead to the emergence of long-range order and thus violation of Mermin-Wagner theory~\cite{tasaki2020hohenberg,andersen2023symmetry}. A recent report has even shown that collection of non-polar active particle can form long-range polar order due to non-reciprocal interactions~\cite{pisegna2024emergent}.
An interesting feature of active systems is the emergence of giant number fluctuations associated with long-range ordering, a phenomenon characterized by an anomalous scaling of the static structure factor at large wavelengths, $S(q) \propto q^{-2}$~\cite{Ramaswamy_2003_Giant_Number_Fluc}. By definition, $S(q \rightarrow 0) = \Delta N^2 / \langle N \rangle$, where $\Delta N$ is the number fluctuation and $\langle N \rangle$ the average number of particles. Assuming the smallest wave vector scales inversely with the system size $V^{-1/d}$, one finds $S(q \rightarrow 0) \sim \langle V \rangle^{2/d} \sim \langle N \rangle^{2/d}$, leading to $\Delta N \sim \langle N \rangle^{\frac{1}{2} + \frac{1}{d}}$. This scaling recovers the equilibrium condition $\Delta N \sim \sqrt{\langle N \rangle}$ only in the limit of infinite dimensions. In two dimensions, both polar and nematic symmetries predict $\Delta N \sim \langle N \rangle^{\alpha}$ with $\alpha \approx 1$~\cite{Ramaswamy_2003_Giant_Number_Fluc,Toner_2019}. Thus, giant number fluctuations are a universal hallmark of active matter, highlighting the long-range correlations absent in passive systems. The existence of giant density fluctuations has been corroborated by experiments on actin filaments~\cite{Densityfluctuations_pnas_2013}, granular rods~\cite{Granular_nem_2007} (Fig.~\ref{fig: GNF}a,b) and bacterial colonies~\cite{bacterial_colony_gnf} (Fig.~\ref{fig: GNF}c,d). However, these systems differ in terms of their symmetries and unlike granular rods, bacterial systems and actin filaments are endowed with polarity at the individual particle level. Despite this, direct quantitative comparisons of the scaling exponents for polar and nematic models, particularly in systems with and without momentum conservation, remains an open area of investigation in continuum models of active matter. 

In addition to large fluctuations, active systems exhibit ordered phases punctuated by topological defects. In polar systems, the polarization field varies smoothly except at defects, where domains of different alignment meet. Topological defects are characterized by an integer charge, representing the net rotation of the polarization field around a closed path. The most energetically favorable polar defects are those with charges $\pm 1$: positive defects ($+1$) can take shapes such as asters, vortices, or spirals, while negative defects ($-1$) exhibit hyperbolic geometries (see Box. 1). 

Nematic active matter, on the other hand, allows for half-integer topological charges, $\pm 1/2$. The $+1/2$ defect has a characteristic shape with broken symmetry, enabling it to self-propel in active systems. In contrast, polar defects with full-integer charges remain symmetric and cannot self-propel; they instead rotate or diffuse passively with the flow field~\cite{rønning2023spontaneousflowsdynamicsfullinteger, Kruse_Rotating_Asters_2004,ardavseva2022topological,vafa2023active}. This distinction has profound consequences for nematic systems, where $+1/2$ defects behave as self-propelled quasi-particles. These motile defects act as organizers of the active flow, influencing mixing rates~\cite{tan2019topological,serra2023defect}, facilitating particle accumulation, and driving three-dimensional structure formation~\cite{meacock2021bacteria,kawaguchi2017topological,MaroudasSacks2020,copenhagen2021topological}.
In biological systems, nematic defects can even localize critical processes, such as cell death and extrusion in epithelial monolayers~\cite{Thuan_Saw_Tissue_Nematics}. For a detailed account of the biological implications of nematic defects, we refer the reader to a recent review~\cite{doostmohammadi2022physics}. The emergence of these nematic defects in bacterial and epithelial cells which posses polarity at the individual particle level is a non-trivial phenomenon, and why such systems do not show polar, integer, topological defects in their bulk remains unresolved.
~\\
~\\
~\\
~\\
~\\
~\\
\begin{framed}
\textbf{Box 2}

\begin{center}
    \includegraphics[width=0.9\linewidth]{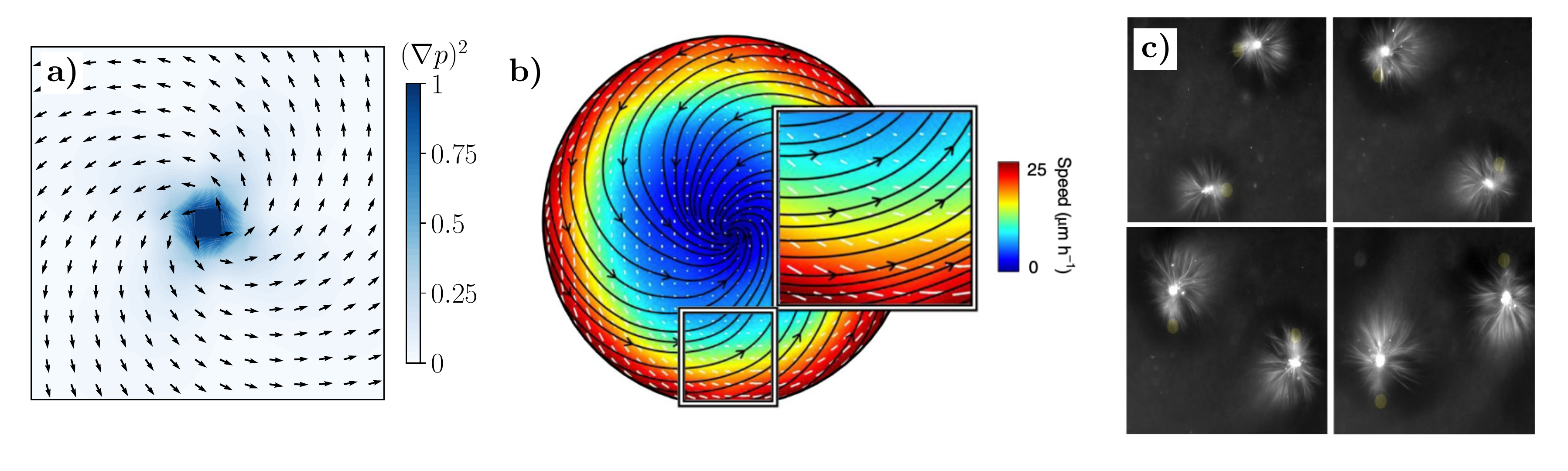}
    \label{fig:activegel}
\end{center}

\noindent\textbf{Figure: Flow fields and defect structures in active polar gels.} . Panel (a) represents the structure of an aster, associated with the polar elastic free energy. Panel (b) shows the flow field of an aster (black streamlines) with its associated velocity magnitude (colorbar) of a confluent C2C12 myoblasts cell culture. (c) Time snapshots of asters, formed by a microtubule-kinesin suspension, that are transported by light activation.  Figures adapted from~\cite{PauGuillamatIntegerTopologicalDefectsTissueMorphogenesis} and~\cite{AstersControllingExp}

Despite this, connections between the two remain poorly understood. Systematic studies are needed to clarify the similarities, differences, and transitions between polar and nematic active matter, particularly under conditions where defects coexist or interact. Establishing such a framework will not only enhance our understanding of active matter but also provide unified principles to guide experimental realizations and applications across diverse systems.

The flow fields and defect structures in active gels are influenced by the interplay between active stresses and the underlying order parameters. Understanding these interactions is key to unravelling the emergent behaviours in active matter systems.
\end{framed}

Despite the ubiquity of polar order in both synthetic and biological active systems, experimental realizations of full-integer defects remain limited. Instances include defects induced by moulds~\cite{FullIntegerTopologicalDefectsCellMonolayersEndresen}, charged boundaries ~\cite{Wioland_2013,PauGuillamatIntegerTopologicalDefectsTissueMorphogenesis}, and depleting microtubule-kinesin mixtures~\cite{Ndlec1997SelforganizationOM}. Confining active particles within a charged boundary induces spontaneous flows \cite{Spontaneous_circulation_as}, initially observed in bacteria \cite{Wioland_2013} and more recently extended to cell assemblies \cite{PauGuillamatIntegerTopologicalDefectsTissueMorphogenesis} (see Box. 2). These experimental realizations underscore the functional role of asters in biological systems, such as driving cell tissue morphogenesis. Further emphasizing the significance of defect control in polar active systems. In this vein, harnessing the movement of topological defects could lead to transformative applications, such as nano-motors or targeted drug delivery systems. While significant progress has been made in active nematics, polar systems present two key findings—one from experimental observations and the other from theoretical studies. In experiments, researchers successfully controlled the movement of asters by using a mixture of microtubules (MT) and light-dimerizable motors~\cite{AstersControllingExp}. This method enables the ordering of asters into arrays and their controlled motion using light as a stimulus. Theoretically, control group theory has been employed to manipulate dynamical activity fields, by tuning the activity level, different states can be achieved, such as altering aster advection, changing the direction of propagating stripes or bands, and transforming bands into asters ~\cite{AsterControlSymTheo}.

The structural differences between polar and nematic active matter—most notably in their distinct topological defect properties—highlight the fundamental incompatibility of using one framework to describe systems governed by the other. Nematic systems, with their half-integer topological defects, exhibit rich dynamics where self-propelling $+1/2$ defects actively shape flow and structure. In contrast, polar systems are characterized by symmetric, full-integer defects that lack intrinsic motility but play distinct roles in rotational or diffusive dynamics. These differences underline that polar and nematic symmetries are not interchangeable and must be treated as separate paradigms when describing active matter. 

\subsubsection{Flow fields, energy spectra, and active Turbulence}
\begin{figure}
    \centering
    \includegraphics[width=1.0\linewidth]{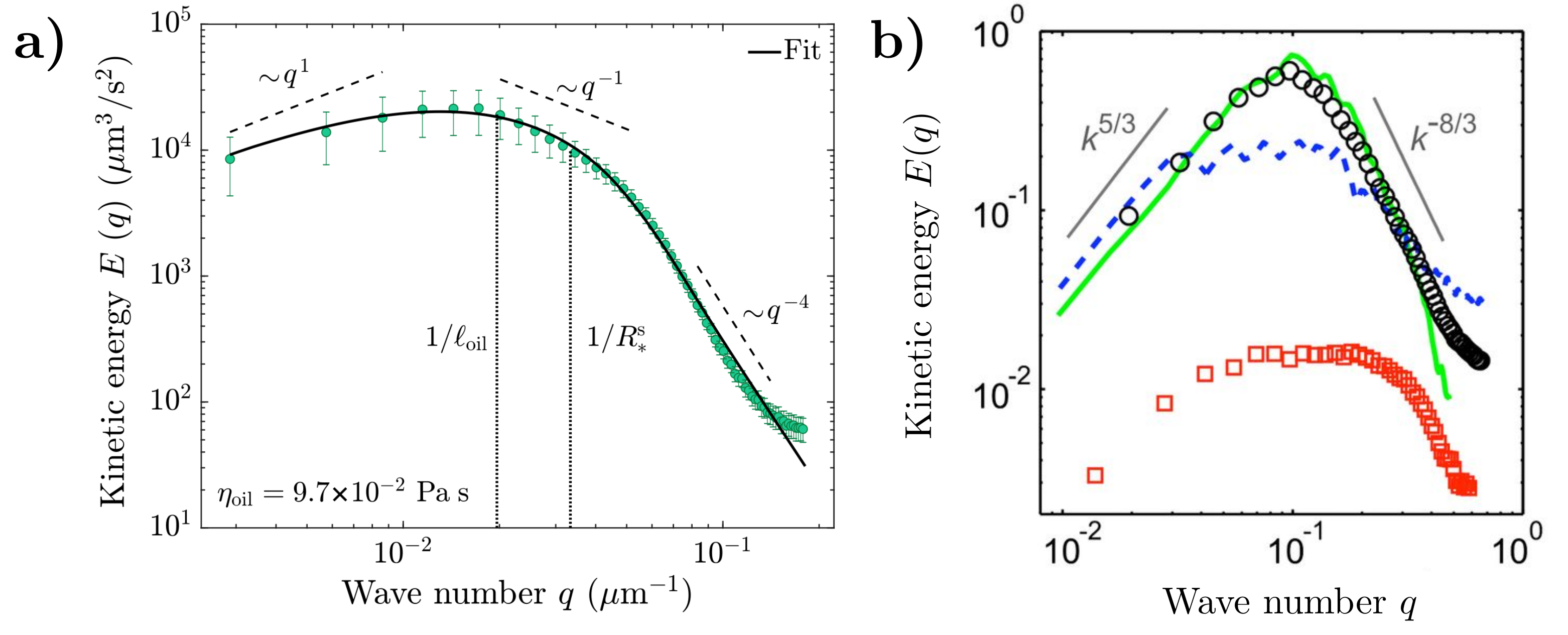}
    \caption{{\bf Distinct energy spectra in polar and nematic active systems.} Illustration of kinetic energy spectra for two experimental realizations: (a) Microtubule-kinesin mixture. Points represent experimental data, while the black line corresponds to the theoretical prediction based on active nematics.(b) \textit{B. subtilis}. Black circles and red squares depict experimental findings for 2D and 3D systems, respectively. The green line shows the expected outcome from the TTSH continuum theory, and the blue line represents predictions from a minimal self-propelled model. Figures adapted from~\cite{Scaling_regimes_active_turbulence} and~\cite{Meso_scale_turbulence}.}
    \label{fig: turbuelt spectra}
\end{figure}

~\\
Emergence of spatiotemporal flows characterized by flow vortices and jets is a  characteristic of many active matter systems where a fluid suspension with small Reynolds number can create chaotic flows, often termed 'active turbulence'. Active nematic theory predicts a scaling, for the kinetic energy, of $E(q) \propto q^{-1}$ for large scales and $E(q) \propto q^{-4}$ for small scales in 2D. Large-scale numerical simulations confirm the large scale scaling at high activities~\cite{rorai2022coexistence}. While experimental realizations such as microtubule suspensions indicate potentially similar scaling (Fig.~\ref{fig: turbuelt spectra}a)~\cite{Scaling_regimes_active_turbulence}, care must be taken in deducing scaling from limited range of wave numbers. Indeed, numerical simulations of active nematics in 2D show that at low-Reynolds numbers an exponential decay can be more likely than any power-law scaling at large wave numbers~\cite{saghatchi2022nematic}. In 3D active nematics for small scales the kinetic energy spectra follows, $E(q) \propto q^{-5}$~\cite{3D_nematic_scaling}. Recent work has found that nematic systems transfer energy across different length scales, specifically in the defect chaos regime~\cite{pearce2024topologicaldefectsleadenergy}, where the energy interchange occurs between the orientation and scalar order nematic parameters during the continuous creation-annihilation of defects. This result has no analogue in polar systems and the features of energy transfer due to defects dynamics is still unknown. Contrary to nematic systems, polar turbulence has been understudied and often polar realizations are encompassed within nematic turbulence~\cite{alert_Turbulence_2022}. Indeed, a recent large-scale numerical study of polar active gels shows $E(q) \propto q^{-1/2}$ for large scales (small $q$) and activity-dependent exponential decay at small scales (large wave numbers $q$)~\cite{andersen2023symmetry}.

Employing hydrodynamic active polar models has had some success, such as in bacterial suspensions of \textit{B. Subtilis}, where a variation of the incompressible dry Toner-Tu equation, the Toner–Tu–Swift–Hohenberg equation (TTSH), aligned well with the experiments and revealed non-universal exponents~\cite{Meso_scale_turbulence, EFrey_New_Class_turbulence} (Fig.~\ref{fig: turbuelt spectra}b). Systems with momentum conservation in polar matter, however, still require further characterization and analytical work. Examples of polar systems where no fit between the active turbulent spectra and theoretical models have been found include: epithelial tissue monolayers~\cite{EpithelialTurbulence}, sperm suspensions~\cite{Turbulence_of_swarming_sperm}, and certain types of bacterial suspensions~\cite{Fluid_Dynamics_Bact_2013}. This discrepancy likely arises from the challenges in describing the exact type of alignment exhibited by certain active particle suspensions, whether it is polar or nematic. The mismatch between theory and experimental realizations could potentially be resolved by introducing mixed symmetry models.

The emergence of spatiotemporal flows, characterized by vortices, jets, and turbulent-like behavior, demonstrates the rich and diverse dynamics of active matter systems. However, the contrasting scaling laws and energy transfer mechanisms in polar and nematic turbulence highlight their fundamental differences. Nematic systems exhibit energy cascades driven by defect dynamics, with unique scaling behaviors and energy exchanges that have no direct analogue in polar systems. Meanwhile, polar turbulence remains understudied, with theoretical models often conflated with nematic frameworks, leading to gaps in understanding and mismatches with experimental results. This lack of clarity is particularly evident in systems where the type of alignment—polar or nematic—is ambiguous, further underscoring the need for models that incorporate mixed or hybrid symmetries. 
\begin{table}[ht]
    \centering
    \caption{Scaling reported for small length limit ($q\rightarrow\infty$) and large length limit ($q\rightarrow 0$)} of the kinetic energy spectra, $E(q)$, for experimental and theoretical systems.
    \begin{tabular}{p{8cm} p{3.3cm} p{3cm}}
        \toprule
        \textbf{System} & \textbf{Small length \newline ($q\rightarrow\infty$) scaling} & \textbf{Large length \newline ($q\rightarrow 0$) scaling}\\
        \midrule
        \textit{B. subtilis} (2D),~\cite{Meso_scale_turbulence}. & $E(q) \propto q^{-8/3}$ & $E(q) \propto q^{5/3}$\\
        Sperm suspension (2D),~\cite{Turbulence_of_swarming_sperm}. & $E(q) \propto q^{-3}$ & --\\
        \textit{E. Coli} (3D),~\cite{Turbulence_E_Coli_3D}. & $E(q) \propto q^{-3}$ & --\\
        Microtubule kinesin suspension(2D),~\cite{Scaling_regimes_active_turbulence}. & $E(q) \propto q^{-4}$ & $E(q) \propto q^{1}$\\
        Epithelial cell monolayers (2D),~\cite{EpithelialTurbulence}. & $E(q) \propto q^{-9/2}$ & --\\
        Mean field continuum active \newline nematics (2D)~\cite{giomi2015geometry}. & $E(q) \propto q^{-4}$ & $E(q) \propto q^{-1}$\\
        Continuum active nematics (3D)~\cite{3D_nematic_scaling}. & $E(q) \propto q^{-5}$ &--\\
        Continuum active polar (2D)~\cite{andersen2023symmetry}. & activity- \newline dependent  & $E(q) \propto q^{-1/2}$ \\
        \bottomrule
    \end{tabular}
\end{table}

Future work must systematically compare polar and nematic turbulence to establish a comprehensive framework that identifies their unique and overlapping features. Such studies are crucial for bridging the gap between theory and experiment, as well as for uncovering universal principles that govern turbulent-like flows in active matter. By addressing these open questions, we can better understand the role of symmetry in shaping the emergent behavior of active systems and unlock their full potential for practical applications.

\subsubsection{Active interfaces in multiphase systems}  
~\\  
Active interfaces in multiphase systems represent a fascinating domain where the interplay between activity, symmetry, and interfacial dynamics leads to emergent behaviours~\cite{doostmohammadi2015celebrating,basan2011undulation,comelles2021epithelial,xu2023geometrical,adkins2022dynamics}. In both active nematic and polar systems, the coupling between interfacial forces and active stresses profoundly influences the morphology, stability, and dynamics of fluid-fluid or fluid-solid boundaries. Despite sharing some common features, the structural and dynamical differences between nematic and polar active matter result in markedly distinct interfacial phenomena, which have implications for both natural and synthetic systems.

In active nematics, interfaces are often governed by the coupling between director alignment and the curvature of the interface. The interplay of active stresses with interfacial tension can lead to phenomena such as spontaneous interface undulations, shape instabilities, and defect-mediated flows~\cite{doostmohammadi2016defect,alert2022fingering}. These dynamics are particularly relevant in systems such as liquid-liquid emulsions, where active nematic layers have been shown to exhibit topological defect dynamics at interfaces, driving self-sustained flows and contributing to phase separation~\cite{blow2014biphasic,caballero2022activity,chaithanya2024transport}. Additionally, stress's as a result of high activities often drive droplet generation at these interface's~\cite{adkins2022dynamics}. Furthermore, recent studies highlight the role of defect formation and annihilation in generating anisotropic stresses, which can deform droplets or thin films and influence their transport properties~\cite{Topology_dyn_an_vesicles,Tunable_structure}. Importantly, nematic activity tends to homogenize interface motion, with defect activity creating localized flows that are critical for interfacial reorganization.

In contrast, polar active interfaces display inherently directional behavior due to the vectorial nature of their order parameter. This polarity introduces a bias in interface dynamics, where active stresses can generate directed motion along the interface, leading to phenomena such as persistent interfacial currents and shape asymmetries~\cite{snezhko2016complex}. For example, in polar active droplets, spontaneous breaking of symmetry has been observed, where interfacial flows align with the internal polarization field, creating persistent motion~\cite{alert2019active}. Similarly, polar order can induce Marangoni-like flows at the interface, enhancing mixing and material transport in a directional manner~\cite{carenza2020soft}. These properties make polar active interfaces particularly suited for applications requiring directed transport or dynamic shape control.

Despite these differences, both nematic and polar systems exhibit emergent properties at active interfaces that challenge conventional paradigms in multiphase fluid dynamics. In nematics, the defect dynamics near interfaces can drive bulk flows~\cite{Nematic_isotropic_interfaces}, In polar active matter systems, interfacial flows can significantly influence bulk polarization and alignment. For instance, studies have shown that adhesive and aligning walls can enhance polarity heterogeneity within the bulk, enabling polar active particles to climb up vertical surfaces~\cite{fins2024steer}. 
Notably, the stability of interfaces in active systems is highly sensitive to the underlying symmetry. In polar systems, interfacial flows can induce bulk condensation, leading to directed instability growth~\cite{kant2024bulk}. Conversely, in nematic systems, the presence of topological defects can mediate equilibration processes, influencing the overall stability and behavior of the interface~\cite{Amin_Active_Nematics_2018}. These findings underscore the critical role of symmetry in determining the interfacial dynamics and stability of active matter systems. 
 
Moreover, continuum models have been particularly successful in characterizing the movement of a confined collective of particles. By increasing the activity parameter, the confined polar particles transition from a stationary droplet to a motile one by creating a vortex dipole~\cite{droplet_2014, symmetry_breaking_droplets} see Box. 3. Recently, the study has been generalized to confined droplets in a channel~\cite{Tiribocchi2023}. Depending on the height of the channel and the activity coefficient, it gives rise to different droplet forms: static, cell-like, and worm-like cells.

These studies have been extended to three dimensions, where it has been shown that a droplet can crawl due to a combination of treadmilling and contractile stresses~\cite{Tjhung_2015}, showing great similarity to different eukaryotic structures such as lamellipodia and pseudopods. 

The dynamics of these droplets are governed by the interplay between active stresses and the surrounding fluid. The active stresses generate internal flows within the droplet, leading to shape deformations and directed motion~\cite{symmetry_breaking_droplets}. Theoretical models have been developed to describe these behaviors, incorporating the effects of active forces, surface tension, and hydrodynamic interactions~\cite{Droplet_active_nematics}. These models provide insights into the mechanisms driving the motility and shape changes of active droplets.

Recent advancements in experimental techniques have allowed for the direct observation of these phenomena in synthetic and biological systems. For instance, experiments with microtubule-kinesin mixtures have demonstrated the formation of motile droplets that exhibit complex behaviors such as crawling and division ~\cite{Tim_Sanchez_Microtubule_Motors,Ndlec1997SelforganizationOM}. These findings highlight the potential of active droplets as model systems for studying the principles of active matter and their applications in designing synthetic active materials.

Future work should aim to bridge the gap between polar and nematic active interfaces, identifying universal principles where possible while carefully characterizing their distinct behaviors. Key open questions include the role of mixed symmetry states at active interfaces, the impact of confinement and external fields, and the interplay between interfacial dynamics and bulk phase transitions. Addressing these challenges will not only deepen our understanding of active matter but also unlock new possibilities for controlling interfacial behavior in soft and living systems.

\begin{framed}
\textbf{Box 3}

\begin{center}
    \includegraphics[width=0.95\linewidth]{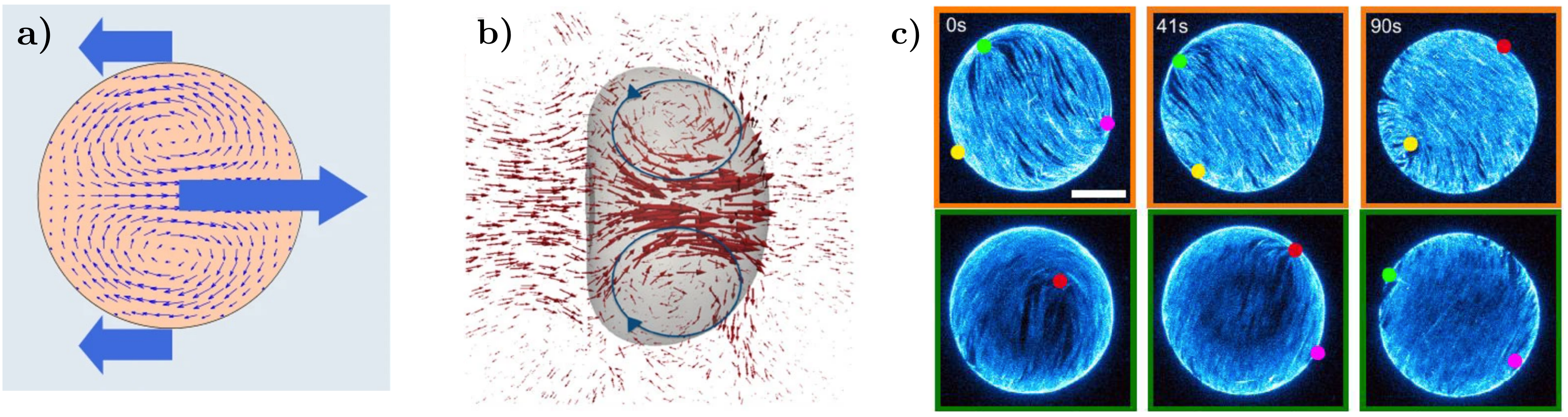}
    \label{fig:activedroplet}
\end{center}
\noindent\textbf{Figure: Polar and nematic active droplets.} Panel (a) shows the schematic picture of the internal flow field (small blue arrows) of a 2D active droplet where the active stress generates internal flows, leading to shape deformations and directed motion. The wider arrows highlight the forces exerted from the droplet on the surrounding medium. Figure (b) illustrates the toroidal fluid flow of a three-dimensional active droplet. (c) temporal hemisphere projections of a nematic active droplet, coloured dots represent the different +1/2 defects . Bar length is $20 \mu m$. Figures adapted from~\cite{droplet_2014}, ~\cite{symmetry_breaking_droplets} and~\cite{Topology_dyn_an_vesicles}.

The dynamics of active droplets are influenced by the interplay between active stresses and the surrounding fluid. Understanding these interactions is key to unravelling the emergent behavior in active matter systems.
\end{framed}

Overall, the interplay of polar and nematic symmetries, along with the continuous input of energy at the microscopic level, gives rise to a rich tapestry of behavior in active matter systems. Each phenomenon—phase separation, active turbulence, topological defects, and active droplets—illustrates the unique and complex physics that arise in these non-equilibrium systems.

\section{Mixed polar and nematic order in active matter}
The review of polar and nematic order in active matter highlights the fundamentally different physics that emerge when these symmetries are explored in isolation, both at the discrete particle level and at the continuum level. As we discussed this distinction becomes even more pronounced when considering phenomena such as phase separation, jamming, active turbulence, topological defects, and active droplets. In real systems, however, polar and nematic symmetries often co-exit.

\subsection{Experimental systems show dual features}

\begin{figure}
    \centering
    \includegraphics[width=1.0\linewidth]{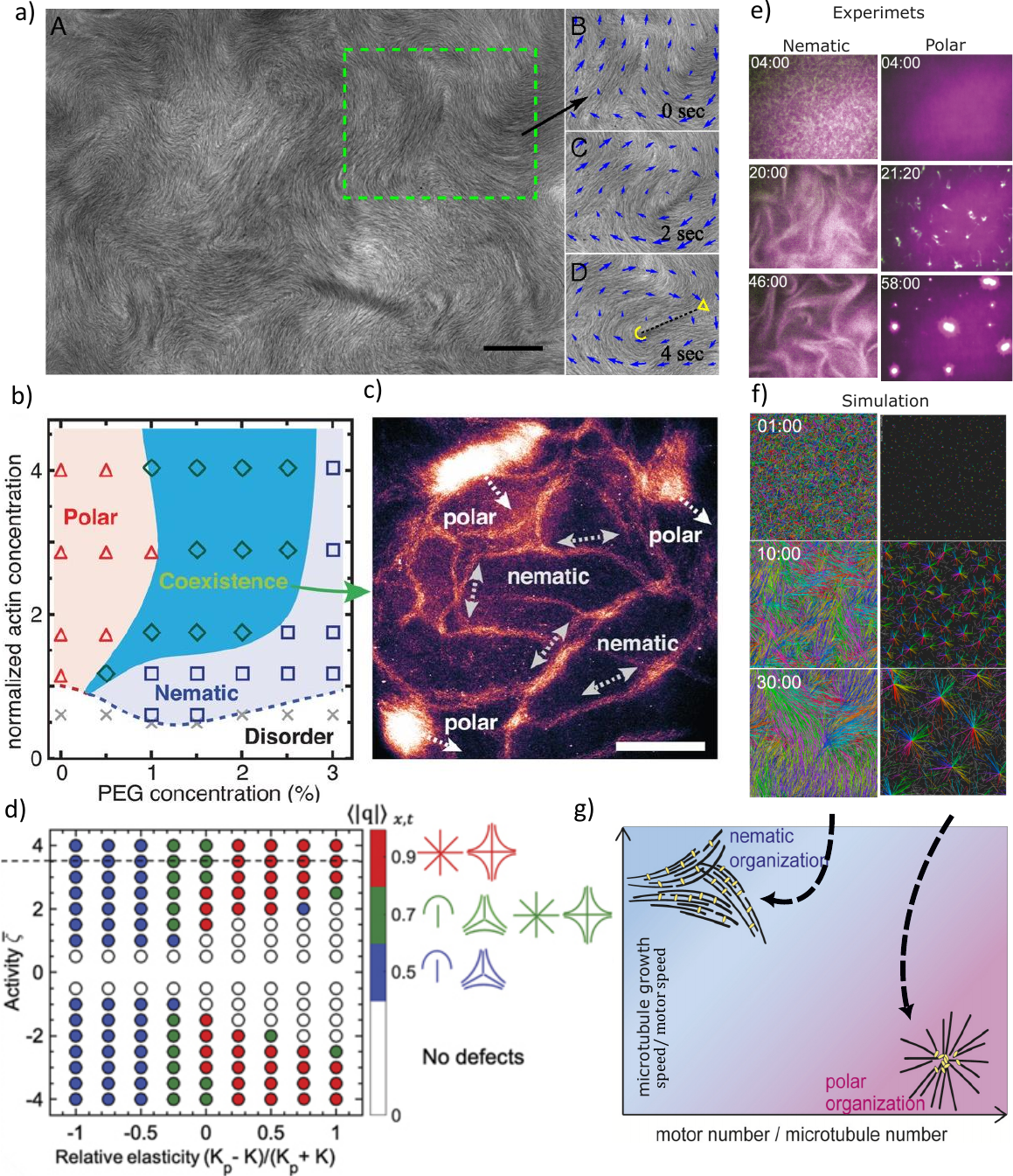}
    \caption{{\bf Mixed polar and nematic order in active systems.} In a) bacteria, often modelled as polar individuals demonstrate half-integer defects, showing in yellow, that are usually associated with a collective of nemetic individuals. Coexistence displayed by (b) the experimental phase diagram and (c) the experimental image of polar clusters and nematic lanes in an actomyosin motility assay. A simulation phase diagram in which coexistence has been studied (d). Microtubule organisation driven by kinesin-5 KIF11 organise as either nematic bands or polar asters (e); simulation of the same system (f) revels two control parameters (g).  Figures adapted from~\cite{Zhou2014} (a),~\cite{Coexisting_ordered_states_Frey} (b)$\&$(c), ~\cite{Amiri_2022}} (d), and~\cite{Roostalu2018} (e), (f) $\&$ (g). 
    \label{fig:exp}
\end{figure}

Experimental systems do not always follow the strict requirements set in models to distinguish polar from nematic symmetries. While phenomenological models require symmetry considerations to be in-built, experimental systems do not have such restrictions. A prime example is bacterial active matter. While an individual bacterium might not in itself have a head-tail symmetry, the hydrodynamic flows generated by the force dipole of a swimming bacterium do have head-tail symmetry. Dense suspensions of bacteria therefore regularly display half-integer defects, a characteristic of 2D active nematics. An example is provided in Fig.~\ref{fig:exp}a~\cite{Zhou2014}. Biological matter additionally display co-existence of integer and half-integer defects within the same system, with different types having been shown to have biological consequences in hydra~\cite{MaroudasSacks2020}, bone marrow cells~\cite{makhija2024topological}, and bacteria that display rosettes~\cite{meacock2021bacteria} through the collision of fast moving +1/2 defects.   %

An important experimental result has been the organization of cytoskeletal networks of microtubules driven by kinesine in either nematic or polar structures (see Fig.~\ref{fig:exp}e).  It was found that two important parameters controlled the formation of nematic bands or polar asters: the motor-to-microtubule ratio, and the relative speeds of microtubule growth and motor movement (see Fig.~\ref{fig:exp}f,g). Additionally, the asymmetry of microtubule growth properties plays a role in polar versus nematic structures; with fast-growing plus-ends and static minus ends favouring nematic networks~\cite{Roostalu2018}.

Defects, however, are not conclusive evidence that systems exhibit dual polar and nematic behavior, as theoretical work has shown that 2D nematic theory can generate integer defects under certain conditions~\cite{Thijssen2020,vafa2022defect}.  Another interesting pattern is shown in Fig.~\ref{fig:exp}b, where the coexistence of fluctuating nematic and polar symmetry phases was observed in an actomyosin motility assay~\cite{Coexisting_ordered_states_Frey}. While some regions showed polar waves, others exhibited nematic lanes. Additionally, the structures were found to be dynamic and stable enough to reform when disrupted. The coexistence of patterns is attributed to sufficiently weak alignment interactions between individual filaments, with increased interaction strength preventing coexistence see Fig.~\ref{fig:exp}c.


Even though active turbulence has shown promising alignment between experiments and theory, such as bacterial suspension being described by the polar Toner-Tu-Swift-Hohenberg equations~\cite{Meso_scale_turbulence} and kinesin microtubule suspensions within the continuum nematic framework~\cite{Scaling_regimes_active_turbulence}, certain experiments realizations remain challenging to encompass within a specific framework~\cite{Turbulence_of_swarming_sperm,EpithelialTurbulence}. For instance, three dimensional bacterial swarms~\cite{Turbulence_E_Coli_3D} which, for an \textit{E.Coli} suspension, exhibits a small-scale energy spectrum scaling of $E(q) \propto q^{-3}$, significantly diverge from the nematic formalism, which predicts $E(q) \propto q^{-5}$~\cite{3D_nematic_scaling} and have neither been encompassed within a polar formalism~\cite{Chatterjee_2021}. Emphasizing the importance of continuing the study of active turbulence from a symmetry perspective to unravel the diverse scaling behaviours observed. The introduction of novel mixed symmetry models could offer deeper insights into both energy scaling and the intrinsic symmetries exhibited by particles at the microscopic level.

\subsection{Modeling mixed symmetries}

An important recent development has been to look at systems with multiple coupled order parameters. Popular systems include coupled hexatic-nematic models in the case of MDCK cells~\cite{Eckert2023}, and ferromagnetic nematic liquid crystals~\cite{Sebastin2023, Rudquist2021}. While conventional nematics have order with two-fold symmetry, the ferromagnetic nematic is characterized and observed to posses the formation, in the absence of applied electric field, of spontaneously polar domains separated by distinct domain boundaries~\cite{Chen2020}.  Models for such materials therefore necessitate a coupling of polar and nematic order. Both the polar and the nematic have fixed magnitudes and the coupling might take place through the gradients~\cite{flexonem3}. These models explicitly consider both order parameters to exist separately though some coupling. However, it is also possible to consider a model with a single order parameter that incorporates the second symmetry via an additional free energy term \cite{marchetti_hydrodynamics_2013}. 

A way to approach such a model was discussed in~\cite{marchetti_hydrodynamics_2013} in the context of self-propelled hard rods. Here coupled equations for the conserved particle density $\rho$ and the two coupled orientational order parameters $\mathbf{P}$ (polar) and $\textbf{Q}$ (nematic) were considered in the absence of hydrodynamics. A comprehensive hydrodynamic theory of $p$-atic liquid crystals have been recently developed, which allows for various forms of coupling between different $p$-atic order parameters~\cite{giomi2022hydrodynamic}.

\subsection{Co-existence of full and half-integer topological defects}

In an attempt to study the presence of half-integer defects in systems comprised of individually polar entities, Ref.~\cite{Amiri_2022} formulated a model that stems from the active polar wet formalism, while adding nematic-like terms into the free energy functional, in the form of the nematic elastic free energy (Eq. \ref{eq: Frank free energy unifying model}).
\begin{equation}
    \label{eq: Frank free energy unifying model}
    \mathcal{F} = \int d \mathbf{r} \left[ A \left( -\frac{\mathbf{P}^2}{2} + \frac{\mathbf{P}^4}{2} \right) + \frac{K_\mathbf{P}}{2} (\nabla \mathbf{P})^2 + \frac{K_n}{2} \left( \nabla \left( \mathbf{P} \mathbf{P}^T - \mathbf{P}^2 \frac{\mathbf{I}}{2} \right) \right)^2 \right].
\end{equation}
On one hand, the $(\nabla \mathbf{P})^2$ term penalizes gradients in the polarity field and therefore is minimized when all the particles have the same direction. Conversely, the last term penalizes gradients in the uniaxial orientations of the polar vector field, and thereby does not account for the direction. Therefore, $K_\mathbf{P}$ aligns the system in a polar manner, while $K_n$ prefers nematic alignment. This addition integrates both types of alignment into a unified formalism, enabling a comprehensive study of the interplay between polar and nematic behaviors.

This model predicts the emergence of full and half-integer defects and is capable of describing systems with mixed symmetry. Excitingly, the model also predicts a region in phase space for the coexistence of half and full-integer defects (see Fig.~\ref{fig:exp}d). This potentially offers a new way to understand experimental realizations of polar particles that create half-integer defects. However, further characterization and comparison to experimental realizations and theory are required.

\subsection{Pattern-induced local symmetry breaking}

Motivated by the experimental realization of synthetic actomyosin motility assay (Fig.~\ref{fig:exp}d))~\cite{Coexisting_ordered_states_Frey}, where polar and nematic patterns interact, a novel model is introduced in~\cite{Symmetry_Breaking_Polar_Nem_Frey}. This model is derived from a kinetic approach, specifically from the Boltzmann equation (Eq. \ref{eq: boltzmann equation}), introducing both the polarization field and the nematic tensor as a function of the one-particle distribution $f(r,\theta,t)$. This model adds a tunable collision rule. Colliding particles align in a polar manner if the angles between their velocity are smaller than $\frac{\pi}{2} + \psi$, and in a nematic way otherwise. The parameter $\psi$ determines the polar bias of the system: $\psi = 0$ corresponds to purely nematic collisions, while $\psi = \frac{\pi}{2}$ results in purely polar collisions.

The interplay between the polar bias and the mean particle density displays intricate behavior. As expected, for the extreme values of $\psi = 0$ and $\frac{\pi}{2}$, the systems exhibit nematic and polar behavior, respectively. This is characterized by the appearance of nematic bands and travelling polar waves. However, for intermediate values of the polar bias, the systems display an entangled behavior where nematic bands and travelling waves can coexist. In that regime, a bistable zone appears where nematic bands are unstable to the polar perturbation and transition to polar travelling waves.

\subsection{Topological defect strings}

A recent work has studied defects in system with coexisting nematic and polar orientational order parameters~\cite{fvafastrings}. Similar to the previous section, the authors use a free energy that incorporates both symmetries. They find three distinct phases - isotropic, nematic, and nematopolar; a phase in which both polar and nematic features are intertwined.  Interestingly, in the nematopolar phase, they find confined strings that connect defects which arise due to the individual particle polarity. 

Half-integer defects connected by strings were also recently observed in an experimental system of endothelial cell~\cite{Ruider2024}. Placing the cell layer under constant shear flow resulted in halg-integer defects being connected by strings. Combining experiments with continuum modeling, the shear flow, which introduces an external polar field to the system, was argued to contribute to the string formation, in analogy with the Zeeman effect in pair-superfluid systems~\cite{james2011phase}. 

Finally, strings were recently also studied in systems that compare polar and nematic topological defects~\cite{strings3}, by using a free energy with both polar and nematic symmetries. As a consequence, a system with both polar and apolar defects is generated. By controlling the relative number and the alignment strength, the authors were able to control the prominence of the strings. 

 The limited number of studies on mixed symmetry systems are already finding rich and new phenomenon. Future work focusing on this might uncover exiting new structures that not only could be relevant as an intellectual curiosity but with biological ramifications, as found in Ref.~\cite{Ruider2024}.

The coexistence of polar and nematic symmetries in active matter challenges the conventional dichotomy between these two forms of order. Experimental observations of hybrid behavior, such as mixed defect types, pattern coexistence, and dual energy scaling, underscore the need for theoretical models that incorporate both symmetries. The distinction between polar and nematic active matter, while foundational at the level of symmetry, is insufficient to describe many real-world systems where these symmetries intertwine. Models that couple polar and nematic order parameters or introduce unified frameworks offer promising approaches to address this complexity. However, the exact mechanisms governing the coexistence of these symmetries remain poorly understood. Further studies should aim to elucidate the role of alignment interactions, defect dynamics, and the coupling between bulk and interfacial phenomena in systems with mixed symmetry. Establishing a deeper understanding of the connections and differences between polar and nematic behavior will be crucial for advancing both the theoretical framework of active matter and its application to biological and synthetic systems.

\section{Current Challenges and Future Directions}

The study of coexisting symmetries in active matter, particularly focusing on polar and nematic order, is an emerging field with increasing evidence pointing to their coexistence and its importance. While significant progress has been made in understanding polar and nematic systems individually, the simultaneous presence of these symmetries in biological systems presents new challenges and opportunities. The coexistence of polar and nematic order is not just a theoretical curiosity but a fundamental aspect of many biological processes, from cellular organization to tissue morphogenesis. Understanding how these symmetries interact and influence each other is crucial for developing a comprehensive framework for active matter.

\subsection{Current Challenges}

\subsubsection*{Complexity of Biological Systems}

Biological systems often exhibit a level of complexity that is difficult to capture with current theoretical models. For instance, the interplay between different types of symmetries and the influence of biochemical signaling pathways are not fully understood. Models that incorporate these complexities are needed to better mimic biological realities. For example, in cellular environments, the simultaneous presence of polar and nematic order is common, yet most models treat these symmetries in isolation, failing to capture the full spectrum of behavior observed in living systems.

\subsubsection*{Experimental Validation}

While theoretical models have predicted various phenomena such as the coexistence of polar and nematic order, experimental validation remains limited. High-resolution imaging and advanced microscopy techniques are required to observe these phenomena in living systems. Quantifying both polarity and nematic order in cells is particularly challenging due to the dynamic and heterogeneous nature of biological tissues. Techniques such as fluorescence polarization microscopy and particle image velocimetry need to be further refined to accurately measure these properties.

\subsubsection*{Integration of Multiple Order Parameters}

Many biological systems exhibit multiple coupled order parameters. Developing models that can accurately describe these interactions is challenging but essential for a comprehensive understanding of active matter. Current models either include multitude of parameters, which makes comparison to experiments non-trivial, or often simplify these interactions, which can lead to discrepancies between theory and experiment. For instance, in epithelial tissues, cells exhibit both nematic alignment due to cell shape and polar alignment due to cell migration, necessitating models that can handle such dual characteristics.

\subsubsection*{Topological Defects}

The behavior of topological defects in active matter, particularly the coexistence of full and half-integer defects, is not fully understood. While some models predict these phenomena, there is a lack of experimental data to support these predictions. Understanding the dynamics of these defects is crucial for applications in material science and biology.

\subsection{Future Directions}

\subsubsection*{Advanced Experimental Techniques}

The development of new experimental techniques, such as super-resolution microscopy and single-molecule tracking, will be crucial for validating theoretical models. These techniques can provide detailed insights into the behavior of active matter at the microscopic level, allowing for more accurate comparisons between theory and experiment. Additionally, the use of optogenetics and microfluidics can help manipulate and control the polarity of biological systems, providing new ways to test theoretical predictions on the role of polar and nematic symmetries in biological systems.

\subsubsection*{Multiscale Modeling}

Bridging the gap between microscopic and macroscopic descriptions of active matter requires multiscale modeling approaches. These models should integrate molecular dynamics simulations with continuum theories to capture the full range of behavior observed in biological systems. For example, combining agent-based models with hydrodynamic simulations can provide a more comprehensive understanding of how polar and nematic symmetry at the individual cell scale affect the collective phenomena at multicellular scales.

\subsubsection*{Minimal Models with Coupled Order Parameters}

Developing minimal models that couple polar and nematic order parameters is essential for understanding the coexistence of these symmetries in biological systems. These models should account for the interactions between different types of order and the resulting emergent behavior. For instance, models that incorporate both polar and nematic alignment can help explain the formation of complex patterns in cell monolayers and the dynamics of cytoskeletal networks.

\subsubsection*{Application to Synthetic Systems}

Understanding the principles of active matter can lead to the design of synthetic systems with tailored properties. For example, creating materials that mimic the adaptive and self-healing properties of biological tissues could have significant implications for material science and engineering. Synthetic active materials that combine polar and nematic order could be used to develop smart materials with programmable behavior, such as shape-changing and self-repairing capabilities.

\subsection*{Bridging Physics and Biology}

Bridging the gap between physics and biology requires a nuanced understanding of the interplay between polar and nematic symmetries in active matter. These distinct yet often coexisting symmetries offer critical insights into biological processes, from cellular motility to tissue morphogenesis. For instance, the co-occurrence of polar and nematic order in systems such as epithelial tissues or bacterial colonies underscores the need for models that capture their complex interactions. Addressing these challenges could enable the design of novel biotechnological applications, such as anisotropic drug delivery systems or bioengineered tissues with controlled mechanical properties. Moreover, understanding the transitions and coexistence between polar and nematic behavior may provide a framework for deciphering symmetry-breaking events in biological systems, offering a deeper connection between physical principles and biological function.

In conclusion, while significant progress has been made in understanding polar and nematic systems individually, the coexistence of these symmetries in biological systems presents new challenges and opportunities. Addressing these challenges will require advanced experimental techniques, multiscale modeling, interdisciplinary collaboration, and the development of minimal models that couple multiple order parameters. By overcoming these obstacles, we can bridge the gaps between physics and biology and unlock new possibilities for research and innovation. This integrated approach will not only enhance our understanding of active matter but also pave the way for novel applications that leverage the unique properties of these systems.

\section*{Acknowledgements}
\noindent
We sincerely apologize to our colleagues whose work could not be cited in this review due to space limitations.
We gratefully acknowledge financial support by the Novo Nordisk Foundation grant no. NNF18SA0035142, NERD grant no. NNF21OC0068687, the Villum Fonden Grant no. 29476, and the European Union via the ERC-Starting Grant PhysCoMeT.

\newpage
\bibliographystyle{naturemag}
\bibliography{list_of_refs/Descrete_polar/discrete, list_of_refs/Continuum_polar/Continuum,list_of_refs/Introduction/introduction, list_of_refs/Exotic/Exotic}

\begin{thebibliography}{100}
\expandafter\ifx\csname url\endcsname\relax
  \def\url#1{\texttt{#1}}\fi
\expandafter\ifx\csname urlprefix\endcsname\relax\def\urlprefix{URL }\fi
\providecommand{\bibinfo}[2]{#2}
\providecommand{\eprint}[2][]{\url{#2}}

\bibitem{Pedley_kessler}
\bibinfo{author}{Pedley, T.~J.} \& \bibinfo{author}{Kessler, J.~O.}
\newblock \bibinfo{title}{Hydrodynamic phenomena in suspensions of swimming microorganisms}.
\newblock \emph{\bibinfo{journal}{Annual Review of Fluid Mechanics}} \textbf{\bibinfo{volume}{24}}, \bibinfo{pages}{313--358} (\bibinfo{year}{1992}).

\bibitem{lauga2009hydrodynamics}
\bibinfo{author}{Lauga, E.} \& \bibinfo{author}{Powers, T.~R.}
\newblock \bibinfo{title}{The hydrodynamics of swimming microorganisms}.
\newblock \emph{\bibinfo{journal}{Reports on Progress in Physics}} \textbf{\bibinfo{volume}{72}}, \bibinfo{pages}{096601} (\bibinfo{year}{2009}).

\bibitem{bechinger2016active}
\bibinfo{author}{Bechinger, C.} \emph{et~al.}
\newblock \bibinfo{title}{Active particles in complex and crowded environments}.
\newblock \emph{\bibinfo{journal}{Reviews of modern physics}} \textbf{\bibinfo{volume}{88}}, \bibinfo{pages}{045006} (\bibinfo{year}{2016}).

\bibitem{gompper20202020}
\bibinfo{author}{Gompper, G.} \emph{et~al.}
\newblock \bibinfo{title}{The 2020 motile active matter roadmap}.
\newblock \emph{\bibinfo{journal}{Journal of Physics: Condensed Matter}} \textbf{\bibinfo{volume}{32}}, \bibinfo{pages}{193001} (\bibinfo{year}{2020}).

\bibitem{marchetti_hydrodynamics_2013}
\bibinfo{author}{Marchetti, M.~C.} \emph{et~al.}
\newblock \bibinfo{title}{Hydrodynamics of soft active matter}.
\newblock \emph{\bibinfo{journal}{Rev. Mod. Phys.}} \textbf{\bibinfo{volume}{85}}, \bibinfo{pages}{1143--1189} (\bibinfo{year}{2013}).

\bibitem{Prost_nature_2015}
\bibinfo{author}{Prost, J.}, \bibinfo{author}{J{\"u}licher, F.} \& \bibinfo{author}{Joanny, J.-F.}
\newblock \bibinfo{title}{Active gel physics}.
\newblock \emph{\bibinfo{journal}{Nature Physics}} \textbf{\bibinfo{volume}{11}}, \bibinfo{pages}{111--117} (\bibinfo{year}{2015}).

\bibitem{Amin_Active_Nematics_2018}
\bibinfo{author}{Doostmohammadi, A.}, \bibinfo{author}{Ign{\'e}s-Mullol, J.}, \bibinfo{author}{Yeomans, J.~M.} \& \bibinfo{author}{Sagu{\'e}s, F.}
\newblock \bibinfo{title}{Active nematics}.
\newblock \emph{\bibinfo{journal}{Nature Communications}} \textbf{\bibinfo{volume}{9}}, \bibinfo{pages}{3246} (\bibinfo{year}{2018}).

\bibitem{Chat2020}
\bibinfo{author}{Chaté, H.}
\newblock \bibinfo{title}{Dry aligning dilute active matter}.
\newblock \emph{\bibinfo{journal}{Annual Review of Condensed Matter Physics}} \textbf{\bibinfo{volume}{11}}, \bibinfo{pages}{189–212} (\bibinfo{year}{2020}).

\bibitem{needleman2017active}
\bibinfo{author}{Needleman, D.} \& \bibinfo{author}{Dogic, Z.}
\newblock \bibinfo{title}{Active matter at the interface between materials science and cell biology}.
\newblock \emph{\bibinfo{journal}{Nature Reviews Materials}} \textbf{\bibinfo{volume}{2}}, \bibinfo{pages}{1--14} (\bibinfo{year}{2017}).

\bibitem{Symmetry_Theormodynamics_2022_Marchetti}
\bibinfo{author}{Bowick, M.~J.}, \bibinfo{author}{Fakhri, N.}, \bibinfo{author}{Marchetti, M.~C.} \& \bibinfo{author}{Ramaswamy, S.}
\newblock \bibinfo{title}{Symmetry, thermodynamics, and topology in active matter}.
\newblock \emph{\bibinfo{journal}{Phys. Rev. X}} \textbf{\bibinfo{volume}{12}}, \bibinfo{pages}{010501} (\bibinfo{year}{2022}).

\bibitem{Kilobots_robotic_swarm}
\bibinfo{author}{Rubenstein, M.}, \bibinfo{author}{Cornejo, A.} \& \bibinfo{author}{Nagpal, R.}
\newblock \bibinfo{title}{Programmable self-assembly in a thousand-robot swarm}.
\newblock \emph{\bibinfo{journal}{Science}} \textbf{\bibinfo{volume}{345}}, \bibinfo{pages}{795--799} (\bibinfo{year}{2014}).

\bibitem{Zttl2016}
\bibinfo{author}{Z\"{o}ttl, A.} \& \bibinfo{author}{Stark, H.}
\newblock \bibinfo{title}{Emergent behavior in active colloids}.
\newblock \emph{\bibinfo{journal}{Journal of Physics: Condensed Matter}} \textbf{\bibinfo{volume}{28}}, \bibinfo{pages}{253001} (\bibinfo{year}{2016}).

\bibitem{vischek_1995}
\bibinfo{author}{Vicsek, T.}, \bibinfo{author}{Czir\'ok, A.}, \bibinfo{author}{Ben-Jacob, E.}, \bibinfo{author}{Cohen, I.} \& \bibinfo{author}{Shochet, O.}
\newblock \bibinfo{title}{Novel type of phase transition in a system of self-driven particles}.
\newblock \emph{\bibinfo{journal}{Phys. Rev. Lett.}} \textbf{\bibinfo{volume}{75}}, \bibinfo{pages}{1226--1229} (\bibinfo{year}{1995}).

\bibitem{Dombrowski_2004}
\bibinfo{author}{Dombrowski, C.}, \bibinfo{author}{Cisneros, L.}, \bibinfo{author}{Chatkaew, S.}, \bibinfo{author}{Goldstein, R.~E.} \& \bibinfo{author}{Kessler, J.~O.}
\newblock \bibinfo{title}{Self-concentration and large-scale coherence in bacterial dynamics}.
\newblock \emph{\bibinfo{journal}{Phys. Rev. Lett.}} \textbf{\bibinfo{volume}{93}}, \bibinfo{pages}{098103} (\bibinfo{year}{2004}).

\bibitem{Tim_Sanchez_Microtubule_Motors}
\bibinfo{author}{Sanchez, T.}, \bibinfo{author}{Chen, D. T.~N.}, \bibinfo{author}{DeCamp, S.~J.}, \bibinfo{author}{Heymann, M.} \& \bibinfo{author}{Dogic, Z.}
\newblock \bibinfo{title}{Spontaneous motion in hierarchically assembled active matter}.
\newblock \emph{\bibinfo{journal}{Nature}} \textbf{\bibinfo{volume}{491}}, \bibinfo{pages}{431--434} (\bibinfo{year}{2012}).

\bibitem{RuiZhang_2018_Actin_Nematic}
\bibinfo{author}{Zhang, R.}, \bibinfo{author}{Kumar, N.}, \bibinfo{author}{Ross, J.~L.}, \bibinfo{author}{Gardel, M.~L.} \& \bibinfo{author}{de~Pablo, J.~J.}
\newblock \bibinfo{title}{Interplay of structure, elasticity, and dynamics in actin-based nematic materials}.
\newblock \emph{\bibinfo{journal}{Proceedings of the National Academy of Sciences}} \textbf{\bibinfo{volume}{115}}, \bibinfo{pages}{E124--E133} (\bibinfo{year}{2018}).

\bibitem{Flocking_Chiral_Activa_Matter_Benno_Liebchen_2017}
\bibinfo{author}{Liebchen, B.} \& \bibinfo{author}{Levis, D.}
\newblock \bibinfo{title}{Collective behavior of chiral active matter: Pattern formation and enhanced flocking}.
\newblock \emph{\bibinfo{journal}{Phys. Rev. Lett.}} \textbf{\bibinfo{volume}{119}}, \bibinfo{pages}{058002} (\bibinfo{year}{2017}).

\bibitem{toner2024physics}
\bibinfo{author}{Toner, J.}
\newblock \emph{\bibinfo{title}{The Physics of Flocking: Birth, Death, and Flight in Active Matter}} (\bibinfo{publisher}{Cambridge University Press}, \bibinfo{year}{2024}).

\bibitem{Swarming_Buhl_2006}
\bibinfo{author}{Buhl, C.} \emph{et~al.}
\newblock \bibinfo{title}{From disorder to order in marching locusts}.
\newblock \emph{\bibinfo{journal}{Science}} \textbf{\bibinfo{volume}{312}}, \bibinfo{pages}{1402--1406} (\bibinfo{year}{2006}).

\bibitem{Cates2015}
\bibinfo{author}{Cates, M.~E.} \& \bibinfo{author}{Tailleur, J.}
\newblock \bibinfo{title}{Motility-induced phase separation}.
\newblock \emph{\bibinfo{journal}{Annual Review of Condensed Matter Physics}} \textbf{\bibinfo{volume}{6}}, \bibinfo{pages}{219–244} (\bibinfo{year}{2015}).

\bibitem{Giomi_2014_defect_dynamics}
\bibinfo{author}{Giomi, L.}, \bibinfo{author}{Bowick, M.~J.}, \bibinfo{author}{Mishra, P.}, \bibinfo{author}{Sknepnek, R.} \& \bibinfo{author}{Cristina~Marchetti, M.}
\newblock \bibinfo{title}{Defect dynamics in active nematics}.
\newblock \emph{\bibinfo{journal}{Philosophical Transactions of the Royal Society A: Mathematical, Physical and Engineering Sciences}} \textbf{\bibinfo{volume}{372}}, \bibinfo{pages}{20130365} (\bibinfo{year}{2014}).

\bibitem{Spontaneous_Flow_Patterns_J_Yeomans_2009}
\bibinfo{author}{Edwards, S.~A.} \& \bibinfo{author}{Yeomans, J.~M.}
\newblock \bibinfo{title}{Spontaneous flow states in active nematics: A unified picture}.
\newblock \emph{\bibinfo{journal}{EPL (Europhysics Letters)}} \textbf{\bibinfo{volume}{85}}, \bibinfo{pages}{18008} (\bibinfo{year}{2009}).

\bibitem{Wioland_2013}
\bibinfo{author}{Wioland, H.}, \bibinfo{author}{Woodhouse, F.~G.}, \bibinfo{author}{Dunkel, J.}, \bibinfo{author}{Kessler, J.~O.} \& \bibinfo{author}{Goldstein, R.~E.}
\newblock \bibinfo{title}{Confinement stabilizes a bacterial suspension into a spiral vortex}.
\newblock \emph{\bibinfo{journal}{Phys. Rev. Lett.}} \textbf{\bibinfo{volume}{110}}, \bibinfo{pages}{268102} (\bibinfo{year}{2013}).

\bibitem{Thuan_Saw_Tissue_Nematics}
\bibinfo{author}{Saw, T.~B.}, \bibinfo{author}{Xi, W.}, \bibinfo{author}{Ladoux, B.} \& \bibinfo{author}{Lim, C.~T.}
\newblock \bibinfo{title}{Biological tissues as active nematic liquid crystals}.
\newblock \emph{\bibinfo{journal}{Advanced Materials}} \textbf{\bibinfo{volume}{30}}, \bibinfo{pages}{1802579} (\bibinfo{year}{2018}).

\bibitem{BenoitLadoux_TissueMorphogenesis_Active_Nematic}
\bibinfo{author}{Balasubramaniam, L.}, \bibinfo{author}{M{\`e}ge, R.-M.} \& \bibinfo{author}{Ladoux, B.}
\newblock \bibinfo{title}{Active nematics across scales from cytoskeleton organization to tissue morphogenesis}.
\newblock \emph{\bibinfo{journal}{Current Opinion in Genetics \& Development}} \textbf{\bibinfo{volume}{73}}, \bibinfo{pages}{101897} (\bibinfo{year}{2022}).

\bibitem{Energy_spectra_3D_bacterial_suspensions}
\bibinfo{author}{Liu, Z.}, \bibinfo{author}{Zeng, W.}, \bibinfo{author}{Ma, X.} \& \bibinfo{author}{Cheng, X.}
\newblock \bibinfo{title}{Density fluctuations and energy spectra of 3d bacterial suspensions}.
\newblock \emph{\bibinfo{journal}{Soft Matter}} \textbf{\bibinfo{volume}{17}}, \bibinfo{pages}{10806--10817} (\bibinfo{year}{2021}).

\bibitem{Spontaneous_flow_nem}
\bibinfo{author}{Duclos, G.} \emph{et~al.}
\newblock \bibinfo{title}{Spontaneous shear flow in confined cellular nematics}.
\newblock \emph{\bibinfo{journal}{Nature Physics}} \textbf{\bibinfo{volume}{14}}, \bibinfo{pages}{728--732} (\bibinfo{year}{2018}).

\bibitem{CellMigration_Doxzen_2013}
\bibinfo{author}{Doxzen, K.} \emph{et~al.}
\newblock \bibinfo{title}{Guidance of collective cell migration by substrate geometry}.
\newblock \emph{\bibinfo{journal}{Integr. Biol.}} \textbf{\bibinfo{volume}{5}}, \bibinfo{pages}{1026--1035} (\bibinfo{year}{2013}).

\bibitem{cell_extrusion}
\bibinfo{author}{Saw, T.~B.} \emph{et~al.}
\newblock \bibinfo{title}{Topological defects in epithelia govern cell death and extrusion}.
\newblock \emph{\bibinfo{journal}{Nature}} \textbf{\bibinfo{volume}{544}}, \bibinfo{pages}{212--216} (\bibinfo{year}{2017}).

\bibitem{Coexisting_ordered_states_Frey}
\bibinfo{author}{Huber, L.}, \bibinfo{author}{Suzuki, R.}, \bibinfo{author}{Kr{\"u}ger, T.}, \bibinfo{author}{Frey, E.} \& \bibinfo{author}{Bausch, A.~R.}
\newblock \bibinfo{title}{Emergence of coexisting ordered states in active matter systems}.
\newblock \emph{\bibinfo{journal}{Science}} \textbf{\bibinfo{volume}{361}}, \bibinfo{pages}{255--258} (\bibinfo{year}{2018}).

\bibitem{ruider2024topological}
\bibinfo{author}{Ruider, I.} \emph{et~al.}
\newblock \bibinfo{title}{Topological excitations govern ordering kinetics in endothelial cell layers}.
\newblock \emph{\bibinfo{journal}{bioRxiv}} \bibinfo{pages}{2024.09.26.615134} (\bibinfo{year}{2024}).

\bibitem{PauGuillamatIntegerTopologicalDefectsTissueMorphogenesis}
\bibinfo{author}{Guillamat, P.}, \bibinfo{author}{Blanch-Mercader, C.}, \bibinfo{author}{Pernollet, G.}, \bibinfo{author}{Kruse, K.} \& \bibinfo{author}{Roux, A.}
\newblock \bibinfo{title}{Integer topological defects organize stresses driving tissue morphogenesis}.
\newblock \emph{\bibinfo{journal}{Nature Materials}} \textbf{\bibinfo{volume}{21}}, \bibinfo{pages}{588--597} (\bibinfo{year}{2022}).

\bibitem{EpithelialTissue_Carlos_2019}
\bibinfo{author}{P{\'e}rez-Gonz{\'a}lez, C.} \emph{et~al.}
\newblock \bibinfo{title}{Active wetting of epithelial tissues}.
\newblock \emph{\bibinfo{journal}{Nature Physics}} \textbf{\bibinfo{volume}{15}}, \bibinfo{pages}{79--88} (\bibinfo{year}{2019}).

\bibitem{Symmetry_Breaking_Polar_Nem_Frey}
\bibinfo{author}{Denk, J.} \& \bibinfo{author}{Frey, E.}
\newblock \bibinfo{title}{Pattern-induced local symmetry breaking in active-matter systems}.
\newblock \emph{\bibinfo{journal}{Proceedings of the National Academy of Sciences}} \textbf{\bibinfo{volume}{117}}, \bibinfo{pages}{31623--31630} (\bibinfo{year}{2020}).

\bibitem{meacock2021bacteria}
\bibinfo{author}{Meacock, O.~J.}, \bibinfo{author}{Doostmohammadi, A.}, \bibinfo{author}{Foster, K.~R.}, \bibinfo{author}{Yeomans, J.~M.} \& \bibinfo{author}{Durham, W.~M.}
\newblock \bibinfo{title}{Bacteria solve the problem of crowding by moving slowly}.
\newblock \emph{\bibinfo{journal}{Nature Physics}} \textbf{\bibinfo{volume}{17}}, \bibinfo{pages}{205--210} (\bibinfo{year}{2021}).

\bibitem{Amiri_2022}
\bibinfo{author}{Amiri, A.}, \bibinfo{author}{Mueller, R.} \& \bibinfo{author}{Doostmohammadi, A.}
\newblock \bibinfo{title}{Unifying polar and nematic active matter: emergence and co-existence of half-integer and full-integer topological defects}.
\newblock \emph{\bibinfo{journal}{Journal of Physics A: Mathematical and Theoretical}} \textbf{\bibinfo{volume}{55}}, \bibinfo{pages}{094002} (\bibinfo{year}{2022}).

\bibitem{han2023local}
\bibinfo{author}{Han, E.} \emph{et~al.}
\newblock \bibinfo{title}{Local polar order controls mechanical stress and triggers layer formation in myxococcus xanthus colonies}.
\newblock \emph{\bibinfo{journal}{Nature Communications}} \textbf{\bibinfo{volume}{16}}, \bibinfo{pages}{952} (\bibinfo{year}{2025}).

\bibitem{wheeler2024individual}
\bibinfo{author}{Wheeler, J.~H.}, \bibinfo{author}{Foster, K.~R.} \& \bibinfo{author}{Durham, W.~M.}
\newblock \bibinfo{title}{Individual bacterial cells can use spatial sensing of chemical gradients to direct chemotaxis on surfaces}.
\newblock \emph{\bibinfo{journal}{Nature Microbiology}} \textbf{\bibinfo{volume}{9}}, \bibinfo{pages}{2308--2322} (\bibinfo{year}{2024}).

\bibitem{Metamaterials}
\bibinfo{author}{Pishvar, M.} \& \bibinfo{author}{Harne, R.~L.}
\newblock \bibinfo{title}{Foundations for soft, smart matter by active mechanical metamaterials}.
\newblock \emph{\bibinfo{journal}{Advanced Science}} \textbf{\bibinfo{volume}{7}}, \bibinfo{pages}{2001384} (\bibinfo{year}{2020}).

\bibitem{robotics1}
\bibinfo{author}{Brandenbourger, M.}, \bibinfo{author}{Scheibner, C.}, \bibinfo{author}{Veenstra, J.}, \bibinfo{author}{Vitelli, V.} \& \bibinfo{author}{Coulais, C.}
\newblock \bibinfo{title}{Limit cycles turn active matter into robots}.
\newblock \emph{\bibinfo{journal}{arXiv preprint arXiv:2108.08837}}  (\bibinfo{year}{2022}).

\bibitem{Harrison2022}
\bibinfo{author}{Harrison, D.}, \bibinfo{author}{Rorot, W.} \& \bibinfo{author}{Laukaityte, U.}
\newblock \bibinfo{title}{Mind the matter: Active matter, soft robotics, and the making of bio-inspired artificial intelligence}.
\newblock \emph{\bibinfo{journal}{Frontiers in Neurorobotics}} \textbf{\bibinfo{volume}{16}}, \bibinfo{pages}{880724} (\bibinfo{year}{2022}).

\bibitem{Shankar_2022}
\bibinfo{author}{Shankar, S.}, \bibinfo{author}{Souslov, A.}, \bibinfo{author}{Bowick, M.~J.}, \bibinfo{author}{Marchetti, M.~C.} \& \bibinfo{author}{Vitelli, V.}
\newblock \bibinfo{title}{Topological active matter}.
\newblock \emph{\bibinfo{journal}{Nature Reviews Physics}} \textbf{\bibinfo{volume}{4}}, \bibinfo{pages}{380–398} (\bibinfo{year}{2022}).

\bibitem{pearce2024topologicaldefectsleadenergy}
\bibinfo{author}{Pearce, D. J.~G.}, \bibinfo{author}{Martínez-Prat, B.}, \bibinfo{author}{Ignés-Mullol, J.} \& \bibinfo{author}{Sagués, F.}
\newblock \bibinfo{title}{Topological defects lead to energy transfer in active nematics}.
\newblock \emph{\bibinfo{journal}{arXiv preprint arXiv:2411.18214}}  (\bibinfo{year}{2024}).

\bibitem{Erdmann2000}
\bibinfo{author}{Erdmann, U.}, \bibinfo{author}{Ebeling, W.}, \bibinfo{author}{Schimansky-Geier, L.} \& \bibinfo{author}{Schweitzer, F.}
\newblock \bibinfo{title}{Brownian particles far from equilibrium}.
\newblock \emph{\bibinfo{journal}{The European Physical Journal B}} \textbf{\bibinfo{volume}{15}}, \bibinfo{pages}{105–113} (\bibinfo{year}{2000}).

\bibitem{cates_diffusive_2012}
\bibinfo{author}{Cates, M.~E.}
\newblock \bibinfo{title}{Diffusive transport without detailed balance in motile bacteria: does microbiology need statistical physics?}
\newblock \emph{\bibinfo{journal}{Reports on Progress in Physics}} \textbf{\bibinfo{volume}{75}}, \bibinfo{pages}{042601} (\bibinfo{year}{2012}).

\bibitem{Basu2018}
\bibinfo{author}{Basu, U.}, \bibinfo{author}{Majumdar, S.~N.}, \bibinfo{author}{Rosso, A.} \& \bibinfo{author}{Schehr, G.}
\newblock \bibinfo{title}{Active brownian motion in two dimensions}.
\newblock \emph{\bibinfo{journal}{Physical Review E}} \textbf{\bibinfo{volume}{98}}, \bibinfo{pages}{062121} (\bibinfo{year}{2018}).

\bibitem{tailleur_statistical_2008}
\bibinfo{author}{Tailleur, J.} \& \bibinfo{author}{Cates, M.~E.}
\newblock \bibinfo{title}{Statistical {Mechanics} of {Interacting} {Run}-and-{Tumble} {Bacteria}}.
\newblock \emph{\bibinfo{journal}{Phys. Rev. Lett.}} \textbf{\bibinfo{volume}{100}}, \bibinfo{pages}{218103} (\bibinfo{year}{2008}).
\newblock \bibinfo{note}{Publisher: American Physical Society}.

\bibitem{Fodor2016}
\bibinfo{author}{Fodor, E.} \emph{et~al.}
\newblock \bibinfo{title}{How far from equilibrium is active matter?}
\newblock \emph{\bibinfo{journal}{Phys. Rev. Lett.}} \textbf{\bibinfo{volume}{117}}, \bibinfo{pages}{038103} (\bibinfo{year}{2016}).

\bibitem{Bonilla2019}
\bibinfo{author}{Bonilla, L.~L.}
\newblock \bibinfo{title}{Active ornstein-uhlenbeck particles}.
\newblock \emph{\bibinfo{journal}{Physical Review E}} \textbf{\bibinfo{volume}{100}}, \bibinfo{pages}{022601} (\bibinfo{year}{2019}).

\bibitem{Martin2021}
\bibinfo{author}{Martin, D.} \emph{et~al.}
\newblock \bibinfo{title}{Statistical mechanics of active ornstein-uhlenbeck particles}.
\newblock \emph{\bibinfo{journal}{Physical Review E}} \textbf{\bibinfo{volume}{103}}, \bibinfo{pages}{032607} (\bibinfo{year}{2021}).

\bibitem{Maggi2015}
\bibinfo{author}{Maggi, C.}, \bibinfo{author}{Marconi, U. M.~B.}, \bibinfo{author}{Gnan, N.} \& \bibinfo{author}{Di~Leonardo, R.}
\newblock \bibinfo{title}{Multidimensional stationary probability distribution for interacting active particles}.
\newblock \emph{\bibinfo{journal}{Scientific Reports}} \textbf{\bibinfo{volume}{5}}, \bibinfo{pages}{10742} (\bibinfo{year}{2015}).

\bibitem{Uhlenbeck1930}
\bibinfo{author}{Uhlenbeck, G.~E.} \& \bibinfo{author}{Ornstein, L.~S.}
\newblock \bibinfo{title}{On the theory of the brownian motion}.
\newblock \emph{\bibinfo{journal}{Physical Review}} \textbf{\bibinfo{volume}{36}}, \bibinfo{pages}{823–841} (\bibinfo{year}{1930}).

\bibitem{Koumakis2014}
\bibinfo{author}{Koumakis, N.}, \bibinfo{author}{Maggi, C.} \& \bibinfo{author}{Di~Leonardo, R.}
\newblock \bibinfo{title}{Directed transport of active particles over asymmetric energy barriers}.
\newblock \emph{\bibinfo{journal}{Soft Matter}} \textbf{\bibinfo{volume}{10}}, \bibinfo{pages}{5695–5701} (\bibinfo{year}{2014}).

\bibitem{Deforet2014}
\bibinfo{author}{Deforet, M.}, \bibinfo{author}{Hakim, V.}, \bibinfo{author}{Yevick, H.}, \bibinfo{author}{Duclos, G.} \& \bibinfo{author}{Silberzan, P.}
\newblock \bibinfo{title}{Emergence of collective modes and tri-dimensional structures from epithelial confinement}.
\newblock \emph{\bibinfo{journal}{Nature Communications}} \textbf{\bibinfo{volume}{5}}, \bibinfo{pages}{3747} (\bibinfo{year}{2014}).

\bibitem{Hakim2017}
\bibinfo{author}{Hakim, V.} \& \bibinfo{author}{Silberzan, P.}
\newblock \bibinfo{title}{Collective cell migration: a physics perspective}.
\newblock \emph{\bibinfo{journal}{Reports on Progress in Physics}} \textbf{\bibinfo{volume}{80}}, \bibinfo{pages}{076601} (\bibinfo{year}{2017}).

\bibitem{Khatami2016}
\bibinfo{author}{Khatami, M.}, \bibinfo{author}{Wolff, K.}, \bibinfo{author}{Pohl, O.}, \bibinfo{author}{Ejtehadi, M.~R.} \& \bibinfo{author}{Stark, H.}
\newblock \bibinfo{title}{Active brownian particles and run-and-tumble particles separate inside a maze}.
\newblock \emph{\bibinfo{journal}{Scientific Reports}} \textbf{\bibinfo{volume}{6}}, \bibinfo{pages}{37670} (\bibinfo{year}{2016}).

\bibitem{Romanczuk2012}
\bibinfo{author}{Romanczuk, P.}, \bibinfo{author}{B\"{a}r, M.}, \bibinfo{author}{Ebeling, W.}, \bibinfo{author}{Lindner, B.} \& \bibinfo{author}{Schimansky-Geier, L.}
\newblock \bibinfo{title}{Active brownian particles: From individual to collective stochastic dynamics}.
\newblock \emph{\bibinfo{journal}{The European Physical Journal Special Topics}} \textbf{\bibinfo{volume}{202}}, \bibinfo{pages}{1–162} (\bibinfo{year}{2012}).

\bibitem{Higdon1979}
\bibinfo{author}{Higdon, J. J.~L.}
\newblock \bibinfo{title}{A hydrodynamic analysis of flagellar propulsion}.
\newblock \emph{\bibinfo{journal}{Journal of Fluid Mechanics}} \textbf{\bibinfo{volume}{90}}, \bibinfo{pages}{685} (\bibinfo{year}{1979}).

\bibitem{Deng2023}
\bibinfo{author}{Deng, J.}, \bibinfo{author}{Molaei, M.}, \bibinfo{author}{Chisholm, N.~G.} \& \bibinfo{author}{Stebe, K.~J.}
\newblock \bibinfo{title}{Interfacial flow around a pusher bacterium}.
\newblock \emph{\bibinfo{journal}{Journal of Fluid Mechanics}} \textbf{\bibinfo{volume}{976}}, \bibinfo{pages}{A18} (\bibinfo{year}{2023}).

\bibitem{Mirzakhanloo2020}
\bibinfo{author}{Mirzakhanloo, M.}, \bibinfo{author}{Esmaeilzadeh, S.} \& \bibinfo{author}{Alam, M.-R.}
\newblock \bibinfo{title}{Active cloaking in stokes flows via reinforcement learning}.
\newblock \emph{\bibinfo{journal}{Journal of Fluid Mechanics}} \textbf{\bibinfo{volume}{903}}, \bibinfo{pages}{A34} (\bibinfo{year}{2020}).

\bibitem{Evans2011}
\bibinfo{author}{Evans, A.~A.}, \bibinfo{author}{Ishikawa, T.}, \bibinfo{author}{Yamaguchi, T.} \& \bibinfo{author}{Lauga, E.}
\newblock \bibinfo{title}{Orientational order in concentrated suspensions of spherical microswimmers}.
\newblock \emph{\bibinfo{journal}{Physics of Fluids}} \textbf{\bibinfo{volume}{23}}, \bibinfo{pages}{111702} (\bibinfo{year}{2011}).

\bibitem{Lighthill1952}
\bibinfo{author}{Lighthill, M.~J.}
\newblock \bibinfo{title}{On the squirming motion of nearly spherical deformable bodies through liquids at very small reynolds numbers}.
\newblock \emph{\bibinfo{journal}{Communications on Pure and Applied Mathematics}} \textbf{\bibinfo{volume}{5}}, \bibinfo{pages}{109–118} (\bibinfo{year}{1952}).

\bibitem{Pessot2018}
\bibinfo{author}{Pessot, G.}, \bibinfo{author}{L\"{o}wen, H.} \& \bibinfo{author}{Menzel, A.~M.}
\newblock \bibinfo{title}{Binary pusher–puller mixtures of active microswimmers and their collective behaviour}.
\newblock \emph{\bibinfo{journal}{Molecular Physics}} \textbf{\bibinfo{volume}{116}}, \bibinfo{pages}{3401–3408} (\bibinfo{year}{2018}).

\bibitem{Makino2004}
\bibinfo{author}{Makino, M.} \& \bibinfo{author}{Doi, M.}
\newblock \bibinfo{title}{Brownian motion of a particle of general shape in newtonian fluid}.
\newblock \emph{\bibinfo{journal}{Journal of the Physical Society of Japan}} \textbf{\bibinfo{volume}{73}}, \bibinfo{pages}{2739–2745} (\bibinfo{year}{2004}).

\bibitem{Reinken2024}
\bibinfo{author}{Reinken, H.}
\newblock \emph{\bibinfo{title}{Derivation of a Continuum Theory for Polar Active Fluids}}, \bibinfo{pages}{61–91} (\bibinfo{publisher}{Springer Nature Switzerland}, \bibinfo{year}{2024}).

\bibitem{Blake1971}
\bibinfo{author}{Blake, J.~R.}
\newblock \bibinfo{title}{A spherical envelope approach to ciliary propulsion}.
\newblock \emph{\bibinfo{journal}{Journal of Fluid Mechanics}} \textbf{\bibinfo{volume}{46}}, \bibinfo{pages}{199–208} (\bibinfo{year}{1971}).

\bibitem{Gtze2010}
\bibinfo{author}{G\"{o}tze, I.~O.} \& \bibinfo{author}{Gompper, G.}
\newblock \bibinfo{title}{Mesoscale simulations of hydrodynamic squirmer interactions}.
\newblock \emph{\bibinfo{journal}{Physical Review E}} \textbf{\bibinfo{volume}{82}}, \bibinfo{pages}{041921} (\bibinfo{year}{2010}).

\bibitem{ISHIKAWA2006}
\bibinfo{author}{Ishikawa, T.}, \bibinfo{author}{Simmonds, M.} \& \bibinfo{author}{Pedley, T.~J.}
\newblock \bibinfo{title}{Hydrodynamic interaction of two swimming model micro-organisms}.
\newblock \emph{\bibinfo{journal}{Journal of Fluid Mechanics}} \textbf{\bibinfo{volume}{568}}, \bibinfo{pages}{119--160} (\bibinfo{year}{2006}).

\bibitem{Long_range_nem_Benoit}
\bibinfo{author}{Mahault, B.} \& \bibinfo{author}{Chat\'e, H.}
\newblock \bibinfo{title}{Long-range nematic order in two-dimensional active matter}.
\newblock \emph{\bibinfo{journal}{Phys. Rev. Lett.}} \textbf{\bibinfo{volume}{127}}, \bibinfo{pages}{048003} (\bibinfo{year}{2021}).

\bibitem{fig3panelab}
\bibinfo{author}{Ngo, S.}, \bibinfo{author}{Ginelli, F.} \& \bibinfo{author}{Chat\'e, H.}
\newblock \bibinfo{title}{Competing ferromagnetic and nematic alignment in self-propelled polar particles}.
\newblock \emph{\bibinfo{journal}{Physical Review E}} \textbf{\bibinfo{volume}{86}}, \bibinfo{pages}{050101} (\bibinfo{year}{2012}).

\bibitem{fig3panelc}
\bibinfo{author}{Solon, A.~P.}, \bibinfo{author}{Chat\'e, H.} \& \bibinfo{author}{Tailleur, J.}
\newblock \bibinfo{title}{From phase to microphase separation in flocking models: The essential role of nonequilibrium fluctuations}.
\newblock \emph{\bibinfo{journal}{Phys. Rev. Lett.}} \textbf{\bibinfo{volume}{114}}, \bibinfo{pages}{068101} (\bibinfo{year}{2015}).

\bibitem{Ngo2014}
\bibinfo{author}{Ngo, S.} \emph{et~al.}
\newblock \bibinfo{title}{Large-scale chaos and fluctuations in active nematics}.
\newblock \emph{\bibinfo{journal}{Phys. Rev. Lett.}} \textbf{\bibinfo{volume}{113}}, \bibinfo{pages}{038302} (\bibinfo{year}{2014}).

\bibitem{Chat2006}
\bibinfo{author}{Chaté, H.}, \bibinfo{author}{Ginelli, F.} \& \bibinfo{author}{Montagne, R.}
\newblock \bibinfo{title}{Simple model for active nematics: Quasi-long-range order and giant fluctuations}.
\newblock \emph{\bibinfo{journal}{Phys. Rev. Lett.}} \textbf{\bibinfo{volume}{96}}, \bibinfo{pages}{180602} (\bibinfo{year}{2006}).

\bibitem{kultty2020}
\bibinfo{author}{Škultéty, V.}, \bibinfo{author}{Nardini, C.}, \bibinfo{author}{Stenhammar, J.}, \bibinfo{author}{Marenduzzo, D.} \& \bibinfo{author}{Morozov, A.}
\newblock \bibinfo{title}{Swimming suppresses correlations in dilute suspensions of pusher microorganisms}.
\newblock \emph{\bibinfo{journal}{Physical Review X}} \textbf{\bibinfo{volume}{10}}, \bibinfo{pages}{031059} (\bibinfo{year}{2020}).

\bibitem{Fily2012}
\bibinfo{author}{Fily, Y.} \& \bibinfo{author}{Marchetti, M.~C.}
\newblock \bibinfo{title}{Athermal phase separation of self-propelled particles with no alignment}.
\newblock \emph{\bibinfo{journal}{Phys. Rev. Lett.}} \textbf{\bibinfo{volume}{108}}, \bibinfo{pages}{235702} (\bibinfo{year}{2012}).

\bibitem{Digregorio2022}
\bibinfo{author}{Digregorio, P.}, \bibinfo{author}{Levis, D.}, \bibinfo{author}{Cugliandolo, L.~F.}, \bibinfo{author}{Gonnella, G.} \& \bibinfo{author}{Pagonabarraga, I.}
\newblock \bibinfo{title}{Unified analysis of topological defects in 2d systems of active and passive disks}.
\newblock \emph{\bibinfo{journal}{Soft Matter}} \textbf{\bibinfo{volume}{18}}, \bibinfo{pages}{566–591} (\bibinfo{year}{2022}).

\bibitem{bialke2015negative}
\bibinfo{author}{Bialk{\'e}, J.}, \bibinfo{author}{Siebert, J.~T.}, \bibinfo{author}{L{\"o}wen, H.} \& \bibinfo{author}{Speck, T.}
\newblock \bibinfo{title}{Negative interfacial tension in phase-separated active brownian particles}.
\newblock \emph{\bibinfo{journal}{Phys. Rev. Lett.}} \textbf{\bibinfo{volume}{115}}, \bibinfo{pages}{098301} (\bibinfo{year}{2015}).

\bibitem{Fausti2021}
\bibinfo{author}{Fausti, G.}, \bibinfo{author}{Tjhung, E.}, \bibinfo{author}{Cates, M.} \& \bibinfo{author}{Nardini, C.}
\newblock \bibinfo{title}{Capillary interfacial tension in active phase separation}.
\newblock \emph{\bibinfo{journal}{Phys. Rev. Lett.}} \textbf{\bibinfo{volume}{127}}, \bibinfo{pages}{068001} (\bibinfo{year}{2021}).

\bibitem{Hermann2019}
\bibinfo{author}{Hermann, S.}, \bibinfo{author}{de~las Heras, D.} \& \bibinfo{author}{Schmidt, M.}
\newblock \bibinfo{title}{Non-negative interfacial tension in phase-separated active brownian particles}.
\newblock \emph{\bibinfo{journal}{Phys. Rev. Lett.}} \textbf{\bibinfo{volume}{123}}, \bibinfo{pages}{268002} (\bibinfo{year}{2019}).

\bibitem{Mandal2019}
\bibinfo{author}{Mandal, S.}, \bibinfo{author}{Liebchen, B.} \& \bibinfo{author}{L\"{o}wen, H.}
\newblock \bibinfo{title}{Motility-induced temperature difference in coexisting phases}.
\newblock \emph{\bibinfo{journal}{Phys. Rev. Lett.}} \textbf{\bibinfo{volume}{123}}, \bibinfo{pages}{228001} (\bibinfo{year}{2019}).

\bibitem{Lee2022}
\bibinfo{author}{Lee, C.~F.}
\newblock \bibinfo{title}{An infinite set of integral formulae for polar, nematic, and higher order structures at the interface of motility-induced phase separation}.
\newblock \emph{\bibinfo{journal}{New Journal of Physics}} \textbf{\bibinfo{volume}{24}}, \bibinfo{pages}{043010} (\bibinfo{year}{2022}).

\bibitem{SesSansa2018}
\bibinfo{author}{Sesé-Sansa, E.}, \bibinfo{author}{Pagonabarraga, I.} \& \bibinfo{author}{Levis, D.}
\newblock \bibinfo{title}{Velocity alignment promotes motility-induced phase separation}.
\newblock \emph{\bibinfo{journal}{EPL (Europhysics Letters)}} \textbf{\bibinfo{volume}{124}}, \bibinfo{pages}{30004} (\bibinfo{year}{2018}).

\bibitem{Spera2024}
\bibinfo{author}{Spera, G.}, \bibinfo{author}{Duclut, C.}, \bibinfo{author}{Durand, M.} \& \bibinfo{author}{Tailleur, J.}
\newblock \bibinfo{title}{Nematic torques in scalar active matter: When fluctuations favor polar order and persistence}.
\newblock \emph{\bibinfo{journal}{Phys. Rev. Lett.}} \textbf{\bibinfo{volume}{132}}, \bibinfo{pages}{078301} (\bibinfo{year}{2024}).

\bibitem{Caprini2023}
\bibinfo{author}{Caprini, L.} \& \bibinfo{author}{L\"{o}wen, H.}
\newblock \bibinfo{title}{Flocking without alignment interactions in attractive active brownian particles}.
\newblock \emph{\bibinfo{journal}{Phys. Rev. Lett.}} \textbf{\bibinfo{volume}{130}}, \bibinfo{pages}{148202} (\bibinfo{year}{2023}).

\bibitem{Kneevi2022}
\bibinfo{author}{Knežević, M.}, \bibinfo{author}{Welker, T.} \& \bibinfo{author}{Stark, H.}
\newblock \bibinfo{title}{Collective motion of active particles exhibiting non-reciprocal orientational interactions}.
\newblock \emph{\bibinfo{journal}{Scientific Reports}} \textbf{\bibinfo{volume}{12}}, \bibinfo{pages}{19437} (\bibinfo{year}{2022}).

\bibitem{Fruchart2021}
\bibinfo{author}{Fruchart, M.}, \bibinfo{author}{Hanai, R.}, \bibinfo{author}{Littlewood, P.~B.} \& \bibinfo{author}{Vitelli, V.}
\newblock \bibinfo{title}{Non-reciprocal phase transitions}.
\newblock \emph{\bibinfo{journal}{Nature}} \textbf{\bibinfo{volume}{592}}, \bibinfo{pages}{363–369} (\bibinfo{year}{2021}).

\bibitem{Dinelli2023}
\bibinfo{author}{Dinelli, A.} \emph{et~al.}
\newblock \bibinfo{title}{Non-reciprocity across scales in active mixtures}.
\newblock \emph{\bibinfo{journal}{Nature Communications}} \textbf{\bibinfo{volume}{14}}, \bibinfo{pages}{7035} (\bibinfo{year}{2023}).

\bibitem{nonreciprocal3}
\bibinfo{author}{Chao, Y.-C.} \emph{et~al.}
\newblock \bibinfo{title}{Selective excitation of work-generating cycles in nonreciprocal living solids}.
\newblock \emph{\bibinfo{journal}{arXiv preprint arXiv:2410.18017}}  (\bibinfo{year}{2024}).

\bibitem{ishikawa2008coherent}
\bibinfo{author}{Ishikawa, T.} \& \bibinfo{author}{Pedley, T.~J.}
\newblock \bibinfo{title}{Coherent structures in monolayers of swimming particles}.
\newblock \emph{\bibinfo{journal}{Phys. Rev. Lett.}} \textbf{\bibinfo{volume}{100}}, \bibinfo{pages}{088103} (\bibinfo{year}{2008}).

\bibitem{Alarc_n_2017}
\bibinfo{author}{Alarcón, F.}, \bibinfo{author}{Valeriani, C.} \& \bibinfo{author}{Pagonabarraga, I.}
\newblock \bibinfo{title}{Morphology of clusters of attractive dry and wet self-propelled spherical particle suspensions}.
\newblock \emph{\bibinfo{journal}{Soft Matter}} \textbf{\bibinfo{volume}{13}}, \bibinfo{pages}{814–826} (\bibinfo{year}{2017}).

\bibitem{Brdfalvy2024}
\bibinfo{author}{Bárdfalvy, D.}, \bibinfo{author}{Škultéty, V.}, \bibinfo{author}{Nardini, C.}, \bibinfo{author}{Morozov, A.} \& \bibinfo{author}{Stenhammar, J.}
\newblock \bibinfo{title}{Collective motion in a sheet of microswimmers}.
\newblock \emph{\bibinfo{journal}{Communications Physics}} \textbf{\bibinfo{volume}{7}}, \bibinfo{pages}{93} (\bibinfo{year}{2024}).

\bibitem{sprimages}
\bibinfo{author}{Shi, X.-q.} \& \bibinfo{author}{Chaté, H.}
\newblock \bibinfo{title}{Self-propelled rods: Linking alignment-dominated and repulsion-dominated active matter}.
\newblock \emph{\bibinfo{journal}{arXiv preprint arXiv:1807.00294}}  (\bibinfo{year}{2018}).

\bibitem{Riedel2005}
\bibinfo{author}{Riedel, I.~H.}, \bibinfo{author}{Kruse, K.} \& \bibinfo{author}{Howard, J.}
\newblock \bibinfo{title}{A self-organized vortex array of hydrodynamically entrained sperm cells}.
\newblock \emph{\bibinfo{journal}{Science}} \textbf{\bibinfo{volume}{309}}, \bibinfo{pages}{300–303} (\bibinfo{year}{2005}).

\bibitem{SelfPropelledRods}
\bibinfo{author}{Bär, M.}, \bibinfo{author}{Großmann, R.}, \bibinfo{author}{Heidenreich, S.} \& \bibinfo{author}{Peruani, F.}
\newblock \bibinfo{title}{Self-propelled rods: Insights and perspectives for active matter}.
\newblock \emph{\bibinfo{journal}{Annual Review of Condensed Matter Physics}} \textbf{\bibinfo{volume}{11}}, \bibinfo{pages}{441–466} (\bibinfo{year}{2020}).

\bibitem{Gromann2020}
\bibinfo{author}{Großmann, R.}, \bibinfo{author}{Aranson, I.~S.} \& \bibinfo{author}{Peruani, F.}
\newblock \bibinfo{title}{A particle-field approach bridges phase separation and collective motion in active matter}.
\newblock \emph{\bibinfo{journal}{Nature Communications}} \textbf{\bibinfo{volume}{11}}, \bibinfo{pages}{5365} (\bibinfo{year}{2020}).

\bibitem{Weitz2015}
\bibinfo{author}{Weitz, S.}, \bibinfo{author}{Deutsch, A.} \& \bibinfo{author}{Peruani, F.}
\newblock \bibinfo{title}{Self-propelled rods exhibit a phase-separated state characterized by the presence of active stresses and the ejection of polar clusters}.
\newblock \emph{\bibinfo{journal}{Physical Review E}} \textbf{\bibinfo{volume}{92}}, \bibinfo{pages}{012322} (\bibinfo{year}{2015}).

\bibitem{Wensink2008}
\bibinfo{author}{Wensink, H.~H.} \& \bibinfo{author}{L\"{o}wen, H.}
\newblock \bibinfo{title}{Aggregation of self-propelled colloidal rods near confining walls}.
\newblock \emph{\bibinfo{journal}{Physical Review E}} \textbf{\bibinfo{volume}{78}}, \bibinfo{pages}{031409} (\bibinfo{year}{2008}).

\bibitem{Abkenar2013}
\bibinfo{author}{Abkenar, M.}, \bibinfo{author}{Marx, K.}, \bibinfo{author}{Auth, T.} \& \bibinfo{author}{Gompper, G.}
\newblock \bibinfo{title}{Collective behavior of penetrable self-propelled rods in two dimensions}.
\newblock \emph{\bibinfo{journal}{Physical Review E}} \textbf{\bibinfo{volume}{88}}, \bibinfo{pages}{062314} (\bibinfo{year}{2013}).

\bibitem{Yang2010}
\bibinfo{author}{Yang, Y.}, \bibinfo{author}{Marceau, V.} \& \bibinfo{author}{Gompper, G.}
\newblock \bibinfo{title}{Swarm behavior of self-propelled rods and swimming flagella}.
\newblock \emph{\bibinfo{journal}{Physical Review E}} \textbf{\bibinfo{volume}{82}}, \bibinfo{pages}{031904} (\bibinfo{year}{2010}).

\bibitem{McCandlish2012}
\bibinfo{author}{McCandlish, S.~R.}, \bibinfo{author}{Baskaran, A.} \& \bibinfo{author}{Hagan, M.~F.}
\newblock \bibinfo{title}{Spontaneous segregation of self-propelled particles with different motilities}.
\newblock \emph{\bibinfo{journal}{Soft Matter}} \textbf{\bibinfo{volume}{8}}, \bibinfo{pages}{2527} (\bibinfo{year}{2012}).

\bibitem{Shi2013}
\bibinfo{author}{Shi, X.-q.} \& \bibinfo{author}{Ma, Y.-q.}
\newblock \bibinfo{title}{Topological structure dynamics revealing collective evolution in active nematics}.
\newblock \emph{\bibinfo{journal}{Nature Communications}} \textbf{\bibinfo{volume}{4}}, \bibinfo{pages}{3013} (\bibinfo{year}{2013}).

\bibitem{Zantop2022}
\bibinfo{author}{Zantop, A.~W.} \& \bibinfo{author}{Stark, H.}
\newblock \bibinfo{title}{Emergent collective dynamics of pusher and puller squirmer rods: swarming, clustering, and turbulence}.
\newblock \emph{\bibinfo{journal}{Soft Matter}} \textbf{\bibinfo{volume}{18}}, \bibinfo{pages}{6179–6191} (\bibinfo{year}{2022}).

\bibitem{Qi2022}
\bibinfo{author}{Qi, K.}, \bibinfo{author}{Westphal, E.}, \bibinfo{author}{Gompper, G.} \& \bibinfo{author}{Winkler, R.~G.}
\newblock \bibinfo{title}{Emergence of active turbulence in microswimmer suspensions due to active hydrodynamic stress and volume exclusion}.
\newblock \emph{\bibinfo{journal}{Communications Physics}} \textbf{\bibinfo{volume}{5}}, \bibinfo{pages}{49} (\bibinfo{year}{2022}).

\bibitem{Venkatesh2022}
\bibinfo{author}{Venkatesh, V.}, \bibinfo{author}{Mondal, C.} \& \bibinfo{author}{Doostmohammadi, A.}
\newblock \bibinfo{title}{Distinct impacts of polar and nematic self-propulsion on active unjamming}.
\newblock \emph{\bibinfo{journal}{The Journal of Chemical Physics}} \textbf{\bibinfo{volume}{157}}, \bibinfo{pages}{164901} (\bibinfo{year}{2022}).

\bibitem{Zhao2024}
\bibinfo{author}{Zhao, C.}, \bibinfo{author}{Yan, R.} \& \bibinfo{author}{Zhao, N.}
\newblock \bibinfo{title}{Collective behavior of active filaments with homogeneous and heterogeneous stiffness}.
\newblock \emph{\bibinfo{journal}{The Journal of Chemical Physics}} \textbf{\bibinfo{volume}{161}}, \bibinfo{pages}{154901} (\bibinfo{year}{2024}).

\bibitem{Ohta2017}
\bibinfo{author}{Ohta, T.}
\newblock \bibinfo{title}{Dynamics of deformable active particles}.
\newblock \emph{\bibinfo{journal}{Journal of the Physical Society of Japan}} \textbf{\bibinfo{volume}{86}}, \bibinfo{pages}{072001} (\bibinfo{year}{2017}).

\bibitem{Joshi2019}
\bibinfo{author}{Joshi, A.}, \bibinfo{author}{Putzig, E.}, \bibinfo{author}{Baskaran, A.} \& \bibinfo{author}{Hagan, M.~F.}
\newblock \bibinfo{title}{The interplay between activity and filament flexibility determines the emergent properties of active nematics}.
\newblock \emph{\bibinfo{journal}{Soft Matter}} \textbf{\bibinfo{volume}{15}}, \bibinfo{pages}{94–101} (\bibinfo{year}{2019}).

\bibitem{Duman2018}
\bibinfo{author}{Duman, O.}, \bibinfo{author}{Isele-Holder, R.~E.}, \bibinfo{author}{Elgeti, J.} \& \bibinfo{author}{Gompper, G.}
\newblock \bibinfo{title}{Collective dynamics of self-propelled semiflexible filaments}.
\newblock \emph{\bibinfo{journal}{Soft Matter}} \textbf{\bibinfo{volume}{14}}, \bibinfo{pages}{4483–4494} (\bibinfo{year}{2018}).

\bibitem{Huber2021}
\bibinfo{author}{Huber, L.}, \bibinfo{author}{Kr\"{u}ger, T.} \& \bibinfo{author}{Frey, E.}
\newblock \bibinfo{title}{Microphase separation in active filament systems maintained by cyclic dynamics of cluster size and order}.
\newblock \emph{\bibinfo{journal}{Physical Review Research}} \textbf{\bibinfo{volume}{3}}, \bibinfo{pages}{013280} (\bibinfo{year}{2021}).

\bibitem{Vliegenthart2020}
\bibinfo{author}{Vliegenthart, G.~A.}, \bibinfo{author}{Ravichandran, A.}, \bibinfo{author}{Ripoll, M.}, \bibinfo{author}{Auth, T.} \& \bibinfo{author}{Gompper, G.}
\newblock \bibinfo{title}{Filamentous active matter: Band formation, bending, buckling, and defects}.
\newblock \emph{\bibinfo{journal}{Science Advances}} \textbf{\bibinfo{volume}{6}}, \bibinfo{pages}{eaaw9975} (\bibinfo{year}{2020}).

\bibitem{Peterson2021}
\bibinfo{author}{Peterson, M. S.~E.}, \bibinfo{author}{Baskaran, A.} \& \bibinfo{author}{Hagan, M.~F.}
\newblock \bibinfo{title}{Vesicle shape transformations driven by confined active filaments}.
\newblock \emph{\bibinfo{journal}{Nature Communications}} \textbf{\bibinfo{volume}{12}}, \bibinfo{pages}{7247} (\bibinfo{year}{2021}).

\bibitem{Dunajova2023}
\bibinfo{author}{Dunajova, Z.} \emph{et~al.}
\newblock \bibinfo{title}{Chiral and nematic phases of flexible active filaments}.
\newblock \emph{\bibinfo{journal}{Nature Physics}} \textbf{\bibinfo{volume}{19}}, \bibinfo{pages}{1916–1926} (\bibinfo{year}{2023}).

\bibitem{Mueller2019}
\bibinfo{author}{Mueller, R.}, \bibinfo{author}{Yeomans, J.~M.} \& \bibinfo{author}{Doostmohammadi, A.}
\newblock \bibinfo{title}{Emergence of active nematic behavior in monolayers of isotropic cells}.
\newblock \emph{\bibinfo{journal}{Phys. Rev. Lett.}} \textbf{\bibinfo{volume}{122}}, \bibinfo{pages}{048004} (\bibinfo{year}{2019}).

\bibitem{ladoux2017mechanobiology}
\bibinfo{author}{Ladoux, B.} \& \bibinfo{author}{M{\`e}ge, R.-M.}
\newblock \bibinfo{title}{Mechanobiology of collective cell behaviours}.
\newblock \emph{\bibinfo{journal}{Nature reviews Molecular cell biology}} \textbf{\bibinfo{volume}{18}}, \bibinfo{pages}{743--757} (\bibinfo{year}{2017}).

\bibitem{balasubramaniam2021investigating}
\bibinfo{author}{Balasubramaniam, L.} \emph{et~al.}
\newblock \bibinfo{title}{Investigating the nature of active forces in tissues reveals how contractile cells can form extensile monolayers}.
\newblock \emph{\bibinfo{journal}{Nature materials}} \textbf{\bibinfo{volume}{20}}, \bibinfo{pages}{1156--1166} (\bibinfo{year}{2021}).

\bibitem{mueller2021phase}
\bibinfo{author}{Mueller, R.} \& \bibinfo{author}{Doostmohammadi, A.}
\newblock \bibinfo{title}{Phase field models of active matter}.
\newblock \emph{\bibinfo{journal}{arXiv preprint arXiv:2102.05557}}  (\bibinfo{year}{2021}).

\bibitem{Zhang2020}
\bibinfo{author}{Zhang, G.}, \bibinfo{author}{Mueller, R.}, \bibinfo{author}{Doostmohammadi, A.} \& \bibinfo{author}{Yeomans, J.~M.}
\newblock \bibinfo{title}{Active inter-cellular forces in collective cell motility}.
\newblock \emph{\bibinfo{journal}{Journal of The Royal Society Interface}} \textbf{\bibinfo{volume}{17}}, \bibinfo{pages}{20200312} (\bibinfo{year}{2020}).

\bibitem{Ardaeva2022}
\bibinfo{author}{Ardaševa, A.}, \bibinfo{author}{Mueller, R.} \& \bibinfo{author}{Doostmohammadi, A.}
\newblock \bibinfo{title}{Bridging microscopic cell dynamics to nematohydrodynamics of cell monolayers}.
\newblock \emph{\bibinfo{journal}{Soft Matter}} \textbf{\bibinfo{volume}{18}}, \bibinfo{pages}{4737–4746} (\bibinfo{year}{2022}).

\bibitem{Hopkins2022}
\bibinfo{author}{Hopkins, A.}, \bibinfo{author}{Chiang, M.}, \bibinfo{author}{Loewe, B.}, \bibinfo{author}{Marenduzzo, D.} \& \bibinfo{author}{Marchetti, M.~C.}
\newblock \bibinfo{title}{Local yield and compliance in active cell monolayers}.
\newblock \emph{\bibinfo{journal}{Phys. Rev. Lett.}} \textbf{\bibinfo{volume}{129}}, \bibinfo{pages}{148101} (\bibinfo{year}{2022}).

\bibitem{Chiang2024}
\bibinfo{author}{Chiang, M.}, \bibinfo{author}{Hopkins, A.}, \bibinfo{author}{Loewe, B.}, \bibinfo{author}{Marchetti, M.~C.} \& \bibinfo{author}{Marenduzzo, D.}
\newblock \bibinfo{title}{Intercellular friction and motility drive orientational order in cell monolayers}.
\newblock \emph{\bibinfo{journal}{Proceedings of the National Academy of Sciences}} \textbf{\bibinfo{volume}{121}}, \bibinfo{pages}{e2319310121} (\bibinfo{year}{2024}).

\bibitem{monfared2023mechanical}
\bibinfo{author}{Monfared, S.}, \bibinfo{author}{Ravichandran, G.}, \bibinfo{author}{Andrade, J.} \& \bibinfo{author}{Doostmohammadi, A.}
\newblock \bibinfo{title}{Mechanical basis and topological routes to cell elimination}.
\newblock \emph{\bibinfo{journal}{Elife}} \textbf{\bibinfo{volume}{12}}, \bibinfo{pages}{e82435} (\bibinfo{year}{2023}).

\bibitem{Tiribocchi2023}
\bibinfo{author}{Tiribocchi, A.} \emph{et~al.}
\newblock \bibinfo{title}{The crucial role of adhesion in the transmigration of active droplets through interstitial orifices}.
\newblock \emph{\bibinfo{journal}{Nature Communications}} \textbf{\bibinfo{volume}{14}}, \bibinfo{pages}{1096} (\bibinfo{year}{2023}).

\bibitem{Zhang2023}
\bibinfo{author}{Zhang, G.} \& \bibinfo{author}{Yeomans, J.~M.}
\newblock \bibinfo{title}{Active forces in confluent cell monolayers}.
\newblock \emph{\bibinfo{journal}{Phys. Rev. Lett.}} \textbf{\bibinfo{volume}{130}}, \bibinfo{pages}{038202} (\bibinfo{year}{2023}).

\bibitem{bird_flocks}
\bibinfo{author}{Cavagna, A.} \& \bibinfo{author}{Giardina, I.}
\newblock \bibinfo{title}{Bird flocks as condensed matter}.
\newblock \emph{\bibinfo{journal}{Annual Review of Condensed Matter Physics}} \textbf{\bibinfo{volume}{5}}, \bibinfo{pages}{183--207} (\bibinfo{year}{2014}).

\bibitem{fish_schools}
\bibinfo{author}{Parrish, J.~K.}, \bibinfo{author}{Viscido, S.~V.} \& \bibinfo{author}{Gr{\"u}nbaum, D.}
\newblock \bibinfo{title}{Self-organized fish schools: An examination of emergent properties}.
\newblock \emph{\bibinfo{journal}{The Biological Bulletin}} \textbf{\bibinfo{volume}{202}}, \bibinfo{pages}{296--305} (\bibinfo{year}{2002}).

\bibitem{migrting_cells_Sknepnek}
\bibinfo{author}{Henkes, S.}, \bibinfo{author}{Kostanjevec, K.}, \bibinfo{author}{Collinson, J.~M.}, \bibinfo{author}{Sknepnek, R.} \& \bibinfo{author}{Bertin, E.}
\newblock \bibinfo{title}{Dense active matter model of motion patterns in confluent cell monolayers}.
\newblock \emph{\bibinfo{journal}{Nature Communications}} \textbf{\bibinfo{volume}{11}}, \bibinfo{pages}{1405} (\bibinfo{year}{2020}).

\bibitem{Vicsek_to_Hydro}
\bibinfo{author}{Bertin, E.}, \bibinfo{author}{Droz, M.} \& \bibinfo{author}{Gr\'egoire, G.}
\newblock \bibinfo{title}{Boltzmann and hydrodynamic description for self-propelled particles}.
\newblock \emph{\bibinfo{journal}{Phys. Rev. E}} \textbf{\bibinfo{volume}{74}}, \bibinfo{pages}{022101} (\bibinfo{year}{2006}).

\bibitem{Bertin_2009_Vicsek_Toner_2009}
\bibinfo{author}{Bertin, E.}, \bibinfo{author}{Droz, M.} \& \bibinfo{author}{Gr{\'e}goire, G.}
\newblock \bibinfo{title}{Hydrodynamic equations for self-propelled particles: microscopic derivation and stability analysis}.
\newblock \emph{\bibinfo{journal}{Journal of Physics A: Mathematical and Theoretical}} \textbf{\bibinfo{volume}{42}}, \bibinfo{pages}{445001} (\bibinfo{year}{2009}).

\bibitem{Toner_1998}
\bibinfo{author}{Toner, J.} \& \bibinfo{author}{Tu, Y.}
\newblock \bibinfo{title}{Flocks, herds, and schools: A quantitative theory of flocking}.
\newblock \emph{\bibinfo{journal}{Physical Review E}} \textbf{\bibinfo{volume}{58}}, \bibinfo{pages}{4828–4858} (\bibinfo{year}{1998}).

\bibitem{Fluid_Dynamics_Bact_2013}
\bibinfo{author}{Dunkel, J.} \emph{et~al.}
\newblock \bibinfo{title}{Fluid dynamics of bacterial turbulence}.
\newblock \emph{\bibinfo{journal}{Phys. Rev. Lett.}} \textbf{\bibinfo{volume}{110}}, \bibinfo{pages}{228102} (\bibinfo{year}{2013}).

\bibitem{Topology_Polar_Ordering_cell_exp}
\bibinfo{author}{L{\aa}ng, E.} \emph{et~al.}
\newblock \bibinfo{title}{Topology-guided polar ordering of collective cell migration}.
\newblock \emph{\bibinfo{journal}{Science Advances}} \textbf{\bibinfo{volume}{10}}, \bibinfo{pages}{eadk4825} (\bibinfo{year}{2024}).

\bibitem{peyret2019sustained}
\bibinfo{author}{Peyret, G.} \emph{et~al.}
\newblock \bibinfo{title}{Sustained oscillations of epithelial cell sheets}.
\newblock \emph{\bibinfo{journal}{Biophysical journal}} \textbf{\bibinfo{volume}{117}}, \bibinfo{pages}{464--478} (\bibinfo{year}{2019}).

\bibitem{Mishra_2010}
\bibinfo{author}{Mishra, S.}, \bibinfo{author}{Baskaran, A.} \& \bibinfo{author}{Marchetti, M.~C.}
\newblock \bibinfo{title}{Fluctuations and pattern formation in self-propelled particles}.
\newblock \emph{\bibinfo{journal}{Physical Review E}} \textbf{\bibinfo{volume}{81}}, \bibinfo{pages}{061916} (\bibinfo{year}{2010}).

\bibitem{Granular_nem_2007}
\bibinfo{author}{Narayan, V.}, \bibinfo{author}{Ramaswamy, S.} \& \bibinfo{author}{Menon, N.}
\newblock \bibinfo{title}{Long-lived giant number fluctuations in a swarming granular nematic}.
\newblock \emph{\bibinfo{journal}{Science}} \textbf{\bibinfo{volume}{317}}, \bibinfo{pages}{105--108} (\bibinfo{year}{2007}).

\bibitem{Hydrodynamics_of_self_propelled_hard_rods}
\bibinfo{author}{Baskaran, A.} \& \bibinfo{author}{Marchetti, M.~C.}
\newblock \bibinfo{title}{Hydrodynamics of self-propelled hard rods}.
\newblock \emph{\bibinfo{journal}{Phys. Rev. E}} \textbf{\bibinfo{volume}{77}}, \bibinfo{pages}{011920} (\bibinfo{year}{2008}).

\bibitem{Self_propelled_rods_2015}
\bibinfo{author}{Bertin, E.}, \bibinfo{author}{Baskaran, A.}, \bibinfo{author}{Chat\'e, H.} \& \bibinfo{author}{Marchetti, M.~C.}
\newblock \bibinfo{title}{Comparison between smoluchowski and boltzmann approaches for self-propelled rods}.
\newblock \emph{\bibinfo{journal}{Phys. Rev. E}} \textbf{\bibinfo{volume}{92}}, \bibinfo{pages}{042141} (\bibinfo{year}{2015}).

\bibitem{Peshkov_SPR}
\bibinfo{author}{Peshkov, A.}, \bibinfo{author}{Aranson, I.~S.}, \bibinfo{author}{Bertin, E.}, \bibinfo{author}{Chat\'e, H.} \& \bibinfo{author}{Ginelli, F.}
\newblock \bibinfo{title}{Nonlinear field equations for aligning self-propelled rods}.
\newblock \emph{\bibinfo{journal}{Phys. Rev. Lett.}} \textbf{\bibinfo{volume}{109}}, \bibinfo{pages}{268701} (\bibinfo{year}{2012}).

\bibitem{SPR_reversal_speed}
\bibinfo{author}{Patelli, A.}, \bibinfo{author}{Djafer-Cherif, I.}, \bibinfo{author}{Aranson, I.~S.}, \bibinfo{author}{Bertin, E.} \& \bibinfo{author}{Chat\'e, H.}
\newblock \bibinfo{title}{Understanding dense active nematics from microscopic models}.
\newblock \emph{\bibinfo{journal}{Phys. Rev. Lett.}} \textbf{\bibinfo{volume}{123}}, \bibinfo{pages}{258001} (\bibinfo{year}{2019}).

\bibitem{IncombressibleTonerTu}
\bibinfo{author}{Chen, L.}, \bibinfo{author}{Lee, C.~F.} \& \bibinfo{author}{Toner, J.}
\newblock \bibinfo{title}{Mapping two-dimensional polar active fluids to two-dimensional soap and one-dimensional sandblasting}.
\newblock \emph{\bibinfo{journal}{Nature Communications}} \textbf{\bibinfo{volume}{7}}, \bibinfo{pages}{12215} (\bibinfo{year}{2016}).

\bibitem{Meso_scale_turbulence}
\bibinfo{author}{Wensink, H.~H.} \emph{et~al.}
\newblock \bibinfo{title}{Meso-scale turbulence in living fluids}.
\newblock \emph{\bibinfo{journal}{Proceedings of the National Academy of Sciences}} \textbf{\bibinfo{volume}{109}}, \bibinfo{pages}{14308--14313} (\bibinfo{year}{2012}).

\bibitem{Santillan_2008}
\bibinfo{author}{Saintillan, D.} \& \bibinfo{author}{Shelley, M.~J.}
\newblock \bibinfo{title}{Instabilities and pattern formation in active particle suspensions: Kinetic theory and continuum simulations}.
\newblock \emph{\bibinfo{journal}{Phys. Rev. Lett.}} \textbf{\bibinfo{volume}{100}}, \bibinfo{pages}{178103} (\bibinfo{year}{2008}).

\bibitem{jeffery1922}
\bibinfo{author}{Jeffery, G.~B.}
\newblock \bibinfo{title}{The motion of ellipsoidal particles immersed in a viscous fluid}.
\newblock \emph{\bibinfo{journal}{Proceedings of the Royal Society of London. Series A}} \textbf{\bibinfo{volume}{102}}, \bibinfo{pages}{161--179} (\bibinfo{year}{1922}).

\bibitem{Ezhilan2013}
\bibinfo{author}{Ezhilan, B.}, \bibinfo{author}{Shelley, M.~J.} \& \bibinfo{author}{Saintillan, D.}
\newblock \bibinfo{title}{Instabilities and nonlinear dynamics of concentrated active suspensions}.
\newblock \emph{\bibinfo{journal}{Physics of Fluids}} \textbf{\bibinfo{volume}{25}}, \bibinfo{pages}{070607} (\bibinfo{year}{2013}).

\bibitem{saintillan2015theory}
\bibinfo{author}{Saintillan, D.} \& \bibinfo{author}{Shelley, M.~J.}
\newblock \bibinfo{title}{Theory of active suspensions}.
\newblock In \emph{\bibinfo{booktitle}{Complex Fluids in Biological Systems}}, \bibinfo{pages}{319--355} (\bibinfo{publisher}{Springer}, \bibinfo{year}{2015}).

\bibitem{JULICHER20073}
\bibinfo{author}{J{\"u}licher, F.}, \bibinfo{author}{Kruse, K.}, \bibinfo{author}{Prost, J.} \& \bibinfo{author}{Joanny, J.-F.}
\newblock \bibinfo{title}{Active behavior of the cytoskeleton}.
\newblock \emph{\bibinfo{journal}{Physics Reports}} \textbf{\bibinfo{volume}{449}}, \bibinfo{pages}{3--28} (\bibinfo{year}{2007}).

\bibitem{Nem_polar_fluid_surf}
\bibinfo{author}{Salbreux, G.}, \bibinfo{author}{J\"ulicher, F.}, \bibinfo{author}{Prost, J.} \& \bibinfo{author}{Callan-Jones, A.}
\newblock \bibinfo{title}{Theory of nematic and polar active fluid surfaces}.
\newblock \emph{\bibinfo{journal}{Phys. Rev. Res.}} \textbf{\bibinfo{volume}{4}}, \bibinfo{pages}{033158} (\bibinfo{year}{2022}).

\bibitem{Jülicher_2018}
\bibinfo{author}{Jülicher, F.}, \bibinfo{author}{Grill, S.~W.} \& \bibinfo{author}{Salbreux, G.}
\newblock \bibinfo{title}{Hydrodynamic theory of active matter}.
\newblock \emph{\bibinfo{journal}{Reports on Progress in Physics}} \textbf{\bibinfo{volume}{81}}, \bibinfo{pages}{076601} (\bibinfo{year}{2018}).

\bibitem{beris1994thermodynamics}
\bibinfo{author}{Beris, A.~N.} \& \bibinfo{author}{Edwards, B.~J.}
\newblock \emph{\bibinfo{title}{Thermodynamics of Flowing Systems: with Internal Microstructure}} (\bibinfo{publisher}{Oxford University Press}, \bibinfo{year}{1994}).

\bibitem{Oza_2016}
\bibinfo{author}{Oza, A.~U.} \& \bibinfo{author}{Dunkel, J.}
\newblock \bibinfo{title}{Antipolar ordering of topological defects in active liquid crystals}.
\newblock \emph{\bibinfo{journal}{New Journal of Physics}} \textbf{\bibinfo{volume}{18}}, \bibinfo{pages}{093006} (\bibinfo{year}{2016}).

\bibitem{Microtubule_nematic}
\bibinfo{author}{Hardo{\"u}in, J.} \emph{et~al.}
\newblock \bibinfo{title}{Reconfigurable flows and defect landscape of confined active nematics}.
\newblock \emph{\bibinfo{journal}{Communications Physics}} \textbf{\bibinfo{volume}{2}}, \bibinfo{pages}{121} (\bibinfo{year}{2019}).

\bibitem{bacterial_colony_gnf}
\bibinfo{author}{Zhang, H.~P.}, \bibinfo{author}{Be'er, A.}, \bibinfo{author}{Florin, E.-L.} \& \bibinfo{author}{Swinney, H.~L.}
\newblock \bibinfo{title}{Collective motion and density fluctuations in bacterial colonies}.
\newblock \emph{\bibinfo{journal}{Proceedings of the National Academy of Sciences}} \textbf{\bibinfo{volume}{107}}, \bibinfo{pages}{13626--13630} (\bibinfo{year}{2010}).

\bibitem{mermin1966absence}
\bibinfo{author}{Mermin, N.~D.} \& \bibinfo{author}{Wagner, H.}
\newblock \bibinfo{title}{Absence of ferromagnetism or antiferromagnetism in one-or two-dimensional isotropic heisenberg models}.
\newblock \emph{\bibinfo{journal}{Phys. Rev. Lett.}} \textbf{\bibinfo{volume}{17}}, \bibinfo{pages}{1133} (\bibinfo{year}{1966}).

\bibitem{tasaki2020hohenberg}
\bibinfo{author}{Tasaki, H.}
\newblock \bibinfo{title}{Hohenberg-mermin-wagner-type theorems for equilibrium models of flocking}.
\newblock \emph{\bibinfo{journal}{Phys. Rev. Lett.}} \textbf{\bibinfo{volume}{125}}, \bibinfo{pages}{220601} (\bibinfo{year}{2020}).

\bibitem{andersen2023symmetry}
\bibinfo{author}{Andersen, B.~H.}, \bibinfo{author}{Renaud, J.}, \bibinfo{author}{R{\o}nning, J.}, \bibinfo{author}{Angheluta, L.} \& \bibinfo{author}{Doostmohammadi, A.}
\newblock \bibinfo{title}{Symmetry-restoring crossover from defect-free to defect-laden turbulence in polar active matter}.
\newblock \emph{\bibinfo{journal}{Physical Review Fluids}} \textbf{\bibinfo{volume}{8}}, \bibinfo{pages}{063101} (\bibinfo{year}{2023}).

\bibitem{pisegna2024emergent}
\bibinfo{author}{Pisegna, G.}, \bibinfo{author}{Saha, S.} \& \bibinfo{author}{Golestanian, R.}
\newblock \bibinfo{title}{Emergent polar order in nonpolar mixtures with nonreciprocal interactions}.
\newblock \emph{\bibinfo{journal}{Proceedings of the National Academy of Sciences}} \textbf{\bibinfo{volume}{121}}, \bibinfo{pages}{e2407705121} (\bibinfo{year}{2024}).

\bibitem{Ramaswamy_2003_Giant_Number_Fluc}
\bibinfo{author}{Ramaswamy, S.}, \bibinfo{author}{Simha, R.~A.} \& \bibinfo{author}{Toner, J.}
\newblock \bibinfo{title}{Active nematics on a substrate: Giant number fluctuations and long-time tails}.
\newblock \emph{\bibinfo{journal}{Europhysics Letters (EPL)}} \textbf{\bibinfo{volume}{62}}, \bibinfo{pages}{196–202} (\bibinfo{year}{2003}).

\bibitem{Toner_2019}
\bibinfo{author}{Toner, J.}
\newblock \bibinfo{title}{Giant number fluctuations in dry active polar fluids: A shocking analogy with lightning rods}.
\newblock \emph{\bibinfo{journal}{The Journal of Chemical Physics}} \textbf{\bibinfo{volume}{150}}, \bibinfo{pages}{154120} (\bibinfo{year}{2019}).

\bibitem{Densityfluctuations_pnas_2013}
\bibinfo{author}{Schaller, V.} \& \bibinfo{author}{Bausch, A.~R.}
\newblock \bibinfo{title}{Topological defects and density fluctuations in collectively moving systems}.
\newblock \emph{\bibinfo{journal}{Proceedings of the National Academy of Sciences}} \textbf{\bibinfo{volume}{110}}, \bibinfo{pages}{4488--4493} (\bibinfo{year}{2013}).

\bibitem{rønning2023spontaneousflowsdynamicsfullinteger}
\bibinfo{author}{R{\o}nning, J.}, \bibinfo{author}{Renaud, J.}, \bibinfo{author}{Doostmohammadi, A.} \& \bibinfo{author}{Angheluta, L.}
\newblock \bibinfo{title}{Spontaneous flows and dynamics of full-integer topological defects in polar active matter}.
\newblock \emph{\bibinfo{journal}{Soft Matter}} \textbf{\bibinfo{volume}{19}}, \bibinfo{pages}{7513--7527} (\bibinfo{year}{2023}).

\bibitem{Kruse_Rotating_Asters_2004}
\bibinfo{author}{Kruse, K.}, \bibinfo{author}{Joanny, J.~F.}, \bibinfo{author}{J\"ulicher, F.}, \bibinfo{author}{Prost, J.} \& \bibinfo{author}{Sekimoto, K.}
\newblock \bibinfo{title}{Asters, vortices, and rotating spirals in active gels of polar filaments}.
\newblock \emph{\bibinfo{journal}{Phys. Rev. Lett.}} \textbf{\bibinfo{volume}{92}}, \bibinfo{pages}{078101} (\bibinfo{year}{2004}).

\bibitem{ardavseva2022topological}
\bibinfo{author}{Arda{\v{s}}eva, A.} \& \bibinfo{author}{Doostmohammadi, A.}
\newblock \bibinfo{title}{Topological defects in biological matter}.
\newblock \emph{\bibinfo{journal}{Nature Reviews Physics}} \textbf{\bibinfo{volume}{4}}, \bibinfo{pages}{354--356} (\bibinfo{year}{2022}).

\bibitem{vafa2023active}
\bibinfo{author}{Vafa, F.}, \bibinfo{author}{Nelson, D.~R.} \& \bibinfo{author}{Doostmohammadi, A.}
\newblock \bibinfo{title}{Active topological defect absorption by a curvature singularity}.
\newblock \emph{\bibinfo{journal}{Journal of Physics: Condensed Matter}} \textbf{\bibinfo{volume}{35}}, \bibinfo{pages}{425101} (\bibinfo{year}{2023}).

\bibitem{tan2019topological}
\bibinfo{author}{Tan, A.~J.} \emph{et~al.}
\newblock \bibinfo{title}{Topological chaos in active nematics}.
\newblock \emph{\bibinfo{journal}{Nature Physics}} \textbf{\bibinfo{volume}{15}}, \bibinfo{pages}{1033--1039} (\bibinfo{year}{2019}).

\bibitem{serra2023defect}
\bibinfo{author}{Serra, M.}, \bibinfo{author}{Lemma, L.}, \bibinfo{author}{Giomi, L.}, \bibinfo{author}{Dogic, Z.} \& \bibinfo{author}{Mahadevan, L.}
\newblock \bibinfo{title}{Defect-mediated dynamics of coherent structures in active nematics}.
\newblock \emph{\bibinfo{journal}{Nature Physics}} \textbf{\bibinfo{volume}{19}}, \bibinfo{pages}{1355--1361} (\bibinfo{year}{2023}).

\bibitem{kawaguchi2017topological}
\bibinfo{author}{Kawaguchi, K.}, \bibinfo{author}{Kageyama, R.} \& \bibinfo{author}{Sano, M.}
\newblock \bibinfo{title}{Topological defects control collective dynamics in neural progenitor cell cultures}.
\newblock \emph{\bibinfo{journal}{Nature}} \textbf{\bibinfo{volume}{545}}, \bibinfo{pages}{327--331} (\bibinfo{year}{2017}).

\bibitem{MaroudasSacks2020}
\bibinfo{author}{Maroudas-Sacks, Y.} \emph{et~al.}
\newblock \bibinfo{title}{Topological defects in the nematic order of actin fibres as organization centres of hydra morphogenesis}.
\newblock \emph{\bibinfo{journal}{Nature Physics}} \textbf{\bibinfo{volume}{17}}, \bibinfo{pages}{251–259} (\bibinfo{year}{2020}).

\bibitem{copenhagen2021topological}
\bibinfo{author}{Copenhagen, K.}, \bibinfo{author}{Alert, R.}, \bibinfo{author}{Wingreen, N.~S.} \& \bibinfo{author}{Shaevitz, J.~W.}
\newblock \bibinfo{title}{Topological defects promote layer formation in myxococcus xanthus colonies}.
\newblock \emph{\bibinfo{journal}{Nature Physics}} \textbf{\bibinfo{volume}{17}}, \bibinfo{pages}{211--215} (\bibinfo{year}{2021}).

\bibitem{doostmohammadi2022physics}
\bibinfo{author}{Doostmohammadi, A.} \& \bibinfo{author}{Ladoux, B.}
\newblock \bibinfo{title}{Physics of liquid crystals in cell biology}.
\newblock \emph{\bibinfo{journal}{Trends in cell biology}} \textbf{\bibinfo{volume}{32}}, \bibinfo{pages}{140--150} (\bibinfo{year}{2022}).

\bibitem{AstersControllingExp}
\bibinfo{author}{Ross, T.} \emph{et~al.}
\newblock \bibinfo{title}{Controlling organization and forces in active matter through optically-defined boundaries}.
\newblock \emph{\bibinfo{journal}{Nature}} \textbf{\bibinfo{volume}{572}}, \bibinfo{pages}{224--229} (\bibinfo{year}{2019}).

\bibitem{FullIntegerTopologicalDefectsCellMonolayersEndresen}
\bibinfo{author}{Endresen, K.~D.}, \bibinfo{author}{Kim, M.}, \bibinfo{author}{Pittman, M.}, \bibinfo{author}{Chen, Y.} \& \bibinfo{author}{Serra, F.}
\newblock \bibinfo{title}{Topological defects of integer charge in cell monolayers}.
\newblock \emph{\bibinfo{journal}{Soft Matter}} \textbf{\bibinfo{volume}{17}}, \bibinfo{pages}{5878--5887} (\bibinfo{year}{2021}).

\bibitem{Ndlec1997SelforganizationOM}
\bibinfo{author}{N{\'e}d{\'e}lec, F.~J.}, \bibinfo{author}{Surrey, T.}, \bibinfo{author}{Maggs, A.~C.} \& \bibinfo{author}{Leibler, S.}
\newblock \bibinfo{title}{Self-organization of microtubules and motors}.
\newblock \emph{\bibinfo{journal}{Nature}} \textbf{\bibinfo{volume}{389}}, \bibinfo{pages}{305--308} (\bibinfo{year}{1997}).

\bibitem{Spontaneous_circulation_as}
\bibinfo{author}{Woodhouse, F.~G.} \& \bibinfo{author}{Goldstein, R.~E.}
\newblock \bibinfo{title}{Spontaneous circulation of confined active suspensions}.
\newblock \emph{\bibinfo{journal}{Phys. Rev. Lett.}} \textbf{\bibinfo{volume}{109}}, \bibinfo{pages}{168105} (\bibinfo{year}{2012}).

\bibitem{AsterControlSymTheo}
\bibinfo{author}{Ghosh, S.}, \bibinfo{author}{Joshi, C.}, \bibinfo{author}{Baskaran, A.} \& \bibinfo{author}{Hagan, M.~F.}
\newblock \bibinfo{title}{Spatiotemporal control of structure and dynamics in a polar active fluid}.
\newblock \emph{\bibinfo{journal}{Soft Matter}} \bibinfo{pages}{7059--7071} (\bibinfo{year}{2024}).

\bibitem{rorai2022coexistence}
\bibinfo{author}{Rorai, C.}, \bibinfo{author}{Toschi, F.} \& \bibinfo{author}{Pagonabarraga, I.}
\newblock \bibinfo{title}{Coexistence of active and hydrodynamic turbulence in two-dimensional active nematics}.
\newblock \emph{\bibinfo{journal}{Phys. Rev. Lett.}} \textbf{\bibinfo{volume}{129}}, \bibinfo{pages}{218001} (\bibinfo{year}{2022}).

\bibitem{Scaling_regimes_active_turbulence}
\bibinfo{author}{Mart\'{\i}nez-Prat, B.} \emph{et~al.}
\newblock \bibinfo{title}{Scaling regimes of active turbulence with external dissipation}.
\newblock \emph{\bibinfo{journal}{Phys. Rev. X}} \textbf{\bibinfo{volume}{11}}, \bibinfo{pages}{031065} (\bibinfo{year}{2021}).

\bibitem{saghatchi2022nematic}
\bibinfo{author}{Saghatchi, R.}, \bibinfo{author}{Yildiz, M.} \& \bibinfo{author}{Doostmohammadi, A.}
\newblock \bibinfo{title}{Nematic order condensation and topological defects in inertial active nematics}.
\newblock \emph{\bibinfo{journal}{Physical Review E}} \textbf{\bibinfo{volume}{106}}, \bibinfo{pages}{014705} (\bibinfo{year}{2022}).

\bibitem{3D_nematic_scaling}
\bibinfo{author}{Krajnik, {\v Z}.}, \bibinfo{author}{Kos, {\v Z}.} \& \bibinfo{author}{Ravnik, M.}
\newblock \bibinfo{title}{Spectral energy analysis of bulk three-dimensional active nematic turbulence}.
\newblock \emph{\bibinfo{journal}{Soft Matter}} \textbf{\bibinfo{volume}{16}}, \bibinfo{pages}{9059--9068} (\bibinfo{year}{2020}).

\bibitem{alert_Turbulence_2022}
\bibinfo{author}{Alert, R.}, \bibinfo{author}{Casademunt, J.} \& \bibinfo{author}{Joanny, J.-F.}
\newblock \bibinfo{title}{Active {Turbulence}}.
\newblock \emph{\bibinfo{journal}{Annual Review of Condensed Matter Physics}} \textbf{\bibinfo{volume}{13}}, \bibinfo{pages}{143--170} (\bibinfo{year}{2022}).

\bibitem{EFrey_New_Class_turbulence}
\bibinfo{author}{Bratanov, V.}, \bibinfo{author}{Jenko, F.} \& \bibinfo{author}{Frey, E.}
\newblock \bibinfo{title}{New class of turbulence in active fluids}.
\newblock \emph{\bibinfo{journal}{Proceedings of the National Academy of Sciences}} \textbf{\bibinfo{volume}{112}}, \bibinfo{pages}{15048--15053} (\bibinfo{year}{2015}).

\bibitem{EpithelialTurbulence}
\bibinfo{author}{Lin, S.-Z.}, \bibinfo{author}{Zhang, W.-Y.}, \bibinfo{author}{Bi, D.}, \bibinfo{author}{Li, B.} \& \bibinfo{author}{Feng, X.-Q.}
\newblock \bibinfo{title}{Energetics of mesoscale cell turbulence in two-dimensional monolayers}.
\newblock \emph{\bibinfo{journal}{Communications Physics}} \textbf{\bibinfo{volume}{4}}, \bibinfo{pages}{21} (\bibinfo{year}{2021}).

\bibitem{Turbulence_of_swarming_sperm}
\bibinfo{author}{Creppy, A.}, \bibinfo{author}{Praud, O.}, \bibinfo{author}{Druart, X.}, \bibinfo{author}{Kohnke, P.~L.} \& \bibinfo{author}{Plourabou\'e, F.}
\newblock \bibinfo{title}{Turbulence of swarming sperm}.
\newblock \emph{\bibinfo{journal}{Phys. Rev. E}} \textbf{\bibinfo{volume}{92}}, \bibinfo{pages}{032722} (\bibinfo{year}{2015}).

\bibitem{Turbulence_E_Coli_3D}
\bibinfo{author}{Peng, Y.}, \bibinfo{author}{Liu, Z.} \& \bibinfo{author}{Cheng, X.}
\newblock \bibinfo{title}{Imaging the emergence of bacterial turbulence: Phase diagram and transition kinetics}.
\newblock \emph{\bibinfo{journal}{Science Advances}} \textbf{\bibinfo{volume}{7}}, \bibinfo{pages}{eabd1240} (\bibinfo{year}{2021}).

\bibitem{giomi2015geometry}
\bibinfo{author}{Giomi, L.}
\newblock \bibinfo{title}{Geometry and topology of turbulence in active nematics}.
\newblock \emph{\bibinfo{journal}{Physical Review X}} \textbf{\bibinfo{volume}{5}}, \bibinfo{pages}{031003} (\bibinfo{year}{2015}).

\bibitem{doostmohammadi2015celebrating}
\bibinfo{author}{Doostmohammadi, A.} \emph{et~al.}
\newblock \bibinfo{title}{Celebrating soft matter's 10th anniversary: Cell division: a source of active stress in cellular monolayers}.
\newblock \emph{\bibinfo{journal}{Soft Matter}} \textbf{\bibinfo{volume}{11}}, \bibinfo{pages}{7328--7336} (\bibinfo{year}{2015}).

\bibitem{basan2011undulation}
\bibinfo{author}{Basan, M.}, \bibinfo{author}{Joanny, J.-F.}, \bibinfo{author}{Prost, J.} \& \bibinfo{author}{Risler, T.}
\newblock \bibinfo{title}{Undulation instability of epithelial tissues}.
\newblock \emph{\bibinfo{journal}{Phys. Rev. Lett.}} \textbf{\bibinfo{volume}{106}}, \bibinfo{pages}{158101} (\bibinfo{year}{2011}).

\bibitem{comelles2021epithelial}
\bibinfo{author}{Comelles, J.} \emph{et~al.}
\newblock \bibinfo{title}{Epithelial colonies in vitro elongate through collective effects}.
\newblock \emph{\bibinfo{journal}{Elife}} \textbf{\bibinfo{volume}{10}}, \bibinfo{pages}{e57730} (\bibinfo{year}{2021}).

\bibitem{xu2023geometrical}
\bibinfo{author}{Xu, H.}, \bibinfo{author}{Nejad, M.~R.}, \bibinfo{author}{Yeomans, J.~M.} \& \bibinfo{author}{Wu, Y.}
\newblock \bibinfo{title}{Geometrical control of interface patterning underlies active matter invasion}.
\newblock \emph{\bibinfo{journal}{Proceedings of the National Academy of Sciences}} \textbf{\bibinfo{volume}{120}}, \bibinfo{pages}{e2219708120} (\bibinfo{year}{2023}).

\bibitem{adkins2022dynamics}
\bibinfo{author}{Adkins, R.} \emph{et~al.}
\newblock \bibinfo{title}{Dynamics of active liquid interfaces}.
\newblock \emph{\bibinfo{journal}{Science}} \textbf{\bibinfo{volume}{377}}, \bibinfo{pages}{768--772} (\bibinfo{year}{2022}).

\bibitem{doostmohammadi2016defect}
\bibinfo{author}{Doostmohammadi, A.}, \bibinfo{author}{Thampi, S.~P.} \& \bibinfo{author}{Yeomans, J.~M.}
\newblock \bibinfo{title}{Defect-mediated morphologies in growing cell colonies}.
\newblock \emph{\bibinfo{journal}{Phys. Rev. Lett.}} \textbf{\bibinfo{volume}{117}}, \bibinfo{pages}{048102} (\bibinfo{year}{2016}).

\bibitem{alert2022fingering}
\bibinfo{author}{Alert, R.}
\newblock \bibinfo{title}{Fingering instability of active nematic droplets}.
\newblock \emph{\bibinfo{journal}{Journal of Physics A: Mathematical and Theoretical}} \textbf{\bibinfo{volume}{55}}, \bibinfo{pages}{234009} (\bibinfo{year}{2022}).

\bibitem{blow2014biphasic}
\bibinfo{author}{Blow, M.~L.}, \bibinfo{author}{Thampi, S.~P.} \& \bibinfo{author}{Yeomans, J.~M.}
\newblock \bibinfo{title}{Biphasic, lyotropic, active nematics}.
\newblock \emph{\bibinfo{journal}{Phys. Rev. Lett.}} \textbf{\bibinfo{volume}{113}}, \bibinfo{pages}{248303} (\bibinfo{year}{2014}).

\bibitem{caballero2022activity}
\bibinfo{author}{Caballero, F.} \& \bibinfo{author}{Marchetti, M.~C.}
\newblock \bibinfo{title}{Activity-suppressed phase separation}.
\newblock \emph{\bibinfo{journal}{Phys. Rev. Lett.}} \textbf{\bibinfo{volume}{129}}, \bibinfo{pages}{268002} (\bibinfo{year}{2022}).

\bibitem{chaithanya2024transport}
\bibinfo{author}{Chaithanya, K.} \emph{et~al.}
\newblock \bibinfo{title}{Transport of topological defects in a biphasic mixture of active and passive nematic fluids}.
\newblock \emph{\bibinfo{journal}{Communications Physics}} \textbf{\bibinfo{volume}{7}}, \bibinfo{pages}{302} (\bibinfo{year}{2024}).

\bibitem{Topology_dyn_an_vesicles}
\bibinfo{author}{Keber, F.~C.} \emph{et~al.}
\newblock \bibinfo{title}{Topology and dynamics of active nematic vesicles}.
\newblock \emph{\bibinfo{journal}{Science}} \textbf{\bibinfo{volume}{345}}, \bibinfo{pages}{1135--1139} (\bibinfo{year}{2014}).

\bibitem{Tunable_structure}
\bibinfo{author}{Kumar, N.}, \bibinfo{author}{Zhang, R.}, \bibinfo{author}{de~Pablo, J.~J.} \& \bibinfo{author}{Gardel, M.~L.}
\newblock \bibinfo{title}{Tunable structure and dynamics of active liquid crystals}.
\newblock \emph{\bibinfo{journal}{Science Advances}} \textbf{\bibinfo{volume}{4}}, \bibinfo{pages}{eaat7779} (\bibinfo{year}{2018}).

\bibitem{snezhko2016complex}
\bibinfo{author}{Snezhko, A.}
\newblock \bibinfo{title}{Complex collective dynamics of active torque-driven colloids at interfaces}.
\newblock \emph{\bibinfo{journal}{Current opinion in colloid \& interface science}} \textbf{\bibinfo{volume}{21}}, \bibinfo{pages}{65--75} (\bibinfo{year}{2016}).

\bibitem{alert2019active}
\bibinfo{author}{Alert, R.}, \bibinfo{author}{Blanch-Mercader, C.} \& \bibinfo{author}{Casademunt, J.}
\newblock \bibinfo{title}{Active fingering instability in tissue spreading}.
\newblock \emph{\bibinfo{journal}{Phys. Rev. Lett.}} \textbf{\bibinfo{volume}{122}}, \bibinfo{pages}{088104} (\bibinfo{year}{2019}).

\bibitem{carenza2020soft}
\bibinfo{author}{Carenza, L.} \emph{et~al.}
\newblock \bibinfo{title}{Soft channel formation and symmetry breaking in exotic active emulsions}.
\newblock \emph{\bibinfo{journal}{Scientific Reports}} \textbf{\bibinfo{volume}{10}}, \bibinfo{pages}{15936} (\bibinfo{year}{2020}).

\bibitem{Nematic_isotropic_interfaces}
\bibinfo{author}{Coelho, R. C.~V.}, \bibinfo{author}{Ara{\'u}jo, N. A.~M.} \& \bibinfo{author}{Telo~da Gama, M.~M.}
\newblock \bibinfo{title}{Active nematic--isotropic interfaces in channels}.
\newblock \emph{\bibinfo{journal}{Soft Matter}} \textbf{\bibinfo{volume}{15}}, \bibinfo{pages}{6819--6829} (\bibinfo{year}{2019}).

\bibitem{fins2024steer}
\bibinfo{author}{Fins~Carreira, A.} \emph{et~al.}
\newblock \bibinfo{title}{How to steer active colloids up a vertical wall}.
\newblock \emph{\bibinfo{journal}{Nature communications}} \textbf{\bibinfo{volume}{15}}, \bibinfo{pages}{1710} (\bibinfo{year}{2024}).

\bibitem{kant2024bulk}
\bibinfo{author}{Kant, R.}, \bibinfo{author}{Gupta, R.~K.}, \bibinfo{author}{Soni, H.}, \bibinfo{author}{Sood, A.} \& \bibinfo{author}{Ramaswamy, S.}
\newblock \bibinfo{title}{Bulk condensation by an active interface}.
\newblock \emph{\bibinfo{journal}{Phys. Rev. Lett.}} \textbf{\bibinfo{volume}{133}}, \bibinfo{pages}{208301} (\bibinfo{year}{2024}).

\bibitem{droplet_2014}
\bibinfo{author}{A.~Whitfield, C.}, \bibinfo{author}{Marenduzzo, D.}, \bibinfo{author}{Voituriez, R.} \& \bibinfo{author}{J.~Hawkins, R.}
\newblock \bibinfo{title}{Active polar fluid flow in finite droplets}.
\newblock \emph{\bibinfo{journal}{The European Physical Journal E}} \textbf{\bibinfo{volume}{37}}, \bibinfo{pages}{8} (\bibinfo{year}{2014}).

\bibitem{symmetry_breaking_droplets}
\bibinfo{author}{Tjhung, E.}, \bibinfo{author}{Marenduzzo, D.} \& \bibinfo{author}{Cates, M.}
\newblock \bibinfo{title}{Spontaneous symmetry breaking in active droplets provides a generic route to motility}.
\newblock \emph{\bibinfo{journal}{Proceedings of the National Academy of Sciences of the United States of America}} \textbf{\bibinfo{volume}{109}}, \bibinfo{pages}{12381--6} (\bibinfo{year}{2012}).

\bibitem{Tjhung_2015}
\bibinfo{author}{Tjhung, E.}, \bibinfo{author}{Tiribocchi, A.}, \bibinfo{author}{Marenduzzo, D.} \& \bibinfo{author}{Cates, M.~E.}
\newblock \bibinfo{title}{A minimal physical model captures the shapes of crawling cells}.
\newblock \emph{\bibinfo{journal}{Nature Communications}} \textbf{\bibinfo{volume}{6}}, \bibinfo{pages}{5420} (\bibinfo{year}{2015}).

\bibitem{Droplet_active_nematics}
\bibinfo{author}{Gao, T.} \& \bibinfo{author}{Li, Z.}
\newblock \bibinfo{title}{Self-driven droplet powered by active nematics}.
\newblock \emph{\bibinfo{journal}{Phys. Rev. Lett.}} \textbf{\bibinfo{volume}{119}}, \bibinfo{pages}{108002} (\bibinfo{year}{2017}).

\bibitem{Zhou2014}
\bibinfo{author}{Zhou, S.}, \bibinfo{author}{Sokolov, A.}, \bibinfo{author}{Lavrentovich, O.~D.} \& \bibinfo{author}{Aranson, I.~S.}
\newblock \bibinfo{title}{Living liquid crystals}.
\newblock \emph{\bibinfo{journal}{Proceedings of the National Academy of Sciences}} \textbf{\bibinfo{volume}{111}}, \bibinfo{pages}{1265–1270} (\bibinfo{year}{2014}).

\bibitem{makhija2024topological}
\bibinfo{author}{Makhija, E.} \emph{et~al.}
\newblock \bibinfo{title}{Topological defects in self-assembled patterns of mesenchymal stromal cells in vitro are predictive attributes of condensation and chondrogenesis}.
\newblock \emph{\bibinfo{journal}{PLoS ONE}} \textbf{\bibinfo{volume}{19}}, \bibinfo{pages}{e0297769} (\bibinfo{year}{2024}).

\bibitem{Roostalu2018}
\bibinfo{author}{Roostalu, J.}, \bibinfo{author}{Rickman, J.}, \bibinfo{author}{Thomas, C.}, \bibinfo{author}{Nédélec, F.} \& \bibinfo{author}{Surrey, T.}
\newblock \bibinfo{title}{Determinants of polar versus nematic organization in networks of dynamic microtubules and mitotic motors}.
\newblock \emph{\bibinfo{journal}{Cell}} \textbf{\bibinfo{volume}{175}}, \bibinfo{pages}{796--808.e14} (\bibinfo{year}{2018}).

\bibitem{Thijssen2020}
\bibinfo{author}{Thijssen, K.} \& \bibinfo{author}{Doostmohammadi, A.}
\newblock \bibinfo{title}{Binding self-propelled topological defects in active turbulence}.
\newblock \emph{\bibinfo{journal}{Physical Review Research}} \textbf{\bibinfo{volume}{2}}, \bibinfo{pages}{042008} (\bibinfo{year}{2020}).

\bibitem{vafa2022defect}
\bibinfo{author}{Vafa, F.}
\newblock \bibinfo{title}{Defect dynamics in active polar fluids vs. active nematics}.
\newblock \emph{\bibinfo{journal}{Soft Matter}} \textbf{\bibinfo{volume}{18}}, \bibinfo{pages}{8087--8097} (\bibinfo{year}{2022}).

\bibitem{Chatterjee_2021}
\bibinfo{author}{Chatterjee, R.}, \bibinfo{author}{Rana, N.}, \bibinfo{author}{Simha, R.~A.}, \bibinfo{author}{Perlekar, P.} \& \bibinfo{author}{Ramaswamy, S.}
\newblock \bibinfo{title}{Inertia drives a flocking phase transition in viscous active fluids}.
\newblock \emph{\bibinfo{journal}{Physical Review X}} \textbf{\bibinfo{volume}{11}}, \bibinfo{pages}{031063} (\bibinfo{year}{2021}).

\bibitem{Eckert2023}
\bibinfo{author}{Eckert, J.}, \bibinfo{author}{Ladoux, B.}, \bibinfo{author}{Mège, R.-M.}, \bibinfo{author}{Giomi, L.} \& \bibinfo{author}{Schmidt, T.}
\newblock \bibinfo{title}{Hexanematic crossover in epithelial monolayers depends on cell adhesion and cell density}.
\newblock \emph{\bibinfo{journal}{Nature Communications}} \textbf{\bibinfo{volume}{14}}, \bibinfo{pages}{5762} (\bibinfo{year}{2023}).

\bibitem{Sebastin2023}
\bibinfo{author}{Sebastián, N.} \emph{et~al.}
\newblock \bibinfo{title}{Polarization patterning in ferroelectric nematic liquids via flexoelectric coupling}.
\newblock \emph{\bibinfo{journal}{Nature Communications}} \textbf{\bibinfo{volume}{14}}, \bibinfo{pages}{3029} (\bibinfo{year}{2023}).

\bibitem{Rudquist2021}
\bibinfo{author}{Rudquist, P.}
\newblock \bibinfo{title}{Revealing the polar nature of a ferroelectric nematic by means of circular alignment}.
\newblock \emph{\bibinfo{journal}{Scientific Reports}} \textbf{\bibinfo{volume}{11}}, \bibinfo{pages}{24411} (\bibinfo{year}{2021}).

\bibitem{Chen2020}
\bibinfo{author}{Chen, X.} \emph{et~al.}
\newblock \bibinfo{title}{First-principles experimental demonstration of ferroelectricity in a thermotropic nematic liquid crystal: Polar domains and striking electro-optics}.
\newblock \emph{\bibinfo{journal}{Proceedings of the National Academy of Sciences}} \textbf{\bibinfo{volume}{117}}, \bibinfo{pages}{14021–14031} (\bibinfo{year}{2020}).

\bibitem{flexonem3}
\bibinfo{author}{Paik, L.} \& \bibinfo{author}{Selinger, J.~V.}
\newblock \bibinfo{title}{Flexoelectricity versus electrostatics in polar nematic liquid crystals}.
\newblock \emph{\bibinfo{journal}{arXiv preprint arXiv:2408.10347}}  (\bibinfo{year}{2024}).

\bibitem{giomi2022hydrodynamic}
\bibinfo{author}{Giomi, L.}, \bibinfo{author}{Toner, J.} \& \bibinfo{author}{Sarkar, N.}
\newblock \bibinfo{title}{Hydrodynamic theory of p-atic liquid crystals}.
\newblock \emph{\bibinfo{journal}{Physical Review E}} \textbf{\bibinfo{volume}{106}}, \bibinfo{pages}{024701} (\bibinfo{year}{2022}).

\bibitem{fvafastrings}
\bibinfo{author}{Vafa, F.} \& \bibinfo{author}{Doostmohammadi, A.}
\newblock \bibinfo{title}{Phase diagram, confining strings, and a new universality class in nematopolar matter}.
\newblock \emph{\bibinfo{journal}{arXiv preprint arXiv:2501.04769}}  (\bibinfo{year}{2025}).

\bibitem{Ruider2024}
\bibinfo{author}{Ruider, I.} \emph{et~al.}
\newblock \bibinfo{title}{Topological excitations govern ordering kinetics in endothelial cell layers}.
\newblock \emph{\bibinfo{journal}{bioRxiv 2024.09.26.615134}}  (\bibinfo{year}{2024}).

\bibitem{james2011phase}
\bibinfo{author}{James, A.} \& \bibinfo{author}{Lamacraft, A.}
\newblock \bibinfo{title}{Phase diagram of two-dimensional polar condensates in a magnetic field}.
\newblock \emph{\bibinfo{journal}{Physical Review Letters}} \textbf{\bibinfo{volume}{106}}, \bibinfo{pages}{140402} (\bibinfo{year}{2011}).

\bibitem{strings3}
\bibinfo{author}{Mishra, P.~K.}, \bibinfo{author}{Mondal, P.~S.}, \bibinfo{author}{Jena, P.} \& \bibinfo{author}{Mishra, S.}
\newblock \bibinfo{title}{String formation and arrested ordering kinetics in nematics induced by polar particles}.
\newblock \emph{\bibinfo{journal}{arXiv preprint arXiv:2502.15477}}  (\bibinfo{year}{2025}).

\end{thebibliography}

\end{document}